\documentclass[floatfix,nofootinbib,twocolumn,preprintnumbers,superscriptaddress,notitlepage,nolongbibliography,aps,prd]{revtex4-2}

\usepackage{times}
\usepackage{xcolor}
\usepackage{listings}
\usepackage{amsfonts}
\usepackage{amsmath, amssymb}
\usepackage{orcidlink}
\usepackage{multirow}
\usepackage{amssymb}
\usepackage{bm}
\usepackage{dcolumn}
\usepackage{enumerate}
\usepackage{epsfig}
\usepackage{graphicx}
\usepackage{graphics}
\usepackage[utf8]{inputenc}
\usepackage{latexsym}
\usepackage{rotating}
\usepackage{hyperref}
\usepackage{cleveref}
\usepackage{subfig}
\usepackage[normalem]{ulem}

\usepackage{booktabs}
\usepackage{multirow}

\usepackage{xspace}

\usepackage[T1]{fontenc}
\usepackage{upquote}
\usepackage{etoolbox} 

\nonstopmode

\definecolor{bg}{rgb}{0.95,0.95,0.95}
\definecolor{keyword}{rgb}{0,0,0}
\definecolor{string}{rgb}{0.8,0,0}
\definecolor{comment}{rgb}{0,0.5,0}

\robustify{\texttt}
\let\originaltexttt\texttt

\begingroup
\catcode`'=\active
\catcode``=\active
\makeatletter
\renewrobustcmd{\texttt}[1]{%
   {%
   \everyeof{\noexpand}\endlinechar-1
   \expandafter\catcode\string``=\active
   \expandafter\catcode\string`'=\active
   \let'\textquotesingle
   \let`\textasciigrave
   \ifx\encodingdefault\upquote@OTone
    \ifx\ttdefault\upquote@cmtt
     \def'{\char13 }\def`{\char18 }%
    \fi
   \fi
   \scantokens{\originaltexttt{#1}}%
   }%
}%
\endgroup

\newcommand{\aap}{Astronomy and Astrophysics}
\newcommand{\aj}{Astronomical Journal}

\newcommand{\cN}{\mathcal{N}}
\newcommand{\cO}{\mathcal{O}}

\newcommand{\cM}{\mathcal{M}}
\newcommand{\cG}{\mathcal{G}}

\newcommand{\Om}{\Omega_{\rm m}}
\newcommand{\Ok}{\Omega_{\rm k}}
\newcommand{\dd}{{\rm d}}

\definecolor{mgreen}{rgb}{0.1,0.7,0.1}

\begin{document}

\title{Shape of U: Measuring the Curvature of the Universe with Gravitational Waves}
\author{Arindam Sharma \orcidlink{0009-0005-6546-5567}}
\email{asharm17@go.olemiss.edu}
\affiliation{Department of Physics and Astronomy, The University of Mississippi, University, Mississippi 38677, USA}

\author{Ish Gupta \orcidlink{0000-0001-6932-8715}}
\email{ishgupta@berkeley.edu}
\affiliation{Department of Physics, University of California, Berkeley, CA 94720, USA}
\affiliation{Department of Physics and Astronomy, Northwestern University, 2145 Sheridan Road, Evanston, IL 60208, USA}
\affiliation{Center for Interdisciplinary Exploration and Research in Astrophysics (CIERA), Northwestern University, 1800 Sherman Ave, Evanston, IL 60201, USA}

\author{Anuradha Gupta \orcidlink{0000-0002-5441-9013}}
\email{agupta1@olemiss.edu}
\affiliation{Department of Physics and Astronomy, The University of Mississippi, University, Mississippi 38677, USA}

\begin{abstract}
Gravitational waves (GWs) from compact binary mergers are standard sirens that can measure distances across the Universe without external calibrators. When an electromagnetic counterpart enables an independent redshift measurement, such ``bright sirens'' can be used to probe the expansion history of the Universe and constrain cosmological models. In this work, we investigate the ability of future GW observatories to measure the spatial curvature parameter, $\Omega_{\rm k}$, in a non-flat $\Lambda$CDM cosmology. We focus on intermediate-mass binary black hole mergers (with masses similar to GW231123) as bright siren sources, motivated by their detectability to high redshifts with next-generation ground-based detectors and by the possibility that mergers in active galactic nucleus disks may produce electromagnetic counterparts. Using Fisher matrix forecasts, we find that a network consisting of two Cosmic Explorer detectors and Einstein Telescope can constrain $\Omega_{\rm k}$ to a $1\sigma$ uncertainty of $0.029$ with these bright sirens. We further show that multiband observations with LISA or the Lunar Gravitational Wave Antenna do not significantly improve these cosmological constraints, because the additional signal-to-noise ratios accumulated in their bands are modest. Further, a population of binary neutron stars as bright sirens provides substantially broader constraints on $\Omega_{\rm k}$, with $1\sigma$ error of $0.055$. Our results show that bright intermediate-mass binary black hole and binary neutron star mergers observed with next-generation GW detectors together can provide an independent and informative probe of spatial curvature, with systematics distinct from those of other cosmological observations.
\end{abstract}

\maketitle

\section{Introduction}
\label{sec:intro}

We are currently in an era of precision cosmology, with multiple large-scale surveys placing strong constraints on the standard Lambda cold dark matter ($\Lambda$CDM) model of cosmology. In this framework, the Universe undergoes an early inflationary phase followed by radiation- and matter-dominated eras, with the present-day expansion driven by dark energy. Dark matter is modeled as a pressureless collisionless component, while dark energy is described by a cosmological constant, $\Lambda$. Measurements of the cosmic microwave background (CMB) from Planck \cite{Planck:2018vyg}, ACT \cite{AtacamaCosmologyTelescope:2025blo}, and SPT-3G \cite{SPT-3G:2025bzu}, baryon acoustic oscillations (BAO) from SDSS \cite{eBOSS:2021pff} and DESI \cite{DESI:2025zgx}, weak-lensing surveys such as KiDS-1000 \cite{KiDS:2020suj}, DES \cite{DES:2021bvc}, and HSC \cite{Li:2023tui}, and Type-Ia supernova observations from SH0ES \cite{Riess:2021jrx} have together constrained several cosmological parameters at the percent level~\cite{DiValentino:2025otz}.

Despite its overall success, the $\Lambda$CDM model exhibits important observational tensions.
The most prominent is the discrepancy between early- and late-Universe measurements of the Hubble constant, $H_0$, which currently stands at the level of $\sim 4\sigma$--$6\sigma$ depending on the datasets considered~\cite{Abdalla:2022yfr,DiValentino:2021izs}. Another tension concerns the amplitude of matter clustering in the late Universe, which is characterized by the $S_8$ parameter. Weak lensing measurements from DES \cite{DES:2021bvc} and HSC \cite{Li:2023tui} have consistently favored a lower value of $S_8$ than those inferred within $\Lambda$CDM from CMB observations, while KiDS-1000 \cite{KiDS:2020suj} and eROSITA \cite{Chiu:2022qgb} find weaker or no evidence for such a discrepancy~\cite{DiValentino:2025otz}. Whether these tensions point to new physics or unresolved systematics remains unclear, which motivates independent probes of cosmology that rely on different observables and have different sources of uncertainty. 

Gravitational waves (GWs) from compact binary mergers provide one such independent way of measuring cosmological parameters. Compact binary mergers are often termed {\it standard sirens} since the observed GW strain amplitude from the source and the rate of change of the signal's frequency can be used to infer both the apparent and intrinsic luminosities of the source and, hence, the luminosity distance $(D_L)$~\cite{Schutz:1986gp, Schutz:2001re, Holz:2005df}. Standard sirens are especially useful because they do not require calibration like the cosmic distance ladder, lowering the probability of systematic errors. 
The main challenge is that the GW signal alone does not provide a direct redshift measurement. However, the redshift, $z$, can be obtained if the merger is associated with an electromagnetic (EM) counterpart that enables host-galaxy identification. Such events are known as {\it bright sirens}.
The classic example of a bright siren is the binary neutron star (BNS) merger GW170817~\cite{LIGOScientific:2017vwq}, whose association with a short gamma-ray burst and a kilonova~\cite{LIGOScientific:2017ync} led to the identification of its host galaxy. This multi-messenger observation provided the first measurement of $H_0$ using GWs, inferring $H_0 = 70^{+12}_{-8}\,\,\mathrm{km}\,\mathrm{s}^{-1}\,\mathrm{Mpc}^{-1}$~\cite{LIGOScientific:2017adf}. 

Binary black hole (BBH) mergers can also act as bright sirens if they occur in environments capable of producing an EM counterpart, for example, within active galactic nucleus (AGN) disks, where a flare could help identify the host galaxy, as has been suggested for GW190521~\cite{LIGOScientific:2020iuh,McKernan:2019hqs}; see, however, Ref.~\cite{Ashton:2020kyr}. In the more common case where no EM counterpart is identified, cosmological information can still be extracted statistically by cross-matching the GW localization volume with galaxy catalogs to construct a redshift distribution for potential hosts. This is the statistical siren, or {\it dark siren + galaxy catalog}, approach~\cite{Schutz:1986gp,Chen:2017rfc,DelPozzo:2011vcw}. It was first applied to GW170817 without using its uniquely identified host galaxy~\cite{LIGOScientific:2018gmd}, obtaining $H_0 = 76^{+48}_{-23}\,\mathrm{km}\,\mathrm{s}^{-1}\,\mathrm{Mpc}^{-1}$. The same statistical method was later applied to the BBH merger GW170814 \cite{LIGOScientific:2017ycc} using redshift information from DES~\cite{DES:2018gui}, yielding $H_0 = 75^{+40}_{-32}\,\mathrm{km}\,\mathrm{s}^{-1}\,\mathrm{Mpc}^{-1}$~\cite{DES:2019ccw}.

The Advanced LIGO~\cite{LIGOScientific:2014pky} and Advanced Virgo~\cite{VIRGO:2014yos} detector network has detected over 200 compact binary coalescences in GWTC-4~\cite{LIGOScientific:2025slb}, with the majority being BBHs.
Using the GLADE+ galaxy catalog~\cite{Dalya:2021ewn}, the LIGO-Virgo-KAGRA (LVK) Collaboration carried out the dark siren analysis with 141 GW candidates in their fiducial FullPop-4.0 population model and obtained $H_0 = 81.6^{+21.5}_{-15.9}\,\mathrm{km}\,\mathrm{s}^{-1}\,\mathrm{Mpc}^{-1}$~\cite{LIGOScientific:2025jau}.
When combined with the bright siren GW170817, the constraints improved to $H_0 = 76.6^{+13.0}_{-9.5}$ km s$^{-1}$ Mpc$^{-1}$, consistent within uncertainties with both the Planck \cite{Planck:2015fie,Planck:2018vyg} and SH0ES \cite{Riess:2020fzl} measurements. While the current GW-based cosmological constraints are still dominated by statistical uncertainty, they already demonstrate the viability of standard sirens as an independent cosmological probe.

This situation is expected to change substantially with next-generation (XG) GW observatories such as the Cosmic Explorer (CE)~\cite{Reitze:2019iox,Evans:2021gyd,Evans:2023euw,Gupta:2023lga} and the Einstein Telescope (ET)~\cite{Punturo:2010zz,Hild:2010id,Branchesi:2023mws,ET:2025xjr}. Owing to their greatly improved broadband and low-frequency sensitivity, these detectors will observe compact binaries to much larger distances and with substantially higher signal-to-noise ratios (SNRs), leading, in turn, to more precise $D_L$ measurements. Forecasts based on BBH dark sirens indicate that a network consisting of two CEs and ET could measure $H_0$ to $0.7\%$ precision and the dark matter density parameter $\Om$ to $9\%$ within one year of GW observation~\cite{Muttoni:2023prw}. A subset of these events will be localized well enough to be associated with a unique host galaxy, thereby functioning as \textit{golden} dark sirens. Such events are expected to individually yield percent-level constraints on $H_0$~\cite{Chen:2016tys,Nishizawa:2016ood,Borhanian:2020vyr,Gupta:2022fwd}. Due to improved localization ability, XG detectors will also significantly enhance the prospects for bright siren cosmology. BNS and neutron star-black hole mergers, with detected EM counterparts, are expected to provide sub-percent measurements of $H_0$~\cite{Chen:2017rfc,Gupta:2022fwd,Chen:2024gdn}.
More importantly, the increased redshift reach of XG observatories will extend standard siren measurements beyond the local Universe. With EM counterparts identified out to higher $z$, standard sirens become sensitive not only to $H_0$ but also to the broader shape of the $D_L$--$z$ relation, enabling constraints on parameters such as $\Om$ and the dark-energy equation of state~\cite{Belgacem:2019tbw,Dhani:2022ulg}. This raises the natural question of whether the same high-$z$ GW observations can also be used to probe the spatial curvature of the Universe.

The spatial curvature parameter $\Ok$ provides a direct test of the global geometry of the Universe and the cosmological model. Using CMB data alone, the Planck collaboration found $\Ok = -0.0106 \pm 0.0065$, while the addition of BAO measurements improved this constraint to $\Ok = 0.0007 \pm 0.0019$~\cite{Planck:2018vyg}, consistent with a flat Universe. Motivated by the high-$z$ reach of future standard siren observations, we investigate whether bright sirens observed with XG ground-based GW detectors such as CE and ET, together with multiband observations involving space-based detectors such as the Laser Interferometer Space Antenna (LISA)~\cite{LISA:2017pwj} and the Lunar Gravitational Wave Antenna (LGWA)~\cite{Ajith:2024mie}, can provide independent and informative constraints on $\Ok$.

 \citet{Califano:2022cmo} show that bright BNS mergers detected with ET can be utilised to measure $\Ok$ with $\mathcal{O}(0.1)$ uncertainty. In this work, we focus primarily on intermediate-mass binary black hole (IMBBH) mergers detected with a network of two CEs and ET, and compare the constraints on $\Ok$ with those obtained from a population of BNS mergers detected by the same network. IMBBHs, with total masses typically in the range $10^2$--$10^5\,M_\odot$, while comparatively louder, merge at substantially lower frequencies than stellar mass binaries. This makes the heaviest IMBBHs difficult to observe with current ground-based GW detectors, whose sensitivity declines rapidly below $\sim 10$ Hz. Future observatories like the CE and ET will improve the low-frequency $(<10\,{\rm Hz})$ sensitivity substantially, bringing a larger fraction of the IMBBH population into band. As a result, IMBBH mergers can be observed to much higher redshifts than is possible today, with the lower-mass end of the population remaining detectable out to $z\sim 10$, while nearer systems can be observed with network SNRs of $\mathcal{O}(10^2$--$10^3)$ and hence very precise luminosity-distance measurements~\cite{Gupta:2023lga,Reali:2024hqf}. Their low merger GW frequencies also cause them to remain in band for much longer in detectors such as LISA and LGWA, making them natural targets for multiband observations~\cite{Jani:2019ffg,Grimm:2020ivq,Muttoni:2021veo,Dong:2025ikq}.

The large masses of IMBBHs are important not only for detectability, but also for parameter inference. In standard siren measurements, $D_L$ is often strongly correlated with the binary's orbit inclination $(\iota)$, particularly for lower mass and nearly symmetric systems. In contrast, for heavier and/or more asymmetric binaries, the higher harmonic content of the signal is more readily measurable~\cite{Gupta:2025paz}, which helps break the $D_L-\iota$ degeneracy and improves the estimation of $D_L$. In addition, if a fraction of IMBBH mergers occur in dense gaseous environments such as AGN disks, they may be accompanied by EM emission and thus serve as bright sirens. The detectability at high redshifts, together with precise $D_L$ measurements and possible redshift information, makes IMBBHs especially promising for curvature parameter measurements.

In this work, we quantify how well bright IMBBH mergers can constrain $\Ok$ with XG GW observations. Motivated by GW231123, an exceptionally massive IMBBH with median total mass $\sim 236\,M_\odot$ detected at redshift $0.40^{+0.27}_{-0.25}$~\cite{LIGOScientific:2025rsn}, we adopt a GW231123-like system as our fiducial IMBBH and simulate a population of such mergers. 
This choice is further motivated by the large uncertainty in the population properties of IMBBHs, such as the mass distribution and merger rates~\cite{LIGOScientific:2021tfm}. We estimate the corresponding measurement uncertainties with the Fisher matrix framework implemented in \textsc{gwfish}~\cite{Dupletsa:2022scg}, considering both a ground-based XG network consisting of two CEs and ET and multiband extensions involving LISA or LGWA. Assuming that these IMBBH mergers act as bright sirens and provide precise redshift information through EM identification, we propagate the forecasted $D_L$ uncertainties into constraints on $H_0$, $\Om$, and $\Ok$. We also compare this performance against that of BNS bright sirens observed with the same ground-based XG network. While the merger rate and redshift distribution of IMBBH mergers is uncertain, our pragmatic estimate indicates that $74$ IMBBH bright sirens observed with two CEs and ET in a year can constrain $H_0$, $\Om$, and $\Ok$ to $0.69\%$, $9.76\%$, and $0.029$, respectively. By contrast, for the BNS population considered in this work, the corresponding uncertainties are $1.65\%$, $20.64\%$, and $0.055$. We further find that adding LISA or LGWA yields only modest gains in SNR and does not significantly improve the cosmological constraints. However, the inclusion of LGWA improves sky localization by roughly an order of magnitude, underscoring its potential importance for EM follow-up and host-galaxy identification. 

Our results show that bright IMBBH mergers detected with XG observatories can provide meaningful and independent constraints on $\Ok$. For the source populations considered here, IMBBH bright sirens constrain $\Ok$ about five times more tightly than BNS bright sirens. However, even with the most optimistic Madau-Dickinson-Belczynski-Ng (MDBN) merger rate~\cite{Madau:2014bja, Ng:2020qpk}, with $\sim400$ IMBBH events observed per year, the forecasted uncertainties in $\Ok$ are greater than current CMB+BAO constraints measured by Planck (2018)~\cite{Planck:2018vyg}. Nonetheless, IMBBH mergers can provide an independent measurement of spatial curvature of the Universe, thus serving an important role in the XG era.

It is important to note that we assume every IMBBH merger and BNS merger in their respective populations have a detectable EM counterpart. Thus, our constraints should be interpreted as those achievable by bright IMBBH/BNS mergers, and not necessarily by an astrophysical population of such binaries. However, GW and EM association becomes more realistic at low redshifts, and we have shown that the dominant contribution to the cosmological constraints comes from these low-redshift events.

The rest of the paper is organized as follows. In Sec.~\ref{sec:cosmology}, we describe the non-flat $\Lambda$CDM cosmology model and introduce its parameters. In Sec.~\ref{sec:networks}, we provide details of the detectors we considered in this study. In Sec.~\ref{sec:sources}, we describe our simulated population of IMBBHs and BNSs, including describing possible EM counterparts of IMBBH mergers in Sec.~\ref{sec:imbh_agn_disk}. Section~\ref{sec:methodology} discusses the Fisher information framework that we used to estimate the measurement uncertainties of the parameters. Section~\ref{sec:results} presents our results, and in Sec.~\ref{sec:concl}, we conclude the paper. 

\section{non-flat $\Lambda$CDM cosmology}
\label{sec:cosmology}

In the $\Lambda$CDM model of cosmology, the spacetime metric is described by the Friedmann-Lemaitre-Robertson-Walker (FLRW) metric,

\begin{equation}
 \dd s^2 = -c^2 \dd t^2 + a^2(t) \left[\frac{\dd r^2}{1 - k r^2} + r^2 \dd \Omega^2 \right],
 \label{eq:flrw}
\end{equation}
where $c$ is the speed of light, $t$ is the coordinate time, $r$ is the co-moving radial coordinate, $\dd\Omega^2 = \dd \theta^2 + \sin^2\theta\,\dd\phi^2$, and $a$ is the scale factor, quantifying the expansion of the Universe. Here, the parameter $k$ describes the geometry of space, which can be flat, spherical, or hyperbolic, corresponding to $k = 0,\ +1, \text{ or }-1$, respectively.

From Einstein's equations, we can derive the Friedmann equations, one of which is given by,
\begin{equation}
    H^2(z) = H_0^2\,[\Om \,(1+z)^3 + \Ok\,(1+z)^2 + \Omega_\Lambda],
    \label{eq:friedmann}
\end{equation}

where $H(z)$ is the Hubble parameter, $\Om$ is the matter density parameter, and $\Omega_{\Lambda}$ is the dark energy density parameter. A positive value of the curvature parameter $\Ok$ corresponds to a hyperbolic Universe, a negative $\Ok$ corresponds to a spherical Universe, and $\Ok=0$ corresponds to a flat Universe. 

Although standard compact binary waveforms are derived within the asymptotically flat and isolated source framework~\cite{Blanchet:2013haa}, they remain valid in a $\Lambda$CDM Universe because the source scale is negligible compared with the cosmological curvature scale. The effect of cosmology, at leading order, on the detected waveform is a propagation effect, encoded in $D_L$ and $z$.

For different values of $\Ok$, $D_L(z)$ is defined as,
\begin{equation}
    D_L(z) = \begin{cases} \frac{c(1+z)}{H_0 \sqrt{\Ok}} \sinh\left(\sqrt{\Ok} \chi(z)\right) & \text{if } \Ok > 0, \\ \frac{c(1+z)\chi(z)}{H_0} & \text{if } \Ok = 0,  \\ \frac{c(1+z)}{H_0 \sqrt{-\Ok}} \sin\left(\sqrt{-\Ok} \chi(z) \right) & \text{if } \Ok < 0,  \end{cases} 
    \label{eq:dL}
\end{equation}
where, $\chi(z) = \int \frac{H_0 \dd z}{H(z)}$. Thus, if the $D_L$ and $z$ are measured independently, constraints on cosmological parameters can be obtained using Eqs.~(\ref{eq:friedmann}) and (\ref{eq:dL}). 

\section{Detector Networks}
\label{sec:networks}
In this section, we introduce the GW detectors considered in this study. The details of detectors such as location, sensitivity range, and power spectral densities are summarized in Table~\ref{tab:detectors}.

\begin{table*}
    \centering
    \begin{ruledtabular}
        \begin{tabular}{l|lllll}
            Detector name & Latitude & Longitude & Orientation & PSD & Frequency range  \\
            \hline
             CE1 & $46^\circ0'0''$ & $-125^\circ0'0''$ & $350^\circ$ & 40 km CBO~\cite{Evans:2021gyd}  & $5-512$ Hz  \\
             CE2 & $29^\circ0'0''$ & $-94^\circ0'0''$ & $290^\circ$ & 20 km CBO~\cite{Evans:2021gyd} & $5-512$ Hz \\
             ET & $40^\circ 31'0''$ & $9^\circ 25'0''$ & $180^\circ$ & ET-D~\cite{Hild:2010id}  & $5-512$ Hz  \\
             LISA & N/A & N/A & N/A  & LISA~\cite{Babak:2021mhe} & $10^{-4} - 0.1$ Hz  \\
             LGWA &  N/A & N/A & N/A & LGWA-Si~\cite{Ajith:2024mie} & $10^{-3} - 4$ Hz  \\
        \end{tabular}
    \end{ruledtabular}
    \caption{Detector characteristics used for the study. The locations of LISA and LGWA are not available as they are non-terrestrial detectors. The orientation is the arm's position measured from North clockwise. Here, CBO stands for compact binary optimized, which is more sensitive at lower frequencies~\cite{Borhanian:2020ypi}. The frequency range listed is the one used in integrals to compute the noise-weighted inner product [Eq.~\eqref{eq:innerproduct}], for instance, and not necessarily the full frequency range of the detector's sensitivity.} 
    \label{tab:detectors}
\end{table*}

First, we describe the proposed terrestrial detectors. Cosmic Explorer is a proposed L-shaped detector with $40$ km arms on one site (referred to as CE1 here) and 20 km arms on the other site (referred to as CE2), to be built in the United States~\cite{Evans:2021gyd,Evans:2023euw}.  
Einstein Telescope, on the other hand, is a next-generation ground-based GW detector, proposed to be built in Europe. In this work, we consider ET to be a triangular-shaped observatory with $10$ km arms and is effectively three V-shaped detectors with an opening angle of $60^{\circ}$ in xylophone design~\cite{Branchesi:2023mws}.
These detectors will have $\sim10$ times better sensitivity than current-generation detectors and will detect mergers up to very high redshifts ($z \gtrsim 10$)~\cite{Gupta:2023lga,ET:2025xjr}. We report our results assuming LISA's mission lifetime to be $4$ years.

Next, we consider non-terrestrial detectors. Laser Interferometric Space Antenna (LISA)~\cite{LISA:2017pwj} is a space-based detector consisting of three spacecrafts arranged in a triangular formation (connected by six laser links) and separated by $2.5$ million km orbiting the Sun, trailing behind Earth's orbit at an inclination of $20^\circ$ with respect to the ecliptic. Set to be launched in 2035, it is sensitive in the frequency range of $10^{-4}-10^{-1}$ Hz and will detect both polarizations of GWs simultaneously. Further, the orbital motion of the spacecraft around the Sun is important for source localization and $D_L$ estimation~\cite{Cutler:1997ta}.

Lunar Gravitational Wave Antenna (LGWA)~\cite{LGWA:2020mma} is a proposed moon-based detector consisting of four stations deployed in a permanent shadow region at one of the lunar poles. It aims to detect GWs by monitoring the vibrational eigenmodes of the moon excited by GWs. It is sensitive in the frequency range of $\sim10^{-2} - 1$ Hz, complementing the sensitivity bands of CE, ET and LISA. 

Although the actual launch dates of all these detectors are somewhat uncertain, we investigate the possibility of observing IMBBHs with a network of two CEs (CE1, CE2) and ET, as well as with multiband observations including either LISA or LGWA. In multiband observation, the same source is observed by multiple detectors operating at completely different frequency ranges. For instance, LISA/LGWA operate at low frequencies and capture the early inspiral of stellar mass compact binaries, enabling precise measurement of the chirp mass (see, e.g.,~\cite{Iacovelli:2025kwn}). On the other hand, ET/CE operate at higher frequencies, (e.g., $5 - 2048$ Hz) and observe the late inspiral and merger of these binaries. Various studies of multiband parameter estimation and tests of general relativity have demonstrated that joint observations can improve measurement precision by several orders of magnitude compared to single-band detections~\cite{Barausse:2016eii,Carson:2019rda,Gnocchi:2019jzp,Grimm:2020ivq,Toubiana:2020vtf,Gupta:2020lxa,Datta:2020vcj}. This improvement arises primarily from the breaking of degeneracies between the chirp mass and other binary parameters when combining observations across bands. In addition, multiband observations are expected to yield improved sky localization due to the longer effective baseline and longer signal duration,
an effect that is clearly reflected in the right panel of Fig.~\ref{fig:snr_sky_locdist}. Hence, in this paper, we also investigate whether multiband observations of IMBBHs could provide improved bounds on cosmological parameters, especially $\Ok$. Specifically, we consider the following three network configurations in our analysis: (i) 2CE+ET, (ii) 2CE+ET+LISA, and (iii) 2CE+ET+LGWA. Throughout, we consider one year of GW observations. Our results can be readily rescaled to account for different observation times.

\section{Simulated Observations}
\label{sec:sources}
\subsection{IMBBH population}
\label{sec:imbh_pop} 
IMBBHs typically have masses in the range $10^2$-$10^5M_\odot$ and are thought to form in the centers of dense stellar clusters, or AGN disks~\cite{Miller:2003sc,2012MNRAS.425..460M, McKernan:2014oxa}. Owing to their large masses (compared to stellar mass BBHs), IMBBHs can be detected out to very high redshifts and therefore have strong potential for constraining cosmological parameters. Furthermore, IMBBHs achieve appreciable SNRs in the LISA- and LGWA-like detectors, making them ideal sources for multiband observations. As discussed in more detail in Sec.~\ref{sec:imbh_agn_disk}, if formed in dense stellar environments, IMBBH mergers may also be associated with EM flares, potentially providing host-galaxy identification and independent $z$ measurements, and hence serving as bright sirens for precision cosmography.

We consider an IMBBH population with the same component masses as that of GW231123, as it is the heaviest binary merger detected so far.
Owing to their suspected dynamical formation, we assume that the IMBBHs are precessing. The spin magnitudes $\chi_1$ and $\chi_2$ are sampled uniformly in the range $[0,\,0.99]$; the cosine of spin tilt angles, $\theta_1$ and $\theta_2$, are sampled uniformly from $[-1,1]$; the difference between the azimuthal angles of the individual spin vectors projected onto the orbital plane, $\phi_{12}$, and the difference between the azimuthal angles of the total and orbital angular momentum angles projected onto the orbital plane, $\phi_{JL}$ are sampled uniformly from $[0,\,2\pi]$. The extrinsic parameters of the binaries, such as right ascension ($\alpha$), declination ($\delta$), inclination angle ($\iota$), and polarization angle ($\psi$), are all sampled uniformly over their respective ranges. The coalescence time and phase, $t_c$ and $\phi_c$, are both set to zero. Table~\ref{tab:imbh_parameters} summarizes the parameters used for the IMBBH population.

To find the number of mergers up to a redshift $z$ or the merger rate, $R(z)$ of a binary population, one integrates the merger rate density, $\dot{n}(z)$ over a co-moving volume, which in the source-frame is multiplied by $(1+z)^{-1}$ to account for the expansion of the Universe,  
\begin{equation}
R(z) = R_0\,\int_0^z \frac{\dot{n}(z')}{\dot{n}(0)}\,\frac{1}{(1+z')}\frac{dV_c}{dz'}\,dz'\,.
\label{eq:merger_rate}
\end{equation}
Here, $R_0$ is the local merger rate and $dV_c/dz = 4\pi c\, d_c^2(z) / H(z)$ is the co-moving volume element, with $d_c$ as the co-moving distance. We use $R_0 = 0.08$ Gpc$^{-3}$ yr$^{-1}$ as reported in Ref.~\cite{LIGOScientific:2021tfm}. 

\begin{figure}
    \centering
    \includegraphics[width=\linewidth]{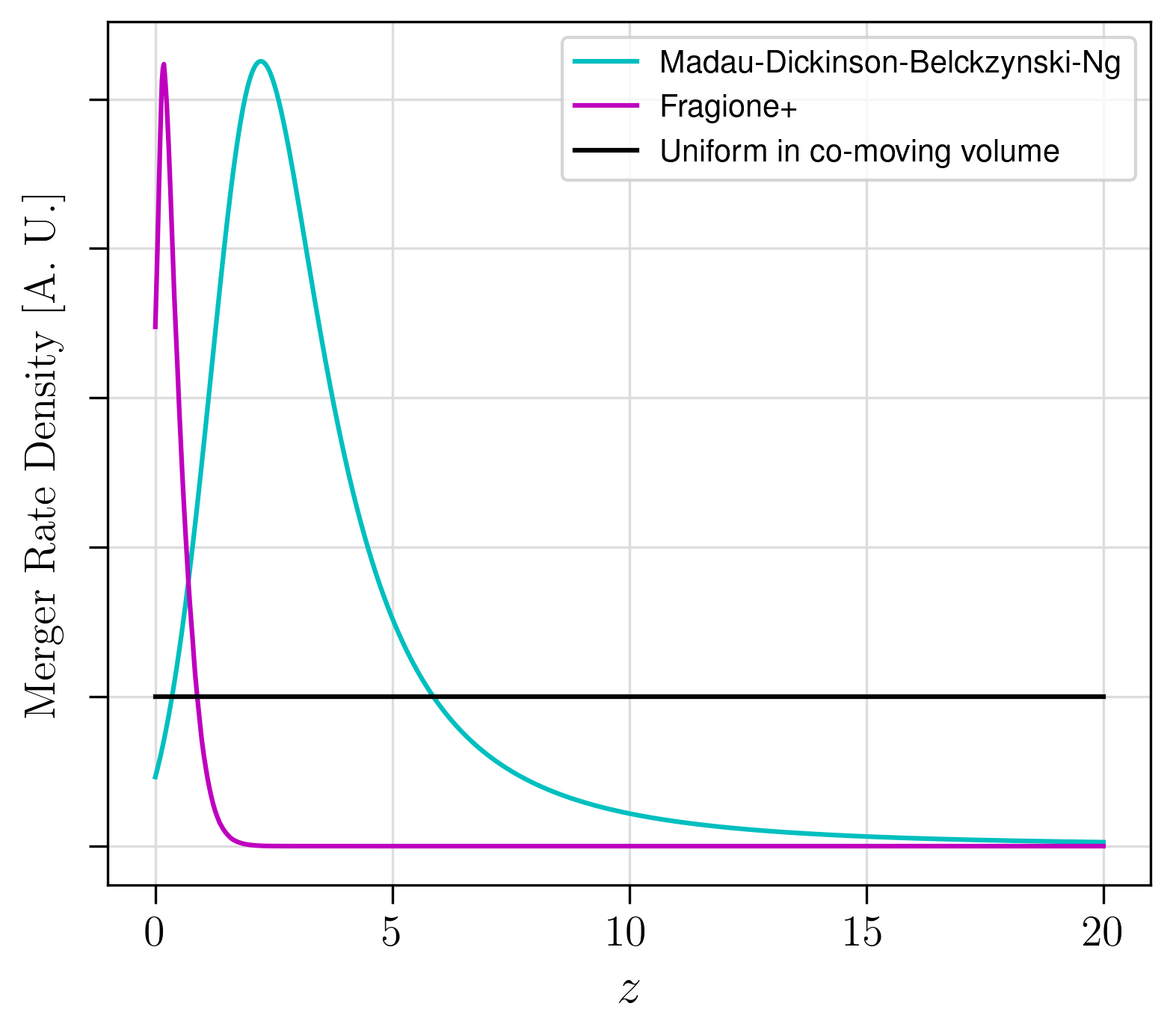}
    \caption{Merger rate densities used in this study.  Rates are shown in arbitrary units (A.U.) as they have been rescaled for illustrative comparison.}
    \label{fig:merger_rates}
\end{figure}

As the redshift distribution of IMBBHs is uncertain due to insufficient detections, we consider three different merger rate densities as illustrated in Fig.~\ref{fig:merger_rates}: 

\begin{itemize}
    \item Uniform in co-moving volume: 
    We assume a uniform merger rate density, ${\dot{n}(z)}/ \dot{n}(0) = 1$. We obtain a total of 76 IMBBH mergers for $z \in [0,\, 20]$ in one year of GW observation. This is our pragmatic scenario.
    \item MDBN~\cite{Madau:2014bja, Ng:2020qpk}: This merger rate density is typically used to describe stellar-mass BBHs, and serves as an optimistic rate for IMBBHs. It is given by,
    \begin{equation}
    \dot{n}(z) = a \frac{(1+z)^{b}}{1 + \left(c\,(1+z) \right)^{d}}\,,
    \end{equation}
    where, $a = 1,\ b = 2.57,\ c = 1/3.36,\ d = 5.83$~\cite{Ng:2020qpk}. We obtain a total of 414 IMBBHs per year in the redshift range of $[0, 20]$. 
    \item Fragione+~\cite{Fragione:2022avp}: Here, the merger rate is obtained using a numerical model of intermediate-mass black hole (IMBH) growth in nuclear star clusters. We obtain the merger rate density by fitting a logarithmic interpolant to the merger rate data of Ref.~\cite{Fragione:2022avp}. This choice corresponds to a pessimistic scenario, with only 4 IMBBH mergers per year in the redshift range of $[0, 20]$.

\end{itemize}

To convert the redshifts to luminosity distance, $D_L$, 
we use a fiducial cosmology from Planck (2018)~\cite{Planck:2018vyg} with $\Om = 0.31,\ H_0 = 67.66\ \text{km s}^{-1}\ \text{Mpc}^{-1},\ \Ok = 0,\ \Omega_\Lambda = 0.69$. 

\begin{table}[t] 
\centering
\caption{Table that lists the properties of the simulated GW231123-like IMBBH population for our study.}
    \begin{ruledtabular}
    \begin{tabular}{c|c}
        \textbf{Parameters} & \textbf{Values} \\
        \midrule
        $m_{1},\ m_{2}$ & $137M_\odot, 101M_\odot$\\
        $\chi_{1},\ \chi_{2}$ & Uniform in [0, 0.99] \\
        $\cos \theta_1, \cos \theta_2$ &  Uniform in [$-1,1$]\\
        $\phi_{12}$, $\phi_{JL}$ & Uniform in [$0,\,2\pi$]\\ 
        $z$ & MDBN, Uniform in co-moving volume, Fragione{+} \\
        $\cos \iota$ & Uniform in $[-1,\,1]$ \\
        $\alpha$ & Uniform in $[0,\,2\pi]$ \\
        $\cos\delta$ & Uniform in $[-1,\,1]$ \\
        $\psi$ & Uniform in $[0,\,2\pi]$ \\
        $t_{c}, \phi_{c}$ & $0,\,0$ \\
    \end{tabular}
    \end{ruledtabular}
\label{tab:imbh_parameters}
\end{table}

\subsection{Potential EM signatures of IMBBH mergers}
\label{sec:imbh_agn_disk}

Dense environments such as AGNs have been proposed as plausible sites where black holes can grow rapidly and form IMBHs~\cite{2012MNRAS.425..460M, McKernan:2014oxa}. A large number of stars and compact objects, known as nuclear cluster objects (NCOs), are expected to cluster around the central supermassive black holes in galaxies due to stellar evolution, dynamical friction, secular evolution, and minor mergers~\cite{2012MNRAS.425..460M}. Depending on the size of the accretion disk, a large fraction of these NCOs will align themselves with the accretion disk and act as IMBH seeds~\cite{syer_1991, Artymowicz:1993xz, Goodman:2003sf, 2012MNRAS.425..460M, McKernan:2014oxa, Yang:2019cbr}. Heavier black holes are also preferentially selected by the AGN. There are three reasons for this: (i) heavier black holes align with the disk more quickly, (ii) they accrete more matter, and (iii) more of them cross the AGN disk plane. Alternatively, some black holes can be formed in the disk itself~\cite{Levin:2006uc, Stone:2016wzz}. As this population of NCOs and IMBH seeds moves around in the disk, they are subjected to two competing effects: orbital excitation due to dynamical heating from cusp stars and orbital damping due to cooling by the gas in the AGN disk. Ref.~\cite{2012MNRAS.425..460M} shows that gas damping is the dominant effect, as a result of which the NCOs migrate inwards into the disk. There are migration traps between $20$ and $300$ times the Schwarzschild radius, where such NCOs can accumulate, accrete, and merge~\cite{Bellovary:2015ifg}. Gas damping also increases the collision cross-section between them, which has been shown to be much higher between NCOs in AGN disks than in star clusters~\cite{2012MNRAS.425..460M, McKernan:2014oxa, Bartos:2016dgn}. Thus, AGN disks are a more efficient environment for IMBH formation than star clusters through hierarchical mergers. Once the IMBH seed merges with nearby NCOs, gas accretion is the primary source of IMBH growth~\cite{2012MNRAS.425..460M}. 

It is reasonable to expect EM counterparts of BBH mergers in the AGN disks due to the interaction of the gas in the disk with the remnant. For instance, it can cause ram pressure stripping of the nearby gas in the disk~\cite{McKernan:2019hqs}. The post-merger black hole can launch outflows or jets as it accretes disk material~\cite{Ma:2024bfo}. Breakout EM emission from radiation-driven outflows has been investigated by Refs.~\cite{Wang:2021tfd, Kimura:2021xxu}, which mainly falls in the soft-X-ray band~\cite{Rodriguez-Ramirez:2024ikd}. AGN-assisted mergers can also produce jets by the Blandford-Znajek process, and their interaction with the AGN disk can produce specific EM signatures in infrared, optical, and X-ray bands~\cite{Tagawa:2023uqa, Rodriguez-Ramirez:2023ejf}. Ref.~\cite{Rodriguez-Ramirez:2024ikd} derived light curves for long-term emission in the ultraviolet and optical, which are produced when the outflow drives disk ejection, which later expands and cools outside the disk. However, it is important to note that AGNs can have their own intrinsic variability, which may be mistaken for an EM counterpart to a compact binary merger.

The IMBBH merger event GW190521 has been tentatively associated with the EM flare ZTF19abanrhr, with an odds ratio between $2-12$ depending on the waveform model used~\cite{Graham:2020gwr, Morton:2023wxg}. The odds ratios are not high enough to confidently associate the GW and EM events~\cite{Ashton:2020kyr}. For comparison, the EM counterpart of GW170817 was associated with an odds ratio of $10^6$~\cite{Ashton:2017ykh}. Although only one IMBBH merger has been tentatively identified as a bright siren candidate so far, future GW detections, particularly with improved localization and sensitivity, offer promising opportunities to identify EM counterparts to IMBBH mergers in dense environments like AGN disks. 

\subsection{BNS population}
\label{sec:bns_pop}

\begin{table}[t]
    \centering
    \caption{BNS injection parameters. We assume the spins of the neutron stars to be positively aligned with the orbital angular momentum.}
    \begin{ruledtabular}
    \begin{tabular}{c|c}
        \textbf{Parameters} & \textbf{Values} \\
        \midrule
        $m_{1},\ m_{2}$ & Double Gaussian with peaks at $1.35M_\odot$ and $1.8M_\odot$ \\
        $\chi_{1},\ \chi_{2}$ & Uniform in $[0,0.1]$\\
        $z$ & Madau-Dickinson star formation rate density \\
        $\cos(\iota)$ & Uniform in $[-1,1]$ \\
        $\alpha$ & Uniform in $[0,2\pi]$ \\
        $\cos(\delta)$ & Uniform in $[-1,1]$ \\
        $\psi$ & Uniform in $[0,2\pi]$ \\
        $t_{c}, \phi_{c}$ & $0,0$ \\
    \end{tabular}
    \end{ruledtabular}
    \label{tab:bns_injection_params}
\end{table}

In addition to a population of IMBBHs, we also consider a population of BNSs and assess its potential in constraining $\Ok$. BNS mergers produce powerful EM counterparts that produce emission in all bands of the EM spectrum~\cite{LIGOScientific:2017ync}, starting with short gamma-ray bursts (sGRBs) seconds after the merger~\cite{LIGOScientific:2017zic}, followed by ultraviolet, optical and infrared transient~\cite{Coulter:2017wya} produced hours and days after the merger, X-rays emitted days and years after the merger~\cite{Margutti:2017cjl, Hajela:2021faz}, and a radio afterglow, weeks, months or years after the merger~\cite{Hallinan:2017woc, Balasubramanian:2022sie}. 

We utilize publicly available data from Refs.~\cite{Gupta:2023lga,gupta_2023_8087733} for our forecasts. Ref.~\cite{Gupta:2023lga} sampled the neutron star masses from a double Gaussian distribution, $p(m) = w \cN(\mu_L, \sigma_L) + (1 - w)\cN(\mu_R, \sigma_R)$, where $\mu_L = 1.35M_\odot$, $\sigma_L = 0.08M_\odot$, $\mu_R = 1.8M_\odot$, $\sigma_R = 0.3M_\odot$, and $w = 0.64$, as reported in~\cite{2020RNAAS...4...65F}. Both normal distributions were independently truncated and normalized in the range $[1,2.2]M_\odot$, where the maximum mass was dictated by their choice of equation of state, i.e., \texttt{APR4}~\cite{Akmal:1998cf}. Similarly, the tidal deformability of neutron stars in the population, $\Lambda_1, \Lambda_2$, were modeled using \texttt{APR4}. The neutron star spins were assumed to be aligned with the orbital angular momentum, with their magnitudes being sampled between [0, 0.1]. Ref.~\cite{Gupta:2023lga} used a redshift distribution in the range $[0,20]$ obtained from the Madau-Dickinson star formation rate density~\cite{Madau:2014bja}\footnote{The functional form of the MDBN and Madau-Dickinson merger rate is the same, with the Madau-Dickinson merger rate having different coefficients: $a=1,\ b=2.7,\ c=1/2.9,\ d=5.6$~\cite{Madau:2014bja}.} that peaks at $z\sim2$, with a local merger rate density of $320$ Gpc$^{-3}$ yr$^{-1}$, as measured in GWTC-2~\cite{LIGOScientific:2020kqk}. If this local merger rate changes significantly in the future, one can scale our bounds as $1/\sqrt{Q}$, where $Q$ is the ratio of the new rate to the one used here. The distribution of other extrinsic parameters of BNSs is summarized in Table~\ref{tab:bns_injection_params}.

The resulting population contained $\sim1.2$ million BNS mergers. However, we consider events only up to $z=5$, where the detectability of BNS mergers is expected to drop to 5\%~\cite{Gupta:2023lga}. Since sGRBs are highly directional, we also set an inclination angle cut of $30^\circ$~\cite{LIGOScientific:2017adf}. After applying these cuts in $z$ and $\iota$, we are left with a population of $\sim 80,000$ BNS mergers. 

We use the same fiducial cosmology from Planck (2018) as mentioned in Sec.~\ref{sec:imbh_pop} to convert redshifts to luminosity distances. 

\section{Methodology}
\label{sec:methodology}

Our analysis utilizes the Fisher matrix formalism \cite{Cutler:1994ys,Poisson:1995}. It provides a fast and computationally inexpensive way to calculate the $1\sigma$ bound on binary parameters, which are related to the Fisher matrix as the square root of the diagonal elements of the inverse Fisher matrix. We use this to first estimate the $1\sigma$ errors on the binary parameters, and then use another Fisher matrix to calculate the bounds on the cosmological parameters. 

First, one simulates the population of IMBBHs or BNSs with the extrinsic and intrinsic parameters, $\lambda_i$, summarized in Tables~\ref{tab:imbh_parameters} and \ref{tab:bns_injection_params}. The strain from the simulated population is calculated using the \texttt{IMRPhenomXPHM}~\cite{Pratten:2020ceb} waveform approximant for IMBBHs and \texttt{IMRPhenomPv2\_NRtidalv2}~\cite{Dietrich:2019kaq} for BNSs. The GW strain for IMBBHs observed in a detector is given by,  
\begin{equation}
    h_I(t, \boldsymbol{\lambda}) = \sum_{A=+,\times} F_A^I(\alpha, \delta, \psi, \boldsymbol{\nu})\, h^I_A(t, \cM, \eta,\chi_1, \chi_2, \iota, D_L ),
    \label{eq:strain}
\end{equation}
where, $F_A^I$ is the antenna pattern function of the $I^{\rm th}$ detector, $\boldsymbol{\nu}$ represents the location of the detector, $\cM$ is the chirp mass of the binary, and $\eta$ is the symmetric mass ratio. Here, the sum is over the two polarizations of GWs, plus ($+$) and cross ($\times$). The strain for BNSs is given by the same expression, but each polarization has additional dependence on $\Lambda_1$ and $ \Lambda_2$.

We next calculate the Fisher matrix, defined as the noise-weighted inner product of the derivative of the strain in the frequency domain with respect to the parameters describing the binary,
\begin{equation}
     \Gamma^{ij}_I = \left\langle \frac{\partial h_I}{\partial\lambda_i} \middle| \frac{\partial h_I}{\partial \lambda_j} \right\rangle,
     \label{eq:fisher1}
\end{equation}
where the noise-weighted inner product is defined as,
\begin{equation}
    \left\langle A \middle| B\right\rangle =  2 \int^{f_{\rm high}}_{f_{\rm low}} \frac{\tilde{A}^*(f) \tilde{B}(f) + \tilde{A}(f)\tilde{B}^*(f)}{S_n(f)}\,{\rm  d}f,
    \label{eq:innerproduct}
\end{equation}
where $\tilde{A}$ and $\tilde{B}$ are the Fourier transforms of $A$ and $B$, respectively, and $f_{\rm low}$ and $f_{\rm high}$ are lower and upper cut-off frequencies of the detector (see Table~\ref{tab:detectors} for their values for different detectors). 
This inner product is used to calculate the SNR, $\rho$, given by
\begin{equation}
    \rho = \sqrt{\langle h | h \rangle}.
    \label{eq:SNR}
\end{equation}

The resultant Fisher matrix for a network of detectors or multibanding is simply the sum of the Fisher matrices of each detector,
\begin{equation}
    \Gamma_{\rm tot}^{ij} = \sum_I \Gamma_{I}^{ij}.
    \label{eq:tot_fisher1}
\end{equation}

This resultant Fisher matrix is then inverted to obtain the covariance matrix, $\Sigma = \Gamma_{\rm tot}^{-1}$ \cite{Poisson:1995}, the diagonal elements of which give the variance in the parameters, $\lambda_i$, including the variance in $D_L$, given by $\sigma^2_{D_L}$.

Once we obtain the $1\sigma$ error on $D_L$, we define another Fisher matrix, $\cG$ to obtain the error forecasts on cosmological parameters, $\boldsymbol{\theta} = \{H_0,\ \Om,\  \Ok\}$~\cite{Gupta:2022fwd, Dhani:2022ulg}, while setting $\Omega_\Lambda = 1 - \Om - \Ok$:
\begin{equation}
    \cG_{ij} = \sum_{k=1}^N \frac{1}{\sigma^2_{D_L}} \frac{\partial D_L}{\partial \theta_i} \frac{\partial D_L}{\partial \theta_j} \Bigg|_k.
    \label{eq:fisher2}
\end{equation}
Here, the sum is over all observed $N$ events. The inverse of this matrix gives us the covariance in cosmological parameters.
\begin{equation}
    \Sigma_{ij} = \cG^{-1}_{ij}.
    \label{eq:cov}
\end{equation}
The constraints on cosmological parameters are obtained from the square root of the diagonal elements of the inverse of this matrix, $\sigma_{\theta_i} = \sqrt{\cG_{ii}^{-1}}$. We report the bounds on cosmological parameters in the following section. 

\section{Results}
\label{sec:results}

\begin{figure*}[t]
    \centering
    \includegraphics[width=0.495\linewidth]{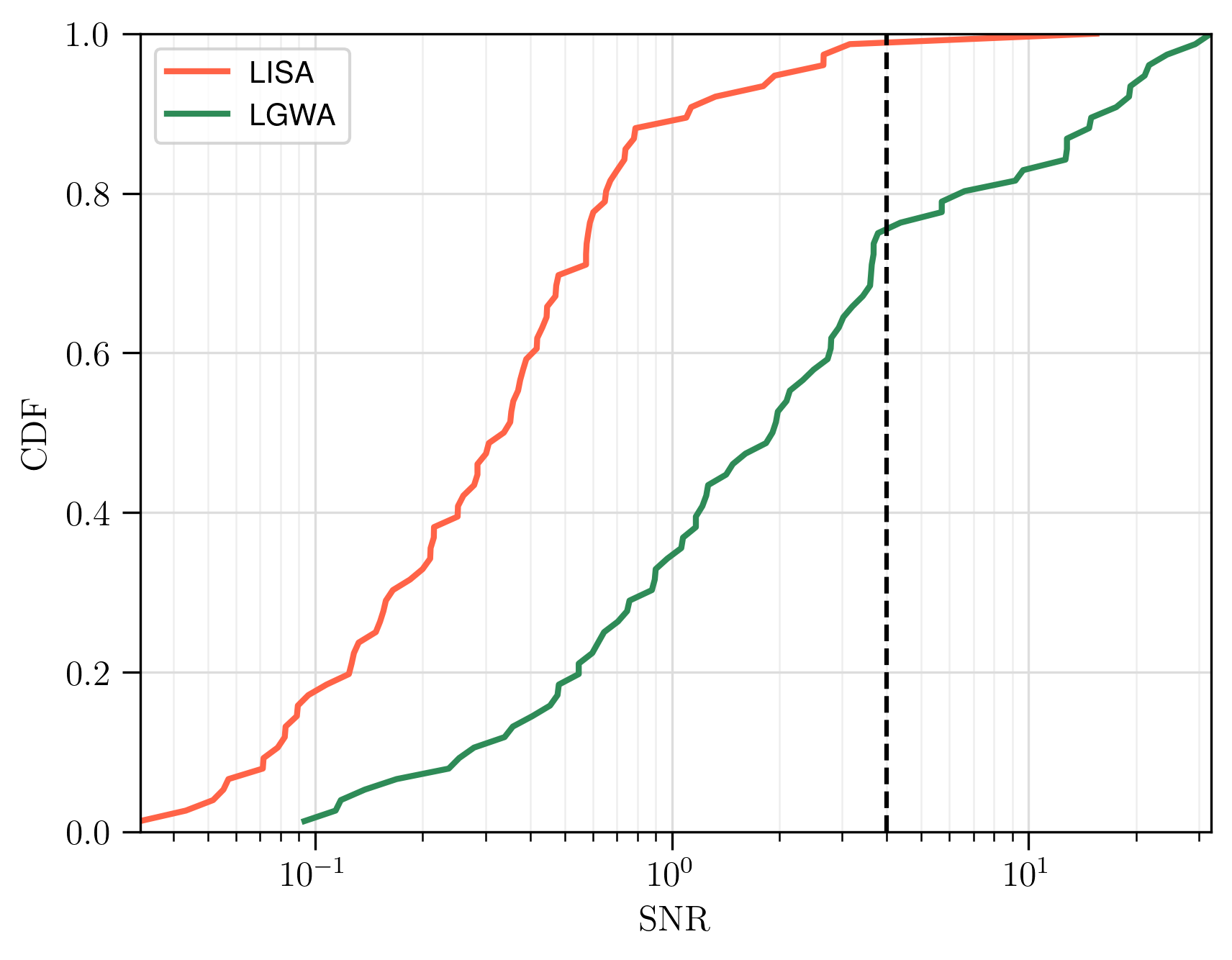}
    \includegraphics[width=0.495\linewidth]{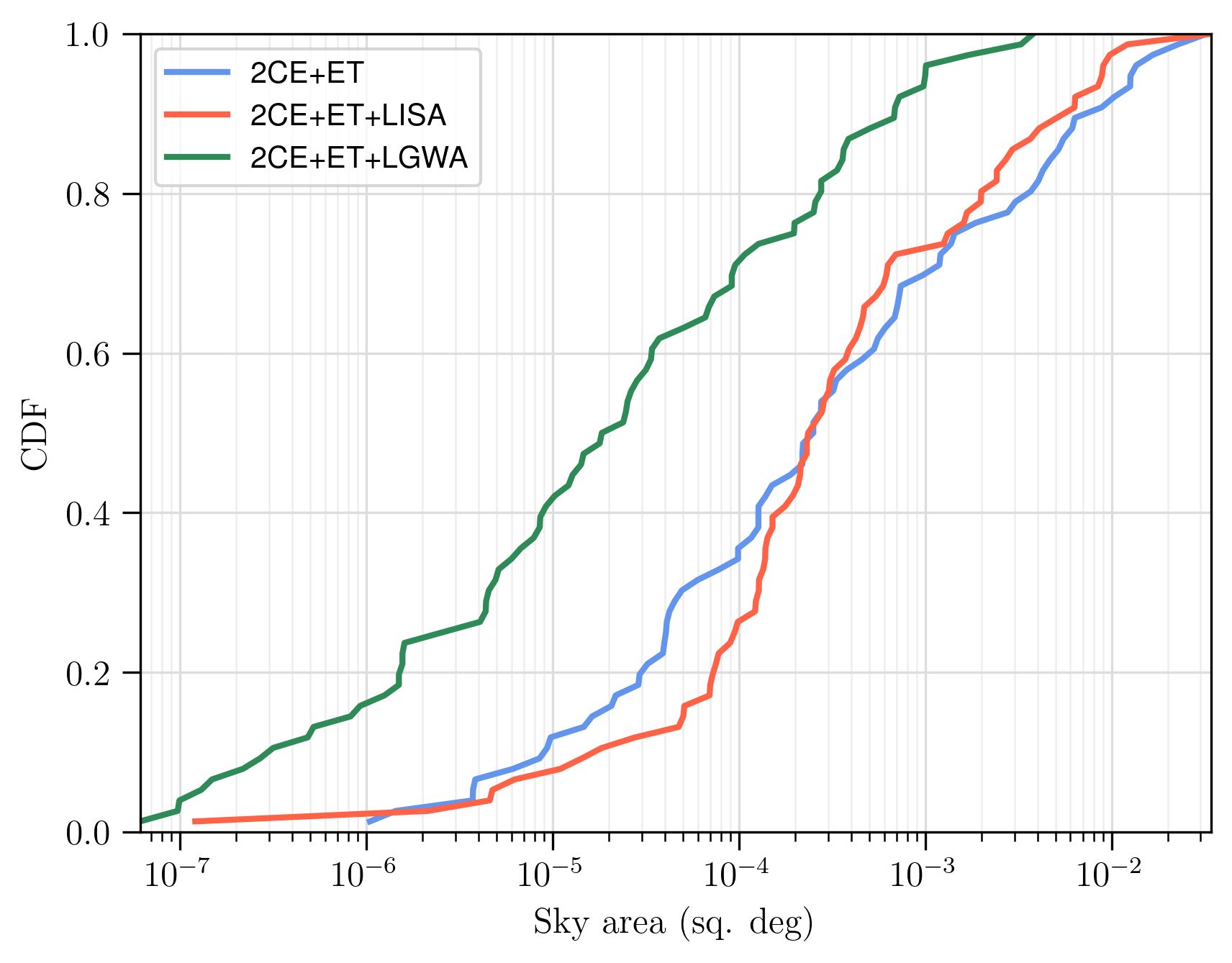}
    \caption{\textit{Left:} Cumulative distribution of SNR of GW231123-like IMBBH events sampled from a uniform in co-moving volume merger rate detected by LISA (red) and LGWA (green) over one year of GW observation. The vertical dashed line represents an SNR threshold of 4. \textit{Right:} Cumulative distribution of sky localization errors for these binaries for three different networks: 2CE+ET (blue), 2CE+ET+LISA (red), 2CE+ET+LGWA (green) for one year of GW observation.}
    \label{fig:snr_sky_locdist}
\end{figure*}

\begin{figure*}[htb]
    \centering
    \includegraphics[width=0.8\textwidth]{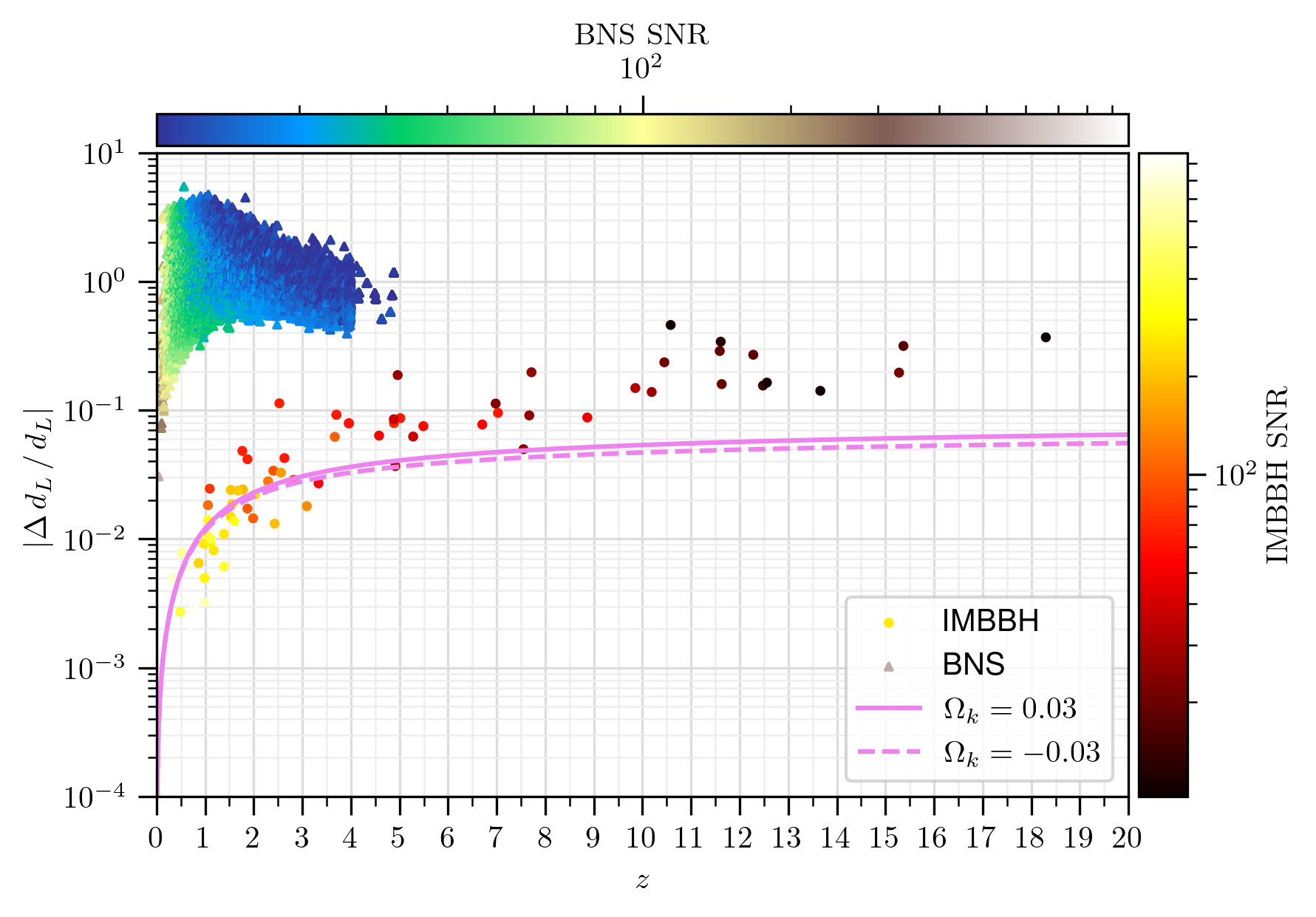}
    \caption{Relative errors in luminosity distance as a function of redshift measured with IMBBHs and BNSs with a network of 2CE+ET for one year of GW observation.} 
    \label{fig:dlvz} 
\end{figure*}
We now discuss the results obtained from our analysis. We use the Fisher matrix code \textsc{gwfish}~\cite{Dupletsa:2022scg} for IMBBHs and Ref.~\cite{Gupta:2023lga} utilizes \textsc{gwbench}~\cite{Borhanian:2020ypi} for BNSs, to compute the SNRs and the $1\sigma$ uncertainties in $D_L$ (along with other binary parameters) for each event. For both IMBBHs and BNSs, we select the events with SNR $>10$ in the ground-based detectors. We obtain $74$, $414$, and $4$ IMBBHs satisfying this SNR cutoff for the pragmatic, optimistic, and pessimistic merger rates, respectively. Similarly, we obtain $\sim60,000$ BNSs after applying this SNR cut-off. As can be seen from the left panel of Fig.~\ref{fig:snr_sky_locdist}, the SNRs in the LISA or LGWA band are very low for the majority of simulated IMBBHs; less than 4 for 98.7\% (75\%) of binaries in the LISA (LGWA) band, implying that most of these events, with the adopted redshift distribution and masses similar to GW231123, will not be detected by non-terrestrial detectors. In contrast, the same binaries have SNRs of ${\mathcal O}(100)$ in the CE1, CE2, and ET bands, as can be seen in Fig.~\ref{fig:dlvz}. To be considered for multiband analysis, we adopt an SNR threshold of 4~\cite{Datta:2020vcj} in the non-terrestrial detectors, so that these sources can be extracted from archival LISA/LGWA data using information from ground-based detectors. Only $1$ $(19)$ out of 76 IMBBH binaries have SNRs above this threshold in the LISA (LGWA) band. We do not consider BNSs for mulibanding analysis due to their small SNRs in the non-terrestrial GW detectors. 

\begin{table}[t]
    \centering
    \begin{ruledtabular}
        \begin{tabular}{c|ccc}
            \textbf{Population}  & $\bm{\sigma_{H_0}}$   & $\bm{\sigma_{\Om}}$   & $\bm{\sigma_{\Ok}}$ \\
            \hline
            \noalign{\vskip 0.1em} 
             IMBBH      &$0.047\ (0.69\%)$ &$0.030\ (9.76\%)$ & $0.029$         \\
             BNS      &$1.12\ (1.65\%)$  &$0.064\ (20.64\%)$  & $0.055$  \\
             IMBBH + BNS&$0.43\ (0.63\%)$  &$0.027\ (8.58\%) $& $0.024$
        \end{tabular}
    \end{ruledtabular}
    \caption{Absolute $1\sigma$ errors in cosmological parameters with one year of GW observation for a particular realization of three different populations observed with a network of 2CE+ET. IMBBHs are sampled from a uniform in co-moving volume merger rate, while BNSs are from the Madau-Dickinson merger rate. The percentage errors for $H_0$  and $\Om$ are shown in parentheses.}
    \label{tab:err_cosm_params}
\end{table}

We use the $1\sigma$ uncertainties in $D_L$ to estimate the $1\sigma$ errors on the cosmological parameters $H_0$, $\Om$, and $\Ok$ using Eq. (\ref{eq:fisher2}). Our results are summarized in Table~\ref{tab:err_cosm_params}.
Figure~\ref{fig:dlvz} shows the fractional errors in $D_L$ for simulated mergers as a function of $z$ with their respective 2CE+ET SNRs on the color axis. The two lines represent the fractional deviation of $D_L$ from a flat $\Lambda$CDM cosmology for two different values of $\Ok: -0.03$ and $0.03$. These values were chosen as these are the predicted errors in $\Ok$ we obtain with a population of IMBBHs observed with a network of 2CE+ET, with a uniform in co-moving volume merger rate.  Firstly, we note that the fractional error in $D_L$ measured with IMBBHs is $\sim1-2$ orders of magnitude lower than BNSs, owing to their large SNRs due to their significantly higher masses. Most of the IMBBHs have SNRs of $\mathcal{O}(100)$, whereas SNRs of BNS mergers are of $\mathcal{O}(10)$. Secondly, 
the fractional errors on $D_L$ measured from a population of BNS are of $\cO(100\%)$, with only three BNS mergers having $1\sigma$ uncertainty below $10\%$, partly due to the low masses of BNSs, and partly due to the distance-inclination degeneracy, which is made worse because we restrict our results to BNSs with small inclination angles ($<30^{\circ}$). No BNS mergers have relative errors in $D_L$ lower than the shown analytic fractional deviation of $D_L$ due to variations in $\Ok$ from the fiducial value of 0, which explains the poor constraints on $\Ok$ and correspondingly $\Om$ due to the strong correlation between the two parameters, as shown in Fig.~\ref{fig:corner}. 

\subsection{IMBBH Population} \label{subsec:IMBBH_res}

We compute the uncertainties on $H_0$, $\Om$, and $\Ok$ for three detector configurations: 2CE+ET, 2CE+ET+LISA, and 2CE+ET+LGWA. The results are reported for one year of GW observation for a GW231123-like IMBBH population that follows a uniform in co-moving volume merger rate, unless stated otherwise. We also compare the results of the GW231123-like population of IMBBHs to a population of GW190521-like IMBBHs in Appendix~\ref{sec:appdx_corner}. 

The $1\sigma$ errors in $H_0$, $\Om$, and $\Ok$ for the 2CE+ET configuration with $74$ detected IMBBHs are $0.69\%$, $9.76\%$, and $0.029$, respectively. With multibanding, we see an improvement in the $1\sigma$ uncertainties on $H_0$, $\Om$, and $\Ok$ by $3.96\%$ $(11.0\%)$, $5.5\%$ $(11.46\%)$,  and $\ 0.17$  $(0.17)$, respectively, when information from LISA (LGWA) is included. These bounds on $H_0$ are comparable to the bounds obtained using stellar mass BBH as dark sirens with the galaxy catalog approach~\cite{Muttoni:2023prw} or the cross-correlation method~\cite{Ferri:2024amc, Afroz:2024joi}, golden dark sirens~\cite{Chen:2024gdn}, and spectral sirens~\cite{Ezquiaga:2022zkx}.

Also, Muttoni \textit{et al.}~\cite{Muttoni:2023prw} predict uncertainties on $\Om$ to be $9.0\%$  with stellar mass BBH dark sirens which agree with our results for an IMBBH population. If we work with a flat $\Lambda$CDM model, i.e., if $\Ok$ is set to be zero, then $H_0$ and $\Om$ are measured with a precision of $0.39\%$, and $2.43\%$, respectively, for a network of 2CE+ET. 

As seen from Fig.~\ref{fig:dlvz}, while there are higher number of IMBBH mergers beyond $z=2$, the fractional error in $D_L$ increases due to lower SNRs at high redshifts. If we select a 
subset of the highest SNR IMBBH events, i.e., events with $z < 2$ (SNR $\gtrsim 200$, $N = 24$), the uncertainties on $H_0$, $\Om$, and $\Ok$ for a network of 2CE+ET are predicted to be $1.12\%,\ 21.81\%$ and $0.076$, respectively. This suggests our constraints are dominated by the high SNR, low $z$ events and not necessarily by the relatively larger number of low SNR, high $z$ events.

\begin{figure*}[htb]
    \centering
\includegraphics[width=0.9\textwidth]{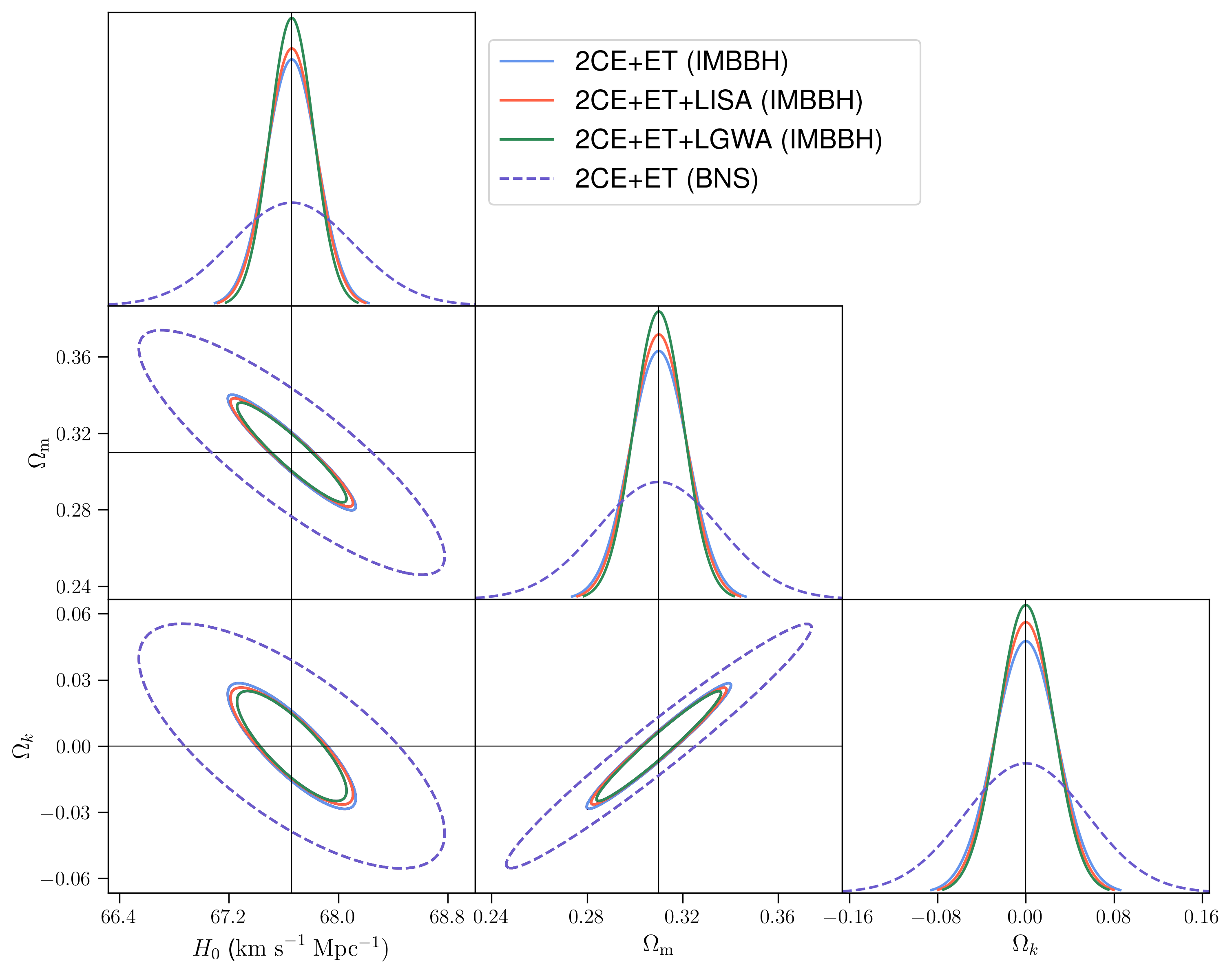}
    \caption{1$\sigma$ corner plots for the cosmological parameters in a non-flat $\Lambda$CDM model for three different networks: 2CE+ET, 2CE+ET+LISA, and 2CE+ET+LGWA. The solid lines represent an IMBBH population, while the dashed line represents the BNS population. The vertical and horizontal lines denote the true values of the parameters from Planck 2018~\cite{Planck:2018vyg}.}
    \label{fig:corner}
\end{figure*}

The right panel of Fig.~\ref{fig:snr_sky_locdist} shows the cumulative distribution of sky localization errors obtained for a population of IMBBHs with the three networks considered. We can see that observations with LGWA in a multiband context can provide nearly two orders of magnitude improvement in sky localization. The median $90\%$ credible sky area is $2 \times 10^{-4}$ deg$^2$ when IMBBHs are observed with the network 2CE+ET, or 2CE+ET+LISA, whereas for the 2CE+ET+LGWA configuration, the median $90\%$ credible sky area is reduced to $2 \times 10^{-5}$ deg$^2$. This is in agreement with the results obtained in Ref.~\cite{Zhang:2025eeh}, which showed that the sky area is confined between $\sim10^{-6}-10^{-10}$ for the moon-based detector Crater Interferometry Gravitational-wave Observatory (CIGO) in LGWA's frequency band of $0.1-1$ Hz. This significant improvement in sky localization highlights the important role that LGWA can play in identifying the host galaxies of IMBBHs and increasing the likelihood of these sources being bright sirens. Unfortunately, the addition of LISA in the multiband analysis does not improve the sky localization. This is expected, as only 1 event contributes to the multiband analysis for LISA in the GW231123-like IMBBH population in our study.

In Fig.~\ref{fig:corner}, we show the joint posterior distribution on cosmological parameters, assuming they follow a multivariate normal distribution with the means obtained from Planck (2018)~\cite{Planck:2018vyg} and the covariance given by the inversion of the Fisher matrix as described in Sec.~\ref{sec:methodology}. We observe a slight improvement in the bounds on cosmological parameters when using multibanding, especially with LGWA. One of the reasons for not seeing any significant improvements with multibanding is because the masses of our simulated events are restricted to that of GW231123, whereas IMBBHs with higher masses that are still observed by ground-based detectors may have higher SNRs in LISA and LGWA bands~\cite{LISA:2017pwj, Ajith:2024mie, Song:2026kii}.
We also observe a strong correlation between the parameters, particularly between $\Om$ and $\Ok$.

\begin{figure}[h]
    \centering
    \includegraphics[width=\linewidth]{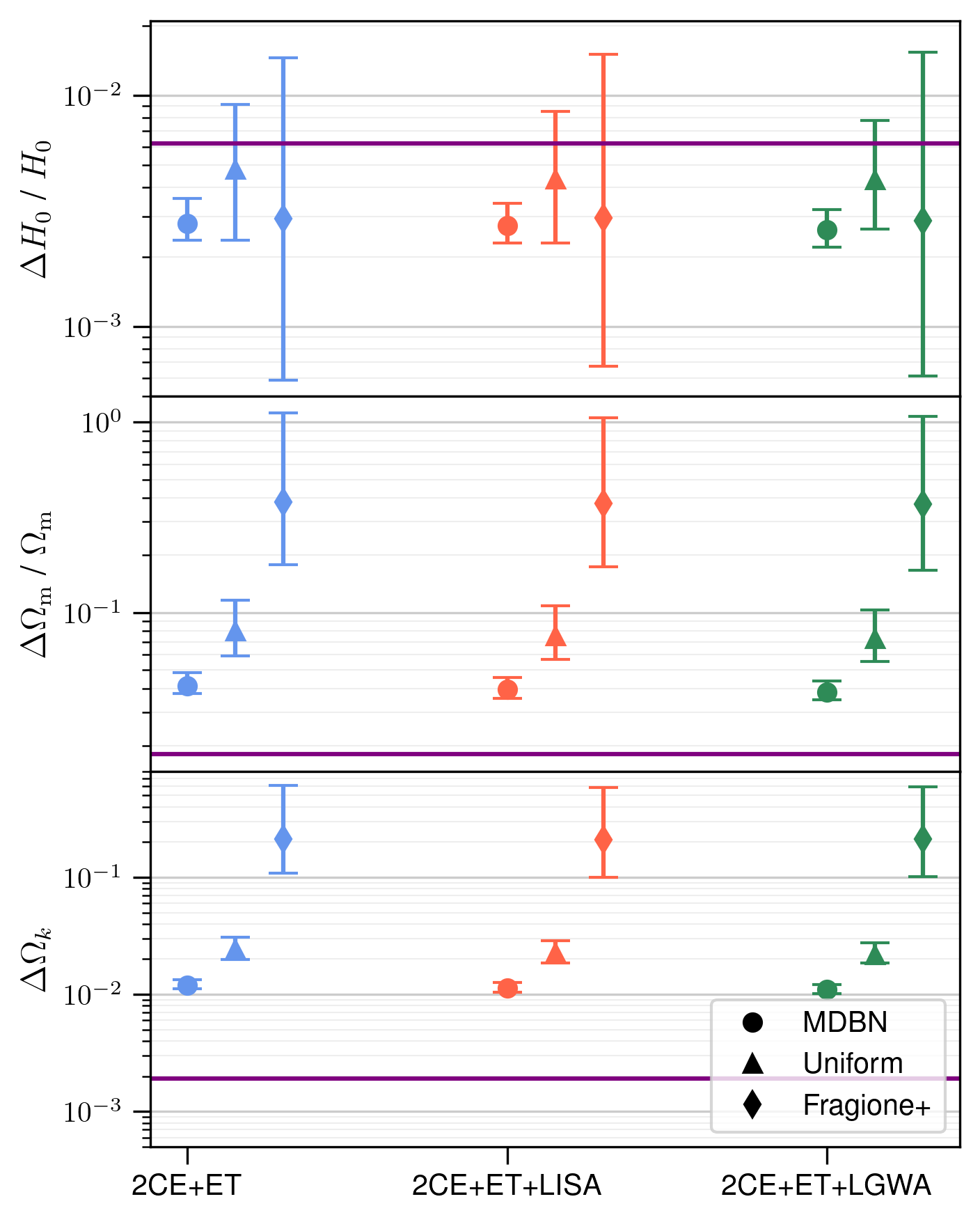}
    \caption{The median fractional errors in the cosmological parameters, $H_0$ and $\Om$, and the median absolute error in $\Ok$ for different detector networks for a population of IMBBHs sampled from three different merger rates observed in a year, represented by a circle, triangle and diamond marker. The error bars represent the $68.3\%$ credible interval of the predicted uncertainties. The colors correspond to the three different detector networks, with 2CE+ET being represented by blue, 2CE+ET+LISA by red, and 2CE+ET+LGWA by green. The purple horizontal lines represent results from Planck (2018)~\cite{Planck:2018vyg}.}
    \label{fig:bootstrap}
\end{figure}

In Fig.~\ref{fig:bootstrap}, we show the fractional uncertainties in $\Om$ and $H_0$, and the absolute uncertainty in $\Ok$ for IMBBH populations with three merger rates: (1) Uniform in co-moving volume, (2) MDBN, and (3) Fragione+, with a network of 2CE+ET.
We generate 100 independent realizations of the IMBBH population to account for statistical fluctuations in the inferred uncertainties of the cosmological parameters. The reported constraints correspond to the mean and standard deviation of these uncertainties across all realizations. The comparatively large uncertainty associated with the Fragione+ merger rate arises from its low predicted event rate of $\sim 4$ mergers per year, making the results susceptible to significant statistical fluctuations.
The purple horizontal line represents the constraints obtained by Planck (2018)~\cite{Planck:2018vyg}. As can be seen, GW observations of an IMBBH population yield comparable constraints on $H_0$, with the median uncertainties being better than Planck~\cite{Planck:2018vyg}. However, GW observations of IMBBHs do not improve the current bounds on $\Om$ and $\Ok$. Currently, $\Om$ is measured with a precision of $1.8\%$ by Planck~\cite{Planck:2018vyg}, which is much better compared to the precision of $4\%$ obtained in this study for the optimistic MDBN merger rate. Furthermore, $\Ok$ is measured with a precision of $0.0019$ by combining Planck CMB~\cite{Planck:2018vyg} and BAO~\cite{eBOSS:2020fvk} data, whereas the best bound on $\Ok$ obtained in this study is $0.01$ for the optimistic MDBN merger rate. 

\subsection{BNS Population}
\label{sec:bns_res}

We present the results for the BNS observations assuming they have an EM counterpart such as a sGRB. As a result, we have precise information on their redshift as they act as bright sirens. 

Table~\ref{tab:err_cosm_params} shows the uncertainties in cosmological parameters obtained with a population of BNSs observed with a network of 2CE+ET. We report an error on $H_0$ and $\Om$ as $1.65\%$ and $20.64\%$ respectively. These bounds are significantly higher than the ones obtained by Dhani \textit{et al.}~\cite{Dhani:2022ulg} and Chen \textit{et al.}~\cite{Chen:2016tys}, who reported a  $0.2\%$, and $0.3\%$ $1\sigma$ error in $H_0$ respectively in a flat $\Lambda$CDM framework, with BNSs detected as bright sirens using a network of 2CE+ET. Moreover, Dhani \textit{et al.} also report a $6\%$ error on $\Om$, which is much lower than our predicted bound of $20.64\%$.

We report $\sigma_{\Ok} = 0.055$ with a population of BNSs as bright sirens observed for one year with 2CE+ET. Califano \textit{et al.}~\cite{Califano:2022cmo} report $\sigma_{\Ok} \approx 0.09$, which is higher than our bounds due to the following factors: (i) We use a network of 2CE+ET whereas they use only ET, (ii) They also account for uncertainties in measuring $D_L$ due to effects of weak lensing distortions and peculiar velocities, which we do not.

We note that the inferred uncertainties on $H_0$ and $\Om$ are larger than those typically reported in the literature, with the discrepancy being particularly pronounced for $\Om$. This can be attributed to the comparatively weak constraints on $\Ok$ obtained from the BNS population, whose uncertainties are nearly a factor of two larger than those from IMBBH systems, which in turn propagate into degraded constraints on the remaining cosmological parameters. The impact is especially severe for $\Om$ due to its strong correlation with $\Ok$. If $\Ok$ is removed from our analysis, i.e., we assume a flat $\Lambda$CDM model, we obtain a $1.2\%$, and $5.0\%$ bound on $H_0$ and $\Om$ respectively, consistent with predictions in the literature.

When combined with an IMBBH population with the same network, we report errors of $0.63\%$, $8.58\%$, and $0.024$ respectively on $H_0$, $\Om$, and $\Ok$, which are similar to the bounds we obtain with just the IMBBH population. This is because of the higher $D_L$ measurement uncertainties with the BNS population as compared to the IMBBH population, as can be seen in Fig.~\ref{fig:dlvz}. This difference is reflected on cosmological constraints, as can be seen in Fig.~\ref{fig:corner}. Thus, the contribution of BNS mergers in the measurement of cosmological parameters will be minimal compared to IMBBH mergers acting as bright sirens. 

\section{Summary and Conclusion}
\label{sec:concl}

In this study, we present forecasts for how well next-generation GW detectors can measure cosmological parameters in a non-flat $\Lambda$CDM model. We simulate a population of IMBBHs with component masses matching those of GW231123~\cite{LIGOScientific:2025rsn}, the heaviest BBH observed so far, adopting three merger rate density models: (1) uniform in comoving volume (pragmatic), (2) MDBN (optimistic), and (3) Fragione+ (pessimistic). We also consider a population of BNSs using a Madau Dickinson merger rate density from Ref.~\cite{Gupta:2023lga}. We assume that all our sources are bright sirens, i.e., they are accompanied by observable EM counterparts with precise redshift measurements, although this is an optimistic assumption. IMBBHs may produce EM signatures if they merge in dense environments such as AGN disks or globular clusters, though such events are expected to be rare. BNS mergers, on the other hand, can produce EM counterparts in the form of sGRBs and kilonovae. 

Even though current GW detectors cannot achieve the precision required to constrain cosmological parameters, future terrestrial and non-terrestrial detectors may do so at the percent level. We consider three GW detector networks: 2CE+ET, 2CE+ET+LISA, and 2CE+ET+LGWA. Using the Fisher matrix formalism, we estimate the luminosity distance uncertainty for each source in the population and propagate these uncertainties to place bounds on the cosmological parameters $H_0$, $\Om$, and $\Ok$.

With one year of GW observations of IMBBHs, the 2CE+ET network yields a $1\sigma$ uncertainty of $0.69\%$ on $H_0$ in the pragmatic scenario, while the optimistic and pessimistic scenarios yield uncertainties of $0.04 - 0.08 \%$ and $0.2 - 1 \%$, respectively. These constraints are comparable to the $0.62\%$ uncertainty on $H_0$ measured by Planck~\cite{Planck:2018vyg} for the pragmatic, and the pessimistic scenarios, while the predicted uncertainty in the optimistic scenario is significantly better than Planck. Similarly, for an IMBBH population observed with the 2CE+ET network, we find that $\Om$ can be constrained to a precision of $\lesssim10\%$ in the optimistic and pragmatic scenarios. In the pessimistic scenario, however, the constraints degrade substantially, with a precision between $20 - 100\%$. These constraints on $\Om$ remain weaker than those from Planck. Finally, the $1\sigma$ errors on $\Ok$ with the 2CE+ET network, using one year of GW observations of an IMBBH population, are $0.01$, $0.02$, and $0.2$ for the optimistic, pragmatic, and pessimistic scenarios, respectively, all of which are weaker than current Planck constraints. 

The majority of simulated IMBBH events detected by CE/ET have SNRs below 4 in LISA or LGWA, and only a small subset can be utilized for multiband constraints. Consequently, multiband observation of IMBBHs improves the $1\sigma$ uncertainties on $H_0$, $\Om$, and $\Ok$ only modestly, by $3.96\%$ $(11.0\%)$, $5.5\%$ $(11.46\%)$, and $0.17$ $(0.17)$, respectively, when information from LISA (LGWA) is included. However, the $90\%$ credible sky-localization area is typically an order of magnitude smaller for 2CE+ET+LGWA as compared to the 2CE+ET network. This significant improvement in sky localization with LGWA could greatly aid EM follow-up observations of the GW signal from IMBBHs.

The constraints obtained from the BNS population are weaker than those from the IMBBH population, primarily because the lower masses of BNS systems yield lower SNRs and larger uncertainties in the luminosity distance. These distance uncertainties are further affected by the distance-inclination degeneracy, which limits the constraining power of BNS bright sirens. Consequently, combining the BNS and IMBBH populations yields only a modest improvement over the IMBBH-only constraints, with the combined result still dominated by the IMBBH contribution. In the optimistic IMBBH scenario, combining constraints from BNSs and IMBBHs constrains $\Ok$ to $\sigma_{\Ok}\simeq 0.01$, which is several times weaker than the constraint $\sigma_{\Ok}=0.0019$ obtained from the combination of Planck CMB and BAO observations~\cite{Planck:2018vyg}. Thus, future GW measurements are unlikely to surpass current CMB+BAO constraints on $\Ok$ in precision. Their value instead lies in providing an independent measurement, with systematics distinct from those affecting EM cosmological probes. A statistically significant deviation of $\Ok$ from its fiducial value of zero, if also robust against GW-specific systematics, would be of considerable interest, as it could indicate physics beyond the standard flat $\Lambda$CDM framework. 

Several caveats should be kept in mind when interpreting these forecasts. First, the inferred constraints on the cosmological parameters depend sensitively on the assumed merger rate and redshift distribution of IMBBHs, both of which remain highly uncertain. Second, our analysis adopts a GW231123-like binary as the fiducial IMBBH system. Since the SNR, the luminosity distance uncertainty, and hence the resulting cosmological constraints depend on the binary masses and other source parameters, the results presented here should not be interpreted as forecasts for a realistic population-averaged IMBBH distribution. Rather, they should be viewed as an assessment of the extent to which sufficiently massive IMBBH mergers can inform measurements of cosmological parameters such as $\Ok$. Third, we assume that IMBBHs act as bright sirens through EM flares generated by interaction with a dense ambient medium. This is an optimistic assumption, particularly at the higher redshifts considered here, where counterpart detectability and robust host-galaxy association may become challenging. Nevertheless, the constraints presented here are driven predominantly by the nearer events. In particular, we find that 24 IMBBH mergers within $z=2$ already constrain $\Ok$ to $\sim 0.08$ (see Sec.~\ref{subsec:IMBBH_res}), implying that even if counterpart identification becomes inefficient at $z>2$, a subset of lower redshift bright IMBBHs could still provide meaningful independent constraints on the spatial curvature parameter.

Finally, we do not include several additional sources of uncertainty that may be considered in standard siren forecasts. For the redshift range most relevant to the constraining power in this work, uncertainties associated with peculiar velocities are expected to be negligible, since they primarily affect constraints from nearby sources~\cite{Tamanini:2016zlh,Gupta:2022fwd}. Weak lensing can, however, induce distance errors at the few-percent level for high-redshift events, with estimates reaching $\sim 5\%$ in some cases~\cite{Holz:2005df}; our forecasts, thus, implicitly assume that delensing methods in the XG era will be sufficiently effective that lensing will not dominate the error budget~\cite{Tamanini:2016zlh}. We also neglect detector calibration systematics, even though present interferometers can have amplitude uncertainties at the several-percent level~\cite{Sun:2020wke}. Since amplitude calibration maps directly onto $D_L$ inference, this is another source of systematic uncertainty that should be incorporated in more realistic studies. Lastly, waveform systematics may also lead to biases in $H_0$ measurements~\cite{Dhani:2024jja, Dhani:2025xgt}. Our results should therefore be interpreted under the assumption that waveform model systematics will be negligible in the XG era. 

In the future, it will be interesting to explore the potential of IMBBH mergers to constrain cosmology within extended frameworks beyond $\Lambda$CDM, including dynamical dark energy models such as $w$CDM and $w_0w_a$CDM. It would also be valuable to assess how these constraints improve with next-generation and multiband detector networks, for example by incorporating LIGO-India~\cite{Saleem:2021iwi}, TianQin~\cite{TianQin:2015yph}, Taiji~\cite{Hu:2017mde}, or B-DECIGO~\cite{Kawamura:2020pcg}. Such studies could reveal the full cosmological potential of IMBBH standard sirens and clarify the gains achievable through expanded global and space-based GW networks.

\acknowledgments
We thank K.~G.~Arun and Daniel Holz for useful discussions. We would also like to thank Arnab Dhani for carefully reviewing this manuscript as part of the LVK internal review process. IG is supported by the NSF grant: PHY-2020275 (Network for Neutrinos, Nuclear Astrophysics, and Symmetries (N3AS)). AG is supported by NSF grants PHY-2308887 and CAREER-2440327. We also acknowledge the use of the Maple (funded by NSF Grant CHE-1338056) and Magnolia clusters at the Mississippi Center for Supercomputing Research, the University of Mississippi.

This study used the software packages matplotlib~\cite{Hunter:2007}, numpy~\cite{Harris:2020xlr}, pandas~\cite{reback2020pandas,mckinney-proc-scipy-2010}, astropy~\cite{astropy:2022, astropy:2018, astropy:2013} and scipy~\cite{2020SciPy-NMeth}, h5py~\cite{collette_python_hdf5_2014}, and \textsc{gwfish}~\cite{Dupletsa:2022scg}. 

This manuscript corresponds to LIGO document number LIGO-P2600234.

\appendix

\section{GW190521-like population}
\label{sec:appdx_corner}

Here we present the results obtained for a population of IMBBH binaries with component masses matching those inferred for GW190521~\cite{LIGOScientific:2020iuh}, and compare them with the corresponding results for GW231123. The other binary parameters are sampled uniformly in their respective ranges, same as the parameters of the GW231123-like population as shown in Table~\ref{tab:imbh_parameters}.

\begin{figure}[h]
    \centering
    \includegraphics[width=0.5 \textwidth]{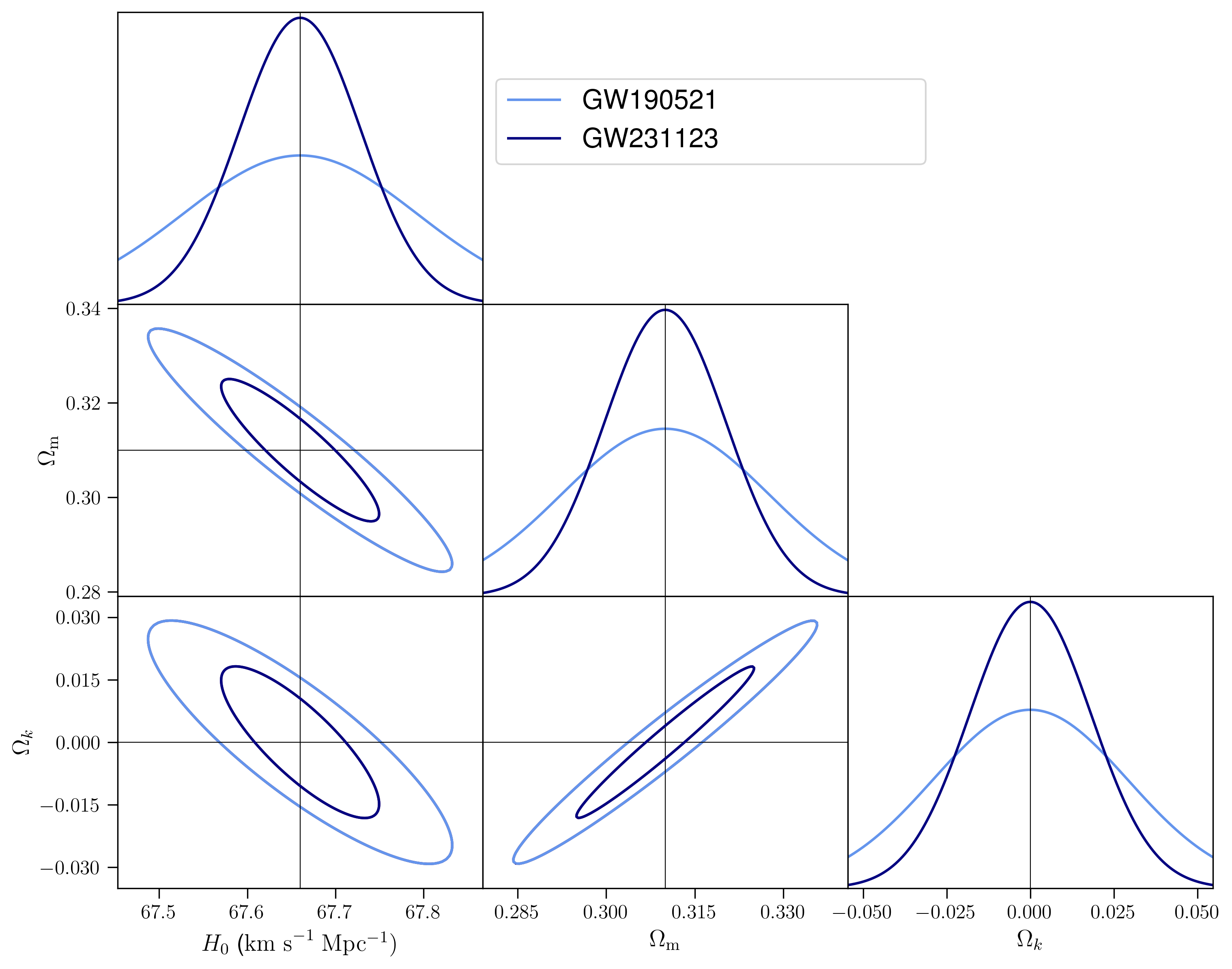}
    \caption{$1\sigma$ corner plots for the cosmological parameters in a non-flat $\Lambda$CDM model with both GW231123- and GW190521-like populations observed with a network of 2CE+ET and sampled from a uniform in co-moving volume merger rate.}
    \label{fig:corner_2IMBH}
\end{figure}

The GW190521-like system has a lower total mass than GW231123, with median component masses of $85M_\odot$ and $66M_\odot$. Owing to its comparatively lower SNR, we expect weaker cosmological constraints than those obtained from GW231123, as shown in Fig.~\ref{fig:corner_2IMBH}. With a 2CE+ET network, we obtain uncertainties of $0.30\%$, $8.3\%$, and $0.029$ on $H_0$, $\Om$, and $\Ok$, respectively, for a GW190521-like IMBBH population sampled from a uniform in co-moving volume merger rate. 

\bibliography{omega_k_refs}

\begin{thebibliography}{135}%
\makeatletter
\providecommand \@ifxundefined [1]{%
 \@ifx{#1\undefined}
}%
\providecommand \@ifnum [1]{%
 \ifnum #1\expandafter \@firstoftwo
 \else \expandafter \@secondoftwo
 \fi
}%
\providecommand \@ifx [1]{%
 \ifx #1\expandafter \@firstoftwo
 \else \expandafter \@secondoftwo
 \fi
}%
\providecommand \natexlab [1]{#1}%
\providecommand \enquote  [1]{``#1''}%
\providecommand \bibnamefont  [1]{#1}%
\providecommand \bibfnamefont [1]{#1}%
\providecommand \citenamefont [1]{#1}%
\providecommand \href@noop [0]{\@secondoftwo}%
\providecommand \href [0]{\begingroup \@sanitize@url \@href}%
\providecommand \@href[1]{\@@startlink{#1}\@@href}%
\providecommand \@@href[1]{\endgroup#1\@@endlink}%
\providecommand \@sanitize@url [0]{\catcode `\\12\catcode `\$12\catcode
  `\&12\catcode `\#12\catcode `\^12\catcode `\_12\catcode `\%12\relax}%
\providecommand \@@startlink[1]{}%
\providecommand \@@endlink[0]{}%
\providecommand \url  [0]{\begingroup\@sanitize@url \@url }%
\providecommand \@url [1]{\endgroup\@href {#1}{\urlprefix }}%
\providecommand \urlprefix  [0]{URL }%
\providecommand \Eprint [0]{\href }%
\providecommand \doibase [0]{https://doi.org/}%
\providecommand \selectlanguage [0]{\@gobble}%
\providecommand \bibinfo  [0]{\@secondoftwo}%
\providecommand \bibfield  [0]{\@secondoftwo}%
\providecommand \translation [1]{[#1]}%
\providecommand \BibitemOpen [0]{}%
\providecommand \bibitemStop [0]{}%
\providecommand \bibitemNoStop [0]{.\EOS\space}%
\providecommand \EOS [0]{\spacefactor3000\relax}%
\providecommand \BibitemShut  [1]{\csname bibitem#1\endcsname}%
\let\auto@bib@innerbib\@empty
\bibitem [{\citenamefont {Aghanim}\ \emph {et~al.}(2020)\citenamefont {Aghanim}
  \emph {et~al.}}]{Planck:2018vyg}%
  \BibitemOpen
  \bibfield  {author} {\bibinfo {author} {\bibfnamefont {N.}~\bibnamefont
  {Aghanim}} \emph {et~al.} (\bibinfo {collaboration} {Planck}),\ }\href
  {https://doi.org/10.1051/0004-6361/201833910} {\bibfield  {journal} {\bibinfo
   {journal} {Astron. Astrophys.}\ }\textbf {\bibinfo {volume} {641}},\
  \bibinfo {pages} {A6} (\bibinfo {year} {2020})},\ \bibinfo {note} {[Erratum:
  Astron.Astrophys. 652, C4 (2021)]},\ \Eprint
  {https://arxiv.org/abs/1807.06209} {arXiv:1807.06209 [astro-ph.CO]}
  \BibitemShut {NoStop}%
\bibitem [{\citenamefont {Louis}\ \emph {et~al.}(2025)\citenamefont {Louis}
  \emph {et~al.}}]{AtacamaCosmologyTelescope:2025blo}%
  \BibitemOpen
  \bibfield  {author} {\bibinfo {author} {\bibfnamefont {T.}~\bibnamefont
  {Louis}} \emph {et~al.} (\bibinfo {collaboration} {Atacama Cosmology
  Telescope}),\ }\href {https://doi.org/10.1088/1475-7516/2025/11/062}
  {\bibfield  {journal} {\bibinfo  {journal} {JCAP}\ }\textbf {\bibinfo
  {volume} {11}},\ \bibinfo {pages} {062}},\ \Eprint
  {https://arxiv.org/abs/2503.14452} {arXiv:2503.14452 [astro-ph.CO]}
  \BibitemShut {NoStop}%
\bibitem [{\citenamefont {Camphuis}\ \emph {et~al.}(2025)\citenamefont
  {Camphuis} \emph {et~al.}}]{SPT-3G:2025bzu}%
  \BibitemOpen
  \bibfield  {author} {\bibinfo {author} {\bibfnamefont {E.}~\bibnamefont
  {Camphuis}} \emph {et~al.} (\bibinfo {collaboration} {SPT-3G}),\ }\Eprint
  {https://arxiv.org/abs/2506.20707} {arXiv:2506.20707 [astro-ph.CO]}
  (\bibinfo {year} {2025})\BibitemShut {NoStop}%
\bibitem [{\citenamefont {Zhao}\ \emph {et~al.}(2022)\citenamefont {Zhao} \emph
  {et~al.}}]{eBOSS:2021pff}%
  \BibitemOpen
  \bibfield  {author} {\bibinfo {author} {\bibfnamefont {C.}~\bibnamefont
  {Zhao}} \emph {et~al.} (\bibinfo {collaboration} {eBOSS}),\ }\href
  {https://doi.org/10.1093/mnras/stac390} {\bibfield  {journal} {\bibinfo
  {journal} {Mon. Not. Roy. Astron. Soc.}\ }\textbf {\bibinfo {volume} {511}},\
  \bibinfo {pages} {5492} (\bibinfo {year} {2022})},\ \Eprint
  {https://arxiv.org/abs/2110.03824} {arXiv:2110.03824 [astro-ph.CO]}
  \BibitemShut {NoStop}%
\bibitem [{\citenamefont {Abdul~Karim}\ \emph {et~al.}(2025)\citenamefont
  {Abdul~Karim} \emph {et~al.}}]{DESI:2025zgx}%
  \BibitemOpen
  \bibfield  {author} {\bibinfo {author} {\bibfnamefont {M.}~\bibnamefont
  {Abdul~Karim}} \emph {et~al.} (\bibinfo {collaboration} {DESI}),\ }\href
  {https://doi.org/10.1103/tr6y-kpc6} {\bibfield  {journal} {\bibinfo
  {journal} {Phys. Rev. D}\ }\textbf {\bibinfo {volume} {112}},\ \bibinfo
  {pages} {083515} (\bibinfo {year} {2025})},\ \Eprint
  {https://arxiv.org/abs/2503.14738} {arXiv:2503.14738 [astro-ph.CO]}
  \BibitemShut {NoStop}%
\bibitem [{\citenamefont {Asgari}\ \emph {et~al.}(2021)\citenamefont {Asgari}
  \emph {et~al.}}]{KiDS:2020suj}%
  \BibitemOpen
  \bibfield  {author} {\bibinfo {author} {\bibfnamefont {M.}~\bibnamefont
  {Asgari}} \emph {et~al.} (\bibinfo {collaboration} {KiDS}),\ }\href
  {https://doi.org/10.1051/0004-6361/202039070} {\bibfield  {journal} {\bibinfo
   {journal} {Astron. Astrophys.}\ }\textbf {\bibinfo {volume} {645}},\
  \bibinfo {pages} {A104} (\bibinfo {year} {2021})},\ \Eprint
  {https://arxiv.org/abs/2007.15633} {arXiv:2007.15633 [astro-ph.CO]}
  \BibitemShut {NoStop}%
\bibitem [{\citenamefont {Amon}\ \emph {et~al.}(2022)\citenamefont {Amon} \emph
  {et~al.}}]{DES:2021bvc}%
  \BibitemOpen
  \bibfield  {author} {\bibinfo {author} {\bibfnamefont {A.}~\bibnamefont
  {Amon}} \emph {et~al.} (\bibinfo {collaboration} {DES}),\ }\href
  {https://doi.org/10.1103/PhysRevD.105.023514} {\bibfield  {journal} {\bibinfo
   {journal} {Phys. Rev. D}\ }\textbf {\bibinfo {volume} {105}},\ \bibinfo
  {pages} {023514} (\bibinfo {year} {2022})},\ \Eprint
  {https://arxiv.org/abs/2105.13543} {arXiv:2105.13543 [astro-ph.CO]}
  \BibitemShut {NoStop}%
\bibitem [{\citenamefont {Li}\ \emph {et~al.}(2023)\citenamefont {Li} \emph
  {et~al.}}]{Li:2023tui}%
  \BibitemOpen
  \bibfield  {author} {\bibinfo {author} {\bibfnamefont {X.}~\bibnamefont {Li}}
  \emph {et~al.},\ }\href {https://doi.org/10.1103/PhysRevD.108.123518}
  {\bibfield  {journal} {\bibinfo  {journal} {Phys. Rev. D}\ }\textbf {\bibinfo
  {volume} {108}},\ \bibinfo {pages} {123518} (\bibinfo {year} {2023})},\
  \Eprint {https://arxiv.org/abs/2304.00702} {arXiv:2304.00702 [astro-ph.CO]}
  \BibitemShut {NoStop}%
\bibitem [{\citenamefont {Riess}\ \emph {et~al.}(2022)\citenamefont {Riess}
  \emph {et~al.}}]{Riess:2021jrx}%
  \BibitemOpen
  \bibfield  {author} {\bibinfo {author} {\bibfnamefont {A.~G.}\ \bibnamefont
  {Riess}} \emph {et~al.},\ }\href {https://doi.org/10.3847/2041-8213/ac5c5b}
  {\bibfield  {journal} {\bibinfo  {journal} {Astrophys. J. Lett.}\ }\textbf
  {\bibinfo {volume} {934}},\ \bibinfo {pages} {L7} (\bibinfo {year} {2022})},\
  \Eprint {https://arxiv.org/abs/2112.04510} {arXiv:2112.04510 [astro-ph.CO]}
  \BibitemShut {NoStop}%
\bibitem [{\citenamefont {Di~Valentino}\ \emph {et~al.}(2025)\citenamefont
  {Di~Valentino}, \citenamefont {Levi~Said},\ and\ \citenamefont
  {Saridakis}}]{DiValentino:2025otz}%
  \BibitemOpen
  \bibfield  {author} {\bibinfo {author} {\bibfnamefont {E.}~\bibnamefont
  {Di~Valentino}}, \bibinfo {author} {\bibfnamefont {J.}~\bibnamefont
  {Levi~Said}},\ and\ \bibinfo {author} {\bibfnamefont {E.~N.}\ \bibnamefont
  {Saridakis}},\ }\Eprint {https://arxiv.org/abs/2509.25288} {arXiv:2509.25288
  [astro-ph.CO]}  (\bibinfo {year} {2025})\BibitemShut {NoStop}%
\bibitem [{\citenamefont {Abdalla}\ \emph {et~al.}(2022)\citenamefont {Abdalla}
  \emph {et~al.}}]{Abdalla:2022yfr}%
  \BibitemOpen
  \bibfield  {author} {\bibinfo {author} {\bibfnamefont {E.}~\bibnamefont
  {Abdalla}} \emph {et~al.},\ }\href
  {https://doi.org/10.1016/j.jheap.2022.04.002} {\bibfield  {journal} {\bibinfo
   {journal} {JHEAp}\ }\textbf {\bibinfo {volume} {34}},\ \bibinfo {pages} {49}
  (\bibinfo {year} {2022})},\ \Eprint {https://arxiv.org/abs/2203.06142}
  {arXiv:2203.06142 [astro-ph.CO]} \BibitemShut {NoStop}%
\bibitem [{\citenamefont {Di~Valentino}\ \emph {et~al.}(2021)\citenamefont
  {Di~Valentino}, \citenamefont {Mena}, \citenamefont {Pan}, \citenamefont
  {Visinelli}, \citenamefont {Yang}, \citenamefont {Melchiorri}, \citenamefont
  {Mota}, \citenamefont {Riess},\ and\ \citenamefont
  {Silk}}]{DiValentino:2021izs}%
  \BibitemOpen
  \bibfield  {author} {\bibinfo {author} {\bibfnamefont {E.}~\bibnamefont
  {Di~Valentino}}, \bibinfo {author} {\bibfnamefont {O.}~\bibnamefont {Mena}},
  \bibinfo {author} {\bibfnamefont {S.}~\bibnamefont {Pan}}, \bibinfo {author}
  {\bibfnamefont {L.}~\bibnamefont {Visinelli}}, \bibinfo {author}
  {\bibfnamefont {W.}~\bibnamefont {Yang}}, \bibinfo {author} {\bibfnamefont
  {A.}~\bibnamefont {Melchiorri}}, \bibinfo {author} {\bibfnamefont {D.~F.}\
  \bibnamefont {Mota}}, \bibinfo {author} {\bibfnamefont {A.~G.}\ \bibnamefont
  {Riess}},\ and\ \bibinfo {author} {\bibfnamefont {J.}~\bibnamefont {Silk}},\
  }\href {https://doi.org/10.1088/1361-6382/ac086d} {\bibfield  {journal}
  {\bibinfo  {journal} {Class. Quant. Grav.}\ }\textbf {\bibinfo {volume}
  {38}},\ \bibinfo {pages} {153001} (\bibinfo {year} {2021})},\ \Eprint
  {https://arxiv.org/abs/2103.01183} {arXiv:2103.01183 [astro-ph.CO]}
  \BibitemShut {NoStop}%
\bibitem [{\citenamefont {Chiu}\ \emph {et~al.}(2023)\citenamefont {Chiu},
  \citenamefont {Klein}, \citenamefont {Mohr},\ and\ \citenamefont
  {Bocquet}}]{Chiu:2022qgb}%
  \BibitemOpen
  \bibfield  {author} {\bibinfo {author} {\bibfnamefont {I.-N.}\ \bibnamefont
  {Chiu}}, \bibinfo {author} {\bibfnamefont {M.}~\bibnamefont {Klein}},
  \bibinfo {author} {\bibfnamefont {J.}~\bibnamefont {Mohr}},\ and\ \bibinfo
  {author} {\bibfnamefont {S.}~\bibnamefont {Bocquet}},\ }\href
  {https://doi.org/10.1093/mnras/stad957} {\bibfield  {journal} {\bibinfo
  {journal} {Mon. Not. Roy. Astron. Soc.}\ }\textbf {\bibinfo {volume} {522}},\
  \bibinfo {pages} {1601} (\bibinfo {year} {2023})},\ \Eprint
  {https://arxiv.org/abs/2207.12429} {arXiv:2207.12429 [astro-ph.CO]}
  \BibitemShut {NoStop}%
\bibitem [{\citenamefont {Schutz}(1986)}]{Schutz:1986gp}%
  \BibitemOpen
  \bibfield  {author} {\bibinfo {author} {\bibfnamefont {B.~F.}\ \bibnamefont
  {Schutz}},\ }\href {https://doi.org/10.1038/323310a0} {\bibfield  {journal}
  {\bibinfo  {journal} {Nature}\ }\textbf {\bibinfo {volume} {323}},\ \bibinfo
  {pages} {310} (\bibinfo {year} {1986})}\BibitemShut {NoStop}%
\bibitem [{\citenamefont {Schutz}(2001)}]{Schutz:2001re}%
  \BibitemOpen
  \bibfield  {author} {\bibinfo {author} {\bibfnamefont {B.~F.}\ \bibnamefont
  {Schutz}},\ }in\ \href {https://doi.org/10.1007/10856495_29} {\emph {\bibinfo
  {booktitle} {{MPA / ESO / MPE / USM Conference on Lighthouses of the
  Universe: The Most Luminous Celestial Objects and their use for
  Cosmology}}}}\ (\bibinfo {year} {2001})\ \Eprint
  {https://arxiv.org/abs/gr-qc/0111095} {arXiv:gr-qc/0111095} \BibitemShut
  {NoStop}%
\bibitem [{\citenamefont {Holz}\ and\ \citenamefont
  {Hughes}(2005)}]{Holz:2005df}%
  \BibitemOpen
  \bibfield  {author} {\bibinfo {author} {\bibfnamefont {D.~E.}\ \bibnamefont
  {Holz}}\ and\ \bibinfo {author} {\bibfnamefont {S.~A.}\ \bibnamefont
  {Hughes}},\ }\href {https://doi.org/10.1086/431341} {\bibfield  {journal}
  {\bibinfo  {journal} {Astrophys. J.}\ }\textbf {\bibinfo {volume} {629}},\
  \bibinfo {pages} {15} (\bibinfo {year} {2005})},\ \Eprint
  {https://arxiv.org/abs/astro-ph/0504616} {arXiv:astro-ph/0504616}
  \BibitemShut {NoStop}%
\bibitem [{\citenamefont {Abbott}\ \emph
  {et~al.}(2017{\natexlab{a}})\citenamefont {Abbott} \emph
  {et~al.}}]{LIGOScientific:2017vwq}%
  \BibitemOpen
  \bibfield  {author} {\bibinfo {author} {\bibfnamefont {B.~P.}\ \bibnamefont
  {Abbott}} \emph {et~al.} (\bibinfo {collaboration} {LIGO Scientific,
  Virgo}),\ }\href {https://doi.org/10.1103/PhysRevLett.119.161101} {\bibfield
  {journal} {\bibinfo  {journal} {Phys. Rev. Lett.}\ }\textbf {\bibinfo
  {volume} {119}},\ \bibinfo {pages} {161101} (\bibinfo {year}
  {2017}{\natexlab{a}})},\ \Eprint {https://arxiv.org/abs/1710.05832}
  {arXiv:1710.05832 [gr-qc]} \BibitemShut {NoStop}%
\bibitem [{\citenamefont {Abbott}\ \emph
  {et~al.}(2017{\natexlab{b}})\citenamefont {Abbott} \emph
  {et~al.}}]{LIGOScientific:2017ync}%
  \BibitemOpen
  \bibfield  {author} {\bibinfo {author} {\bibfnamefont {B.~P.}\ \bibnamefont
  {Abbott}} \emph {et~al.} (\bibinfo {collaboration} {LIGO Scientific, Virgo,
  Fermi GBM, INTEGRAL, IceCube, AstroSat Cadmium Zinc Telluride Imager Team,
  IPN, Insight-Hxmt, ANTARES, Swift, AGILE Team, 1M2H Team, Dark Energy Camera
  GW-EM, DES, DLT40, GRAWITA, Fermi-LAT, ATCA, ASKAP, Las Cumbres Observatory
  Group, OzGrav, DWF (Deeper Wider Faster Program), AST3, CAASTRO, VINROUGE,
  MASTER, J-GEM, GROWTH, JAGWAR, CaltechNRAO, TTU-NRAO, NuSTAR, Pan-STARRS,
  MAXI Team, TZAC Consortium, KU, Nordic Optical Telescope, ePESSTO, GROND,
  Texas Tech University, SALT Group, TOROS, BOOTES, MWA, CALET, IKI-GW
  Follow-up, H.E.S.S., LOFAR, LWA, HAWC, Pierre Auger, ALMA, Euro VLBI Team, Pi
  of Sky, Chandra Team at McGill University, DFN, ATLAS Telescopes, High Time
  Resolution Universe Survey, RIMAS, RATIR, SKA South Africa/MeerKAT}),\ }\href
  {https://doi.org/10.3847/2041-8213/aa91c9} {\bibfield  {journal} {\bibinfo
  {journal} {Astrophys. J. Lett.}\ }\textbf {\bibinfo {volume} {848}},\
  \bibinfo {pages} {L12} (\bibinfo {year} {2017}{\natexlab{b}})},\ \Eprint
  {https://arxiv.org/abs/1710.05833} {arXiv:1710.05833 [astro-ph.HE]}
  \BibitemShut {NoStop}%
\bibitem [{\citenamefont {Abbott}\ \emph
  {et~al.}(2017{\natexlab{c}})\citenamefont {Abbott} \emph
  {et~al.}}]{LIGOScientific:2017adf}%
  \BibitemOpen
  \bibfield  {author} {\bibinfo {author} {\bibfnamefont {B.~P.}\ \bibnamefont
  {Abbott}} \emph {et~al.} (\bibinfo {collaboration} {LIGO Scientific, Virgo,
  1M2H, Dark Energy Camera GW-E, DES, DLT40, Las Cumbres Observatory, VINROUGE,
  MASTER}),\ }\href {https://doi.org/10.1038/nature24471} {\bibfield  {journal}
  {\bibinfo  {journal} {Nature}\ }\textbf {\bibinfo {volume} {551}},\ \bibinfo
  {pages} {85} (\bibinfo {year} {2017}{\natexlab{c}})},\ \Eprint
  {https://arxiv.org/abs/1710.05835} {arXiv:1710.05835 [astro-ph.CO]}
  \BibitemShut {NoStop}%
\bibitem [{\citenamefont {Abbott}\ \emph {et~al.}(2020)\citenamefont {Abbott}
  \emph {et~al.}}]{LIGOScientific:2020iuh}%
  \BibitemOpen
  \bibfield  {author} {\bibinfo {author} {\bibfnamefont {R.}~\bibnamefont
  {Abbott}} \emph {et~al.} (\bibinfo {collaboration} {LIGO Scientific,
  Virgo}),\ }\href {https://doi.org/10.1103/PhysRevLett.125.101102} {\bibfield
  {journal} {\bibinfo  {journal} {Phys. Rev. Lett.}\ }\textbf {\bibinfo
  {volume} {125}},\ \bibinfo {pages} {101102} (\bibinfo {year} {2020})},\
  \Eprint {https://arxiv.org/abs/2009.01075} {arXiv:2009.01075 [gr-qc]}
  \BibitemShut {NoStop}%
\bibitem [{\citenamefont {McKernan}\ \emph {et~al.}(2019)\citenamefont
  {McKernan}, \citenamefont {Ford}, \citenamefont {Bartos}, \citenamefont
  {Graham}, \citenamefont {Lyra}, \citenamefont {Marka}, \citenamefont {Marka},
  \citenamefont {Ross}, \citenamefont {Stern},\ and\ \citenamefont
  {Yang}}]{McKernan:2019hqs}%
  \BibitemOpen
  \bibfield  {author} {\bibinfo {author} {\bibfnamefont {B.}~\bibnamefont
  {McKernan}}, \bibinfo {author} {\bibfnamefont {K.~E.~S.}\ \bibnamefont
  {Ford}}, \bibinfo {author} {\bibfnamefont {I.}~\bibnamefont {Bartos}},
  \bibinfo {author} {\bibfnamefont {M.~J.}\ \bibnamefont {Graham}}, \bibinfo
  {author} {\bibfnamefont {W.}~\bibnamefont {Lyra}}, \bibinfo {author}
  {\bibfnamefont {S.}~\bibnamefont {Marka}}, \bibinfo {author} {\bibfnamefont
  {Z.}~\bibnamefont {Marka}}, \bibinfo {author} {\bibfnamefont {N.~P.}\
  \bibnamefont {Ross}}, \bibinfo {author} {\bibfnamefont {D.}~\bibnamefont
  {Stern}},\ and\ \bibinfo {author} {\bibfnamefont {Y.}~\bibnamefont {Yang}},\
  }\href {https://doi.org/10.3847/2041-8213/ab4886} {\bibfield  {journal}
  {\bibinfo  {journal} {Astrophys. J. Lett.}\ }\textbf {\bibinfo {volume}
  {884}},\ \bibinfo {pages} {L50} (\bibinfo {year} {2019})},\ \Eprint
  {https://arxiv.org/abs/1907.03746} {arXiv:1907.03746 [astro-ph.HE]}
  \BibitemShut {NoStop}%
\bibitem [{\citenamefont {Ashton}\ \emph {et~al.}(2021)\citenamefont {Ashton},
  \citenamefont {Ackley}, \citenamefont {Hernandez},\ and\ \citenamefont
  {Piotrzkowski}}]{Ashton:2020kyr}%
  \BibitemOpen
  \bibfield  {author} {\bibinfo {author} {\bibfnamefont {G.}~\bibnamefont
  {Ashton}}, \bibinfo {author} {\bibfnamefont {K.}~\bibnamefont {Ackley}},
  \bibinfo {author} {\bibfnamefont {I.~M.}\ \bibnamefont {Hernandez}},\ and\
  \bibinfo {author} {\bibfnamefont {B.}~\bibnamefont {Piotrzkowski}},\ }\href
  {https://doi.org/10.1088/1361-6382/ac33bb} {\bibfield  {journal} {\bibinfo
  {journal} {Class. Quant. Grav.}\ }\textbf {\bibinfo {volume} {38}},\ \bibinfo
  {pages} {235004} (\bibinfo {year} {2021})},\ \Eprint
  {https://arxiv.org/abs/2009.12346} {arXiv:2009.12346 [astro-ph.HE]}
  \BibitemShut {NoStop}%
\bibitem [{\citenamefont {Chen}\ \emph {et~al.}(2018)\citenamefont {Chen},
  \citenamefont {Fishbach},\ and\ \citenamefont {Holz}}]{Chen:2017rfc}%
  \BibitemOpen
  \bibfield  {author} {\bibinfo {author} {\bibfnamefont {H.-Y.}\ \bibnamefont
  {Chen}}, \bibinfo {author} {\bibfnamefont {M.}~\bibnamefont {Fishbach}},\
  and\ \bibinfo {author} {\bibfnamefont {D.~E.}\ \bibnamefont {Holz}},\ }\href
  {https://doi.org/10.1038/s41586-018-0606-0} {\bibfield  {journal} {\bibinfo
  {journal} {Nature}\ }\textbf {\bibinfo {volume} {562}},\ \bibinfo {pages}
  {545} (\bibinfo {year} {2018})},\ \Eprint {https://arxiv.org/abs/1712.06531}
  {arXiv:1712.06531 [astro-ph.CO]} \BibitemShut {NoStop}%
\bibitem [{\citenamefont {Del~Pozzo}(2012)}]{DelPozzo:2011vcw}%
  \BibitemOpen
  \bibfield  {author} {\bibinfo {author} {\bibfnamefont {W.}~\bibnamefont
  {Del~Pozzo}},\ }\href {https://doi.org/10.1103/PhysRevD.86.043011} {\bibfield
   {journal} {\bibinfo  {journal} {Phys. Rev. D}\ }\textbf {\bibinfo {volume}
  {86}},\ \bibinfo {pages} {043011} (\bibinfo {year} {2012})},\ \Eprint
  {https://arxiv.org/abs/1108.1317} {arXiv:1108.1317 [astro-ph.CO]}
  \BibitemShut {NoStop}%
\bibitem [{\citenamefont {Fishbach}\ \emph {et~al.}(2019)\citenamefont
  {Fishbach} \emph {et~al.}}]{LIGOScientific:2018gmd}%
  \BibitemOpen
  \bibfield  {author} {\bibinfo {author} {\bibfnamefont {M.}~\bibnamefont
  {Fishbach}} \emph {et~al.} (\bibinfo {collaboration} {LIGO Scientific,
  Virgo}),\ }\href {https://doi.org/10.3847/2041-8213/aaf96e} {\bibfield
  {journal} {\bibinfo  {journal} {Astrophys. J. Lett.}\ }\textbf {\bibinfo
  {volume} {871}},\ \bibinfo {pages} {L13} (\bibinfo {year} {2019})},\ \Eprint
  {https://arxiv.org/abs/1807.05667} {arXiv:1807.05667 [astro-ph.CO]}
  \BibitemShut {NoStop}%
\bibitem [{\citenamefont {Abbott}\ \emph
  {et~al.}(2017{\natexlab{d}})\citenamefont {Abbott} \emph
  {et~al.}}]{LIGOScientific:2017ycc}%
  \BibitemOpen
  \bibfield  {author} {\bibinfo {author} {\bibfnamefont {B.~P.}\ \bibnamefont
  {Abbott}} \emph {et~al.} (\bibinfo {collaboration} {LIGO Scientific,
  Virgo}),\ }\href {https://doi.org/10.1103/PhysRevLett.119.141101} {\bibfield
  {journal} {\bibinfo  {journal} {Phys. Rev. Lett.}\ }\textbf {\bibinfo
  {volume} {119}},\ \bibinfo {pages} {141101} (\bibinfo {year}
  {2017}{\natexlab{d}})},\ \Eprint {https://arxiv.org/abs/1709.09660}
  {arXiv:1709.09660 [gr-qc]} \BibitemShut {NoStop}%
\bibitem [{\citenamefont {Abbott}\ \emph {et~al.}(2018)\citenamefont {Abbott}
  \emph {et~al.}}]{DES:2018gui}%
  \BibitemOpen
  \bibfield  {author} {\bibinfo {author} {\bibfnamefont {T.~M.~C.}\
  \bibnamefont {Abbott}} \emph {et~al.} (\bibinfo {collaboration} {DES, NOAO
  Data Lab}),\ }\href {https://doi.org/10.3847/1538-4365/aae9f0} {\bibfield
  {journal} {\bibinfo  {journal} {Astrophys. J. Suppl.}\ }\textbf {\bibinfo
  {volume} {239}},\ \bibinfo {pages} {18} (\bibinfo {year} {2018})},\ \Eprint
  {https://arxiv.org/abs/1801.03181} {arXiv:1801.03181 [astro-ph.IM]}
  \BibitemShut {NoStop}%
\bibitem [{\citenamefont {Soares-Santos}\ \emph {et~al.}(2019)\citenamefont
  {Soares-Santos} \emph {et~al.}}]{DES:2019ccw}%
  \BibitemOpen
  \bibfield  {author} {\bibinfo {author} {\bibfnamefont {M.}~\bibnamefont
  {Soares-Santos}} \emph {et~al.} (\bibinfo {collaboration} {DES, LIGO
  Scientific, Virgo}),\ }\href {https://doi.org/10.3847/2041-8213/ab14f1}
  {\bibfield  {journal} {\bibinfo  {journal} {Astrophys. J. Lett.}\ }\textbf
  {\bibinfo {volume} {876}},\ \bibinfo {pages} {L7} (\bibinfo {year} {2019})},\
  \Eprint {https://arxiv.org/abs/1901.01540} {arXiv:1901.01540 [astro-ph.CO]}
  \BibitemShut {NoStop}%
\bibitem [{\citenamefont {Aasi}\ \emph {et~al.}(2015)\citenamefont {Aasi} \emph
  {et~al.}}]{LIGOScientific:2014pky}%
  \BibitemOpen
  \bibfield  {author} {\bibinfo {author} {\bibfnamefont {J.}~\bibnamefont
  {Aasi}} \emph {et~al.} (\bibinfo {collaboration} {LIGO Scientific}),\ }\href
  {https://doi.org/10.1088/0264-9381/32/7/074001} {\bibfield  {journal}
  {\bibinfo  {journal} {Class. Quant. Grav.}\ }\textbf {\bibinfo {volume}
  {32}},\ \bibinfo {pages} {074001} (\bibinfo {year} {2015})},\ \Eprint
  {https://arxiv.org/abs/1411.4547} {arXiv:1411.4547 [gr-qc]} \BibitemShut
  {NoStop}%
\bibitem [{\citenamefont {Acernese}\ \emph {et~al.}(2015)\citenamefont
  {Acernese} \emph {et~al.}}]{VIRGO:2014yos}%
  \BibitemOpen
  \bibfield  {author} {\bibinfo {author} {\bibfnamefont {F.}~\bibnamefont
  {Acernese}} \emph {et~al.} (\bibinfo {collaboration} {VIRGO}),\ }\href
  {https://doi.org/10.1088/0264-9381/32/2/024001} {\bibfield  {journal}
  {\bibinfo  {journal} {Class. Quant. Grav.}\ }\textbf {\bibinfo {volume}
  {32}},\ \bibinfo {pages} {024001} (\bibinfo {year} {2015})},\ \Eprint
  {https://arxiv.org/abs/1408.3978} {arXiv:1408.3978 [gr-qc]} \BibitemShut
  {NoStop}%
\bibitem [{\citenamefont {Abac}\ \emph
  {et~al.}(2025{\natexlab{a}})\citenamefont {Abac} \emph
  {et~al.}}]{LIGOScientific:2025slb}%
  \BibitemOpen
  \bibfield  {author} {\bibinfo {author} {\bibfnamefont {A.~G.}\ \bibnamefont
  {Abac}} \emph {et~al.} (\bibinfo {collaboration} {LIGO Scientific, VIRGO,
  KAGRA}),\ }\Eprint {https://arxiv.org/abs/2508.18082} {arXiv:2508.18082
  [gr-qc]}  (\bibinfo {year} {2025}{\natexlab{a}})\BibitemShut {NoStop}%
\bibitem [{\citenamefont {D{\'a}lya}\ \emph {et~al.}(2022)\citenamefont
  {D{\'a}lya} \emph {et~al.}}]{Dalya:2021ewn}%
  \BibitemOpen
  \bibfield  {author} {\bibinfo {author} {\bibfnamefont {G.}~\bibnamefont
  {D{\'a}lya}} \emph {et~al.},\ }\href {https://doi.org/10.1093/mnras/stac1443}
  {\bibfield  {journal} {\bibinfo  {journal} {Mon. Not. Roy. Astron. Soc.}\
  }\textbf {\bibinfo {volume} {514}},\ \bibinfo {pages} {1403} (\bibinfo {year}
  {2022})},\ \Eprint {https://arxiv.org/abs/2110.06184} {arXiv:2110.06184
  [astro-ph.CO]} \BibitemShut {NoStop}%
\bibitem [{\citenamefont {Abac}\ \emph
  {et~al.}(2025{\natexlab{b}})\citenamefont {Abac} \emph
  {et~al.}}]{LIGOScientific:2025jau}%
  \BibitemOpen
  \bibfield  {author} {\bibinfo {author} {\bibfnamefont {A.~G.}\ \bibnamefont
  {Abac}} \emph {et~al.} (\bibinfo {collaboration} {LIGO Scientific, VIRGO,
  KAGRA}),\ }\href@noop {} {\bibinfo {title} {{GWTC-4.0: Constraints on the
  Cosmic Expansion Rate and Modified Gravitational-wave Propagation}}}
  (\bibinfo {year} {2025}{\natexlab{b}}),\ \Eprint
  {https://arxiv.org/abs/2509.04348} {arXiv:2509.04348 [astro-ph.CO]}
  \BibitemShut {NoStop}%
\bibitem [{\citenamefont {Ade}\ \emph {et~al.}(2016)\citenamefont {Ade} \emph
  {et~al.}}]{Planck:2015fie}%
  \BibitemOpen
  \bibfield  {author} {\bibinfo {author} {\bibfnamefont {P.~A.~R.}\
  \bibnamefont {Ade}} \emph {et~al.} (\bibinfo {collaboration} {Planck}),\
  }\href {https://doi.org/10.1051/0004-6361/201525830} {\bibfield  {journal}
  {\bibinfo  {journal} {Astron. Astrophys.}\ }\textbf {\bibinfo {volume}
  {594}},\ \bibinfo {pages} {A13} (\bibinfo {year} {2016})},\ \Eprint
  {https://arxiv.org/abs/1502.01589} {arXiv:1502.01589 [astro-ph.CO]}
  \BibitemShut {NoStop}%
\bibitem [{\citenamefont {Riess}\ \emph {et~al.}(2021)\citenamefont {Riess},
  \citenamefont {Casertano}, \citenamefont {Yuan}, \citenamefont {Bowers},
  \citenamefont {Macri}, \citenamefont {Zinn},\ and\ \citenamefont
  {Scolnic}}]{Riess:2020fzl}%
  \BibitemOpen
  \bibfield  {author} {\bibinfo {author} {\bibfnamefont {A.~G.}\ \bibnamefont
  {Riess}}, \bibinfo {author} {\bibfnamefont {S.}~\bibnamefont {Casertano}},
  \bibinfo {author} {\bibfnamefont {W.}~\bibnamefont {Yuan}}, \bibinfo {author}
  {\bibfnamefont {J.~B.}\ \bibnamefont {Bowers}}, \bibinfo {author}
  {\bibfnamefont {L.}~\bibnamefont {Macri}}, \bibinfo {author} {\bibfnamefont
  {J.~C.}\ \bibnamefont {Zinn}},\ and\ \bibinfo {author} {\bibfnamefont
  {D.}~\bibnamefont {Scolnic}},\ }\href
  {https://doi.org/10.3847/2041-8213/abdbaf} {\bibfield  {journal} {\bibinfo
  {journal} {Astrophys. J. Lett.}\ }\textbf {\bibinfo {volume} {908}},\
  \bibinfo {pages} {L6} (\bibinfo {year} {2021})},\ \Eprint
  {https://arxiv.org/abs/2012.08534} {arXiv:2012.08534 [astro-ph.CO]}
  \BibitemShut {NoStop}%
\bibitem [{\citenamefont {Reitze}\ \emph {et~al.}(2019)\citenamefont {Reitze}
  \emph {et~al.}}]{Reitze:2019iox}%
  \BibitemOpen
  \bibfield  {author} {\bibinfo {author} {\bibfnamefont {D.}~\bibnamefont
  {Reitze}} \emph {et~al.},\ }\href@noop {} {\bibfield  {journal} {\bibinfo
  {journal} {Bull. Am. Astron. Soc.}\ }\textbf {\bibinfo {volume} {51}},\
  \bibinfo {pages} {035} (\bibinfo {year} {2019})},\ \Eprint
  {https://arxiv.org/abs/1907.04833} {arXiv:1907.04833 [astro-ph.IM]}
  \BibitemShut {NoStop}%
\bibitem [{\citenamefont {Evans}\ \emph {et~al.}(2021)\citenamefont {Evans}
  \emph {et~al.}}]{Evans:2021gyd}%
  \BibitemOpen
  \bibfield  {author} {\bibinfo {author} {\bibfnamefont {M.}~\bibnamefont
  {Evans}} \emph {et~al.},\ }\href@noop {} {\bibinfo {title} {{A Horizon Study
  for Cosmic Explorer: Science, Observatories, and Community}}} (\bibinfo
  {year} {2021}),\ \Eprint {https://arxiv.org/abs/2109.09882} {arXiv:2109.09882
  [astro-ph.IM]} \BibitemShut {NoStop}%
\bibitem [{\citenamefont {Evans}\ \emph {et~al.}(2023)\citenamefont {Evans}
  \emph {et~al.}}]{Evans:2023euw}%
  \BibitemOpen
  \bibfield  {author} {\bibinfo {author} {\bibfnamefont {M.}~\bibnamefont
  {Evans}} \emph {et~al.},\ }\Eprint {https://arxiv.org/abs/2306.13745}
  {arXiv:2306.13745 [astro-ph.IM]}  (\bibinfo {year} {2023})\BibitemShut
  {NoStop}%
\bibitem [{\citenamefont {Gupta}\ \emph {et~al.}(2024)\citenamefont {Gupta}
  \emph {et~al.}}]{Gupta:2023lga}%
  \BibitemOpen
  \bibfield  {author} {\bibinfo {author} {\bibfnamefont {I.}~\bibnamefont
  {Gupta}} \emph {et~al.},\ }\href {https://doi.org/10.1088/1361-6382/ad7b99}
  {\bibfield  {journal} {\bibinfo  {journal} {Class. Quant. Grav.}\ }\textbf
  {\bibinfo {volume} {41}},\ \bibinfo {pages} {245001} (\bibinfo {year}
  {2024})},\ \Eprint {https://arxiv.org/abs/2307.10421} {arXiv:2307.10421
  [gr-qc]} \BibitemShut {NoStop}%
\bibitem [{\citenamefont {Punturo}\ \emph {et~al.}(2010)\citenamefont {Punturo}
  \emph {et~al.}}]{Punturo:2010zz}%
  \BibitemOpen
  \bibfield  {author} {\bibinfo {author} {\bibfnamefont {M.}~\bibnamefont
  {Punturo}} \emph {et~al.},\ }\href
  {https://doi.org/10.1088/0264-9381/27/19/194002} {\bibfield  {journal}
  {\bibinfo  {journal} {Class. Quant. Grav.}\ }\textbf {\bibinfo {volume}
  {27}},\ \bibinfo {pages} {194002} (\bibinfo {year} {2010})}\BibitemShut
  {NoStop}%
\bibitem [{\citenamefont {Hild}\ \emph {et~al.}(2011)\citenamefont {Hild} \emph
  {et~al.}}]{Hild:2010id}%
  \BibitemOpen
  \bibfield  {author} {\bibinfo {author} {\bibfnamefont {S.}~\bibnamefont
  {Hild}} \emph {et~al.},\ }\href
  {https://doi.org/10.1088/0264-9381/28/9/094013} {\bibfield  {journal}
  {\bibinfo  {journal} {Class. Quant. Grav.}\ }\textbf {\bibinfo {volume}
  {28}},\ \bibinfo {pages} {094013} (\bibinfo {year} {2011})},\ \Eprint
  {https://arxiv.org/abs/1012.0908} {arXiv:1012.0908 [gr-qc]} \BibitemShut
  {NoStop}%
\bibitem [{\citenamefont {Branchesi}\ \emph {et~al.}(2023)\citenamefont
  {Branchesi} \emph {et~al.}}]{Branchesi:2023mws}%
  \BibitemOpen
  \bibfield  {author} {\bibinfo {author} {\bibfnamefont {M.}~\bibnamefont
  {Branchesi}} \emph {et~al.},\ }\href
  {https://doi.org/10.1088/1475-7516/2023/07/068} {\bibfield  {journal}
  {\bibinfo  {journal} {JCAP}\ }\textbf {\bibinfo {volume} {07}},\ \bibinfo
  {pages} {068}},\ \Eprint {https://arxiv.org/abs/2303.15923} {arXiv:2303.15923
  [gr-qc]} \BibitemShut {NoStop}%
\bibitem [{\citenamefont {Abac}\ \emph {et~al.}(2026)\citenamefont {Abac} \emph
  {et~al.}}]{ET:2025xjr}%
  \BibitemOpen
  \bibfield  {author} {\bibinfo {author} {\bibfnamefont {A.}~\bibnamefont
  {Abac}} \emph {et~al.} (\bibinfo {collaboration} {ET}),\ }\href
  {https://doi.org/10.1088/1475-7516/2026/03/081} {\bibfield  {journal}
  {\bibinfo  {journal} {JCAP}\ }\textbf {\bibinfo {volume} {03}},\ \bibinfo
  {pages} {081}},\ \Eprint {https://arxiv.org/abs/2503.12263} {arXiv:2503.12263
  [gr-qc]} \BibitemShut {NoStop}%
\bibitem [{\citenamefont {Muttoni}\ \emph {et~al.}(2023)\citenamefont
  {Muttoni}, \citenamefont {Laghi}, \citenamefont {Tamanini}, \citenamefont
  {Marsat},\ and\ \citenamefont {Izquierdo-Villalba}}]{Muttoni:2023prw}%
  \BibitemOpen
  \bibfield  {author} {\bibinfo {author} {\bibfnamefont {N.}~\bibnamefont
  {Muttoni}}, \bibinfo {author} {\bibfnamefont {D.}~\bibnamefont {Laghi}},
  \bibinfo {author} {\bibfnamefont {N.}~\bibnamefont {Tamanini}}, \bibinfo
  {author} {\bibfnamefont {S.}~\bibnamefont {Marsat}},\ and\ \bibinfo {author}
  {\bibfnamefont {D.}~\bibnamefont {Izquierdo-Villalba}},\ }\href
  {https://doi.org/10.1103/PhysRevD.108.043543} {\bibfield  {journal} {\bibinfo
   {journal} {Phys. Rev. D}\ }\textbf {\bibinfo {volume} {108}},\ \bibinfo
  {pages} {043543} (\bibinfo {year} {2023})},\ \Eprint
  {https://arxiv.org/abs/2303.10693} {arXiv:2303.10693 [astro-ph.CO]}
  \BibitemShut {NoStop}%
\bibitem [{\citenamefont {Chen}\ and\ \citenamefont
  {Holz}(2016)}]{Chen:2016tys}%
  \BibitemOpen
  \bibfield  {author} {\bibinfo {author} {\bibfnamefont {H.-Y.}\ \bibnamefont
  {Chen}}\ and\ \bibinfo {author} {\bibfnamefont {D.~E.}\ \bibnamefont
  {Holz}},\ }\Eprint {https://arxiv.org/abs/1612.01471} {arXiv:1612.01471
  [astro-ph.HE]}  (\bibinfo {year} {2016})\BibitemShut {NoStop}%
\bibitem [{\citenamefont {Nishizawa}(2017)}]{Nishizawa:2016ood}%
  \BibitemOpen
  \bibfield  {author} {\bibinfo {author} {\bibfnamefont {A.}~\bibnamefont
  {Nishizawa}},\ }\href {https://doi.org/10.1103/PhysRevD.96.101303} {\bibfield
   {journal} {\bibinfo  {journal} {Phys. Rev. D}\ }\textbf {\bibinfo {volume}
  {96}},\ \bibinfo {pages} {101303} (\bibinfo {year} {2017})},\ \Eprint
  {https://arxiv.org/abs/1612.06060} {arXiv:1612.06060 [astro-ph.CO]}
  \BibitemShut {NoStop}%
\bibitem [{\citenamefont {Borhanian}\ \emph {et~al.}(2020)\citenamefont
  {Borhanian}, \citenamefont {Dhani}, \citenamefont {Gupta}, \citenamefont
  {Arun},\ and\ \citenamefont {Sathyaprakash}}]{Borhanian:2020vyr}%
  \BibitemOpen
  \bibfield  {author} {\bibinfo {author} {\bibfnamefont {S.}~\bibnamefont
  {Borhanian}}, \bibinfo {author} {\bibfnamefont {A.}~\bibnamefont {Dhani}},
  \bibinfo {author} {\bibfnamefont {A.}~\bibnamefont {Gupta}}, \bibinfo
  {author} {\bibfnamefont {K.~G.}\ \bibnamefont {Arun}},\ and\ \bibinfo
  {author} {\bibfnamefont {B.~S.}\ \bibnamefont {Sathyaprakash}},\ }\href
  {https://doi.org/10.3847/2041-8213/abcaf5} {\bibfield  {journal} {\bibinfo
  {journal} {Astrophys. J. Lett.}\ }\textbf {\bibinfo {volume} {905}},\
  \bibinfo {pages} {L28} (\bibinfo {year} {2020})},\ \Eprint
  {https://arxiv.org/abs/2007.02883} {arXiv:2007.02883 [astro-ph.CO]}
  \BibitemShut {NoStop}%
\bibitem [{\citenamefont {Gupta}(2023{\natexlab{a}})}]{Gupta:2022fwd}%
  \BibitemOpen
  \bibfield  {author} {\bibinfo {author} {\bibfnamefont {I.}~\bibnamefont
  {Gupta}},\ }\href {https://doi.org/10.1093/mnras/stad2115} {\bibfield
  {journal} {\bibinfo  {journal} {Mon. Not. Roy. Astron. Soc.}\ }\textbf
  {\bibinfo {volume} {524}},\ \bibinfo {pages} {3537} (\bibinfo {year}
  {2023}{\natexlab{a}})},\ \Eprint {https://arxiv.org/abs/2212.00163}
  {arXiv:2212.00163 [gr-qc]} \BibitemShut {NoStop}%
\bibitem [{\citenamefont {Chen}\ \emph {et~al.}(2024)\citenamefont {Chen},
  \citenamefont {Ezquiaga},\ and\ \citenamefont {Gupta}}]{Chen:2024gdn}%
  \BibitemOpen
  \bibfield  {author} {\bibinfo {author} {\bibfnamefont {H.-Y.}\ \bibnamefont
  {Chen}}, \bibinfo {author} {\bibfnamefont {J.~M.}\ \bibnamefont {Ezquiaga}},\
  and\ \bibinfo {author} {\bibfnamefont {I.}~\bibnamefont {Gupta}},\ }\href
  {https://doi.org/10.1088/1361-6382/ad424f} {\bibfield  {journal} {\bibinfo
  {journal} {Class. Quant. Grav.}\ }\textbf {\bibinfo {volume} {41}},\ \bibinfo
  {pages} {125004} (\bibinfo {year} {2024})},\ \Eprint
  {https://arxiv.org/abs/2402.03120} {arXiv:2402.03120 [gr-qc]} \BibitemShut
  {NoStop}%
\bibitem [{\citenamefont {Belgacem}\ \emph {et~al.}(2019)\citenamefont
  {Belgacem}, \citenamefont {Dirian}, \citenamefont {Foffa}, \citenamefont
  {Howell}, \citenamefont {Maggiore},\ and\ \citenamefont
  {Regimbau}}]{Belgacem:2019tbw}%
  \BibitemOpen
  \bibfield  {author} {\bibinfo {author} {\bibfnamefont {E.}~\bibnamefont
  {Belgacem}}, \bibinfo {author} {\bibfnamefont {Y.}~\bibnamefont {Dirian}},
  \bibinfo {author} {\bibfnamefont {S.}~\bibnamefont {Foffa}}, \bibinfo
  {author} {\bibfnamefont {E.~J.}\ \bibnamefont {Howell}}, \bibinfo {author}
  {\bibfnamefont {M.}~\bibnamefont {Maggiore}},\ and\ \bibinfo {author}
  {\bibfnamefont {T.}~\bibnamefont {Regimbau}},\ }\href
  {https://doi.org/10.1088/1475-7516/2019/08/015} {\bibfield  {journal}
  {\bibinfo  {journal} {JCAP}\ }\textbf {\bibinfo {volume} {08}},\ \bibinfo
  {pages} {015}},\ \Eprint {https://arxiv.org/abs/1907.01487} {arXiv:1907.01487
  [astro-ph.CO]} \BibitemShut {NoStop}%
\bibitem [{\citenamefont {Dhani}\ \emph {et~al.}(2022)\citenamefont {Dhani},
  \citenamefont {Borhanian}, \citenamefont {Gupta},\ and\ \citenamefont
  {Sathyaprakash}}]{Dhani:2022ulg}%
  \BibitemOpen
  \bibfield  {author} {\bibinfo {author} {\bibfnamefont {A.}~\bibnamefont
  {Dhani}}, \bibinfo {author} {\bibfnamefont {S.}~\bibnamefont {Borhanian}},
  \bibinfo {author} {\bibfnamefont {A.}~\bibnamefont {Gupta}},\ and\ \bibinfo
  {author} {\bibfnamefont {B.}~\bibnamefont {Sathyaprakash}},\ }\href@noop {}
  {\bibinfo {title} {{Cosmography with bright and Love sirens}}} (\bibinfo
  {year} {2022}),\ \Eprint {https://arxiv.org/abs/2212.13183} {arXiv:2212.13183
  [gr-qc]} \BibitemShut {NoStop}%
\bibitem [{\citenamefont {Amaro-Seoane}\ \emph {et~al.}(2017)\citenamefont
  {Amaro-Seoane} \emph {et~al.}}]{LISA:2017pwj}%
  \BibitemOpen
  \bibfield  {author} {\bibinfo {author} {\bibfnamefont {P.}~\bibnamefont
  {Amaro-Seoane}} \emph {et~al.},\ }\href@noop {} {\bibinfo {title} {{Laser
  Interferometer Space Antenna}}} (\bibinfo {year} {2017}),\ \Eprint
  {https://arxiv.org/abs/1702.00786} {arXiv:1702.00786 [astro-ph.IM]}
  \BibitemShut {NoStop}%
\bibitem [{\citenamefont {Ajith}\ \emph {et~al.}(2025)\citenamefont {Ajith}
  \emph {et~al.}}]{Ajith:2024mie}%
  \BibitemOpen
  \bibfield  {author} {\bibinfo {author} {\bibfnamefont {P.}~\bibnamefont
  {Ajith}} \emph {et~al.},\ }\href
  {https://doi.org/10.1088/1475-7516/2025/01/108} {\bibfield  {journal}
  {\bibinfo  {journal} {JCAP}\ }\textbf {\bibinfo {volume} {01}},\ \bibinfo
  {pages} {108}},\ \Eprint {https://arxiv.org/abs/2404.09181} {arXiv:2404.09181
  [gr-qc]} \BibitemShut {NoStop}%
\bibitem [{\citenamefont {Califano}\ \emph {et~al.}(2022)\citenamefont
  {Califano}, \citenamefont {de~Martino}, \citenamefont {Vernieri},\ and\
  \citenamefont {Capozziello}}]{Califano:2022cmo}%
  \BibitemOpen
  \bibfield  {author} {\bibinfo {author} {\bibfnamefont {M.}~\bibnamefont
  {Califano}}, \bibinfo {author} {\bibfnamefont {I.}~\bibnamefont
  {de~Martino}}, \bibinfo {author} {\bibfnamefont {D.}~\bibnamefont
  {Vernieri}},\ and\ \bibinfo {author} {\bibfnamefont {S.}~\bibnamefont
  {Capozziello}},\ }\href {https://doi.org/10.1093/mnras/stac3230} {\bibfield
  {journal} {\bibinfo  {journal} {Mon. Not. Roy. Astron. Soc.}\ }\textbf
  {\bibinfo {volume} {518}},\ \bibinfo {pages} {3372} (\bibinfo {year}
  {2022})},\ \Eprint {https://arxiv.org/abs/2205.11221} {arXiv:2205.11221
  [astro-ph.CO]} \BibitemShut {NoStop}%
\bibitem [{\citenamefont {Reali}\ \emph {et~al.}(2024)\citenamefont {Reali},
  \citenamefont {Cotesta}, \citenamefont {Antonelli}, \citenamefont {Kritos},
  \citenamefont {Strokov},\ and\ \citenamefont {Berti}}]{Reali:2024hqf}%
  \BibitemOpen
  \bibfield  {author} {\bibinfo {author} {\bibfnamefont {L.}~\bibnamefont
  {Reali}}, \bibinfo {author} {\bibfnamefont {R.}~\bibnamefont {Cotesta}},
  \bibinfo {author} {\bibfnamefont {A.}~\bibnamefont {Antonelli}}, \bibinfo
  {author} {\bibfnamefont {K.}~\bibnamefont {Kritos}}, \bibinfo {author}
  {\bibfnamefont {V.}~\bibnamefont {Strokov}},\ and\ \bibinfo {author}
  {\bibfnamefont {E.}~\bibnamefont {Berti}},\ }\href
  {https://doi.org/10.1103/PhysRevD.110.103002} {\bibfield  {journal} {\bibinfo
   {journal} {Phys. Rev. D}\ }\textbf {\bibinfo {volume} {110}},\ \bibinfo
  {pages} {103002} (\bibinfo {year} {2024})},\ \Eprint
  {https://arxiv.org/abs/2406.01687} {arXiv:2406.01687 [gr-qc]} \BibitemShut
  {NoStop}%
\bibitem [{\citenamefont {Jani}\ \emph {et~al.}(2019)\citenamefont {Jani},
  \citenamefont {Shoemaker},\ and\ \citenamefont {Cutler}}]{Jani:2019ffg}%
  \BibitemOpen
  \bibfield  {author} {\bibinfo {author} {\bibfnamefont {K.}~\bibnamefont
  {Jani}}, \bibinfo {author} {\bibfnamefont {D.}~\bibnamefont {Shoemaker}},\
  and\ \bibinfo {author} {\bibfnamefont {C.}~\bibnamefont {Cutler}},\ }\href
  {https://doi.org/10.1038/s41550-019-0932-7} {\bibfield  {journal} {\bibinfo
  {journal} {Nature Astron.}\ }\textbf {\bibinfo {volume} {4}},\ \bibinfo
  {pages} {260} (\bibinfo {year} {2019})},\ \Eprint
  {https://arxiv.org/abs/1908.04985} {arXiv:1908.04985 [gr-qc]} \BibitemShut
  {NoStop}%
\bibitem [{\citenamefont {Grimm}\ and\ \citenamefont
  {Harms}(2020)}]{Grimm:2020ivq}%
  \BibitemOpen
  \bibfield  {author} {\bibinfo {author} {\bibfnamefont {S.}~\bibnamefont
  {Grimm}}\ and\ \bibinfo {author} {\bibfnamefont {J.}~\bibnamefont {Harms}},\
  }\href {https://doi.org/10.1103/PhysRevD.102.022007} {\bibfield  {journal}
  {\bibinfo  {journal} {Phys. Rev. D}\ }\textbf {\bibinfo {volume} {102}},\
  \bibinfo {pages} {022007} (\bibinfo {year} {2020})},\ \Eprint
  {https://arxiv.org/abs/2004.01434} {arXiv:2004.01434 [gr-qc]} \BibitemShut
  {NoStop}%
\bibitem [{\citenamefont {Muttoni}\ \emph {et~al.}(2022)\citenamefont
  {Muttoni}, \citenamefont {Mangiagli}, \citenamefont {Sesana}, \citenamefont
  {Laghi}, \citenamefont {Del~Pozzo}, \citenamefont {Izquierdo-Villalba},\ and\
  \citenamefont {Rosati}}]{Muttoni:2021veo}%
  \BibitemOpen
  \bibfield  {author} {\bibinfo {author} {\bibfnamefont {N.}~\bibnamefont
  {Muttoni}}, \bibinfo {author} {\bibfnamefont {A.}~\bibnamefont {Mangiagli}},
  \bibinfo {author} {\bibfnamefont {A.}~\bibnamefont {Sesana}}, \bibinfo
  {author} {\bibfnamefont {D.}~\bibnamefont {Laghi}}, \bibinfo {author}
  {\bibfnamefont {W.}~\bibnamefont {Del~Pozzo}}, \bibinfo {author}
  {\bibfnamefont {D.}~\bibnamefont {Izquierdo-Villalba}},\ and\ \bibinfo
  {author} {\bibfnamefont {M.}~\bibnamefont {Rosati}},\ }\href
  {https://doi.org/10.1103/PhysRevD.105.043509} {\bibfield  {journal} {\bibinfo
   {journal} {Phys. Rev. D}\ }\textbf {\bibinfo {volume} {105}},\ \bibinfo
  {pages} {043509} (\bibinfo {year} {2022})},\ \Eprint
  {https://arxiv.org/abs/2109.13934} {arXiv:2109.13934 [astro-ph.CO]}
  \BibitemShut {NoStop}%
\bibitem [{\citenamefont {Dong}\ \emph {et~al.}(2025)\citenamefont {Dong},
  \citenamefont {Song}, \citenamefont {Zhang},\ and\ \citenamefont
  {Zhang}}]{Dong:2025ikq}%
  \BibitemOpen
  \bibfield  {author} {\bibinfo {author} {\bibfnamefont {Y.-Y.}\ \bibnamefont
  {Dong}}, \bibinfo {author} {\bibfnamefont {J.-Y.}\ \bibnamefont {Song}},
  \bibinfo {author} {\bibfnamefont {J.-F.}\ \bibnamefont {Zhang}},\ and\
  \bibinfo {author} {\bibfnamefont {X.}~\bibnamefont {Zhang}},\ }\Eprint
  {https://arxiv.org/abs/2507.10165} {arXiv:2507.10165 [gr-qc]}  (\bibinfo
  {year} {2025})\BibitemShut {NoStop}%
\bibitem [{\citenamefont {Gupta}\ \emph {et~al.}(2025)\citenamefont {Gupta},
  \citenamefont {Narayan}, \citenamefont {London}, \citenamefont {Tiwari},\
  and\ \citenamefont {Sathyaprakash}}]{Gupta:2025paz}%
  \BibitemOpen
  \bibfield  {author} {\bibinfo {author} {\bibfnamefont {I.}~\bibnamefont
  {Gupta}}, \bibinfo {author} {\bibfnamefont {P.}~\bibnamefont {Narayan}},
  \bibinfo {author} {\bibfnamefont {L.}~\bibnamefont {London}}, \bibinfo
  {author} {\bibfnamefont {S.}~\bibnamefont {Tiwari}},\ and\ \bibinfo {author}
  {\bibfnamefont {B.}~\bibnamefont {Sathyaprakash}},\ }\href@noop {} {\bibinfo
  {title} {{Testing general relativity with amplitudes of subdominant
  gravitational-wave modes}}} (\bibinfo {year} {2025}),\ \Eprint
  {https://arxiv.org/abs/2511.11886} {arXiv:2511.11886 [gr-qc]} \BibitemShut
  {NoStop}%
\bibitem [{\citenamefont {Abac}\ \emph
  {et~al.}(2025{\natexlab{c}})\citenamefont {Abac} \emph
  {et~al.}}]{LIGOScientific:2025rsn}%
  \BibitemOpen
  \bibfield  {author} {\bibinfo {author} {\bibfnamefont {A.~G.}\ \bibnamefont
  {Abac}} \emph {et~al.} (\bibinfo {collaboration} {LIGO Scientific, VIRGO,
  KAGRA}),\ }\href {https://doi.org/10.3847/2041-8213/ae0c9c} {\bibfield
  {journal} {\bibinfo  {journal} {Astrophys. J. Lett.}\ }\textbf {\bibinfo
  {volume} {993}},\ \bibinfo {pages} {L25} (\bibinfo {year}
  {2025}{\natexlab{c}})},\ \Eprint {https://arxiv.org/abs/2507.08219}
  {arXiv:2507.08219 [astro-ph.HE]} \BibitemShut {NoStop}%
\bibitem [{\citenamefont {Abbott}\ \emph {et~al.}(2022)\citenamefont {Abbott}
  \emph {et~al.}}]{LIGOScientific:2021tfm}%
  \BibitemOpen
  \bibfield  {author} {\bibinfo {author} {\bibfnamefont {R.}~\bibnamefont
  {Abbott}} \emph {et~al.} (\bibinfo {collaboration} {LIGO Scientific, VIRGO,
  KAGRA}),\ }\href {https://doi.org/10.1051/0004-6361/202141452} {\bibfield
  {journal} {\bibinfo  {journal} {Astron. Astrophys.}\ }\textbf {\bibinfo
  {volume} {659}},\ \bibinfo {pages} {A84} (\bibinfo {year} {2022})},\ \Eprint
  {https://arxiv.org/abs/2105.15120} {arXiv:2105.15120 [astro-ph.HE]}
  \BibitemShut {NoStop}%
\bibitem [{\citenamefont {Dupletsa}\ \emph {et~al.}(2023)\citenamefont
  {Dupletsa}, \citenamefont {Harms}, \citenamefont {Banerjee}, \citenamefont
  {Branchesi}, \citenamefont {Goncharov}, \citenamefont {Maselli},
  \citenamefont {Oliveira}, \citenamefont {Ronchini},\ and\ \citenamefont
  {Tissino}}]{Dupletsa:2022scg}%
  \BibitemOpen
  \bibfield  {author} {\bibinfo {author} {\bibfnamefont {U.}~\bibnamefont
  {Dupletsa}}, \bibinfo {author} {\bibfnamefont {J.}~\bibnamefont {Harms}},
  \bibinfo {author} {\bibfnamefont {B.}~\bibnamefont {Banerjee}}, \bibinfo
  {author} {\bibfnamefont {M.}~\bibnamefont {Branchesi}}, \bibinfo {author}
  {\bibfnamefont {B.}~\bibnamefont {Goncharov}}, \bibinfo {author}
  {\bibfnamefont {A.}~\bibnamefont {Maselli}}, \bibinfo {author} {\bibfnamefont
  {A.~C.~S.}\ \bibnamefont {Oliveira}}, \bibinfo {author} {\bibfnamefont
  {S.}~\bibnamefont {Ronchini}},\ and\ \bibinfo {author} {\bibfnamefont
  {J.}~\bibnamefont {Tissino}},\ }\href
  {https://doi.org/10.1016/j.ascom.2022.100671} {\bibfield  {journal} {\bibinfo
   {journal} {Astron. Comput.}\ }\textbf {\bibinfo {volume} {42}},\ \bibinfo
  {pages} {100671} (\bibinfo {year} {2023})},\ \Eprint
  {https://arxiv.org/abs/2205.02499} {arXiv:2205.02499 [gr-qc]} \BibitemShut
  {NoStop}%
\bibitem [{\citenamefont {Madau}\ and\ \citenamefont
  {Dickinson}(2014)}]{Madau:2014bja}%
  \BibitemOpen
  \bibfield  {author} {\bibinfo {author} {\bibfnamefont {P.}~\bibnamefont
  {Madau}}\ and\ \bibinfo {author} {\bibfnamefont {M.}~\bibnamefont
  {Dickinson}},\ }\href {https://doi.org/10.1146/annurev-astro-081811-125615}
  {\bibfield  {journal} {\bibinfo  {journal} {Ann. Rev. Astron. Astrophys.}\
  }\textbf {\bibinfo {volume} {52}},\ \bibinfo {pages} {415} (\bibinfo {year}
  {2014})},\ \Eprint {https://arxiv.org/abs/1403.0007} {arXiv:1403.0007
  [astro-ph.CO]} \BibitemShut {NoStop}%
\bibitem [{\citenamefont {Ng}\ \emph {et~al.}(2021)\citenamefont {Ng},
  \citenamefont {Vitale}, \citenamefont {Farr},\ and\ \citenamefont
  {Rodriguez}}]{Ng:2020qpk}%
  \BibitemOpen
  \bibfield  {author} {\bibinfo {author} {\bibfnamefont {K.~K.~Y.}\
  \bibnamefont {Ng}}, \bibinfo {author} {\bibfnamefont {S.}~\bibnamefont
  {Vitale}}, \bibinfo {author} {\bibfnamefont {W.~M.}\ \bibnamefont {Farr}},\
  and\ \bibinfo {author} {\bibfnamefont {C.~L.}\ \bibnamefont {Rodriguez}},\
  }\href {https://doi.org/10.3847/2041-8213/abf8be} {\bibfield  {journal}
  {\bibinfo  {journal} {Astrophys. J. Lett.}\ }\textbf {\bibinfo {volume}
  {913}},\ \bibinfo {pages} {L5} (\bibinfo {year} {2021})},\ \Eprint
  {https://arxiv.org/abs/2012.09876} {arXiv:2012.09876 [astro-ph.CO]}
  \BibitemShut {NoStop}%
\bibitem [{\citenamefont {Blanchet}(2014)}]{Blanchet:2013haa}%
  \BibitemOpen
  \bibfield  {author} {\bibinfo {author} {\bibfnamefont {L.}~\bibnamefont
  {Blanchet}},\ }\href {https://doi.org/10.12942/lrr-2014-2} {\bibfield
  {journal} {\bibinfo  {journal} {Living Rev. Rel.}\ }\textbf {\bibinfo
  {volume} {17}},\ \bibinfo {pages} {2} (\bibinfo {year} {2014})},\ \Eprint
  {https://arxiv.org/abs/1310.1528} {arXiv:1310.1528 [gr-qc]} \BibitemShut
  {NoStop}%
\bibitem [{\citenamefont {Babak}\ \emph {et~al.}(2021)\citenamefont {Babak},
  \citenamefont {Petiteau},\ and\ \citenamefont {Hewitson}}]{Babak:2021mhe}%
  \BibitemOpen
  \bibfield  {author} {\bibinfo {author} {\bibfnamefont {S.}~\bibnamefont
  {Babak}}, \bibinfo {author} {\bibfnamefont {A.}~\bibnamefont {Petiteau}},\
  and\ \bibinfo {author} {\bibfnamefont {M.}~\bibnamefont {Hewitson}},\
  }\href@noop {} {\bibinfo {title} {Lisa sensitivity and snr calculations}}
  (\bibinfo {year} {2021}),\ \Eprint {https://arxiv.org/abs/2108.01167}
  {arXiv:2108.01167 [astro-ph.IM]} \BibitemShut {NoStop}%
\bibitem [{\citenamefont {Borhanian}(2021)}]{Borhanian:2020ypi}%
  \BibitemOpen
  \bibfield  {author} {\bibinfo {author} {\bibfnamefont {S.}~\bibnamefont
  {Borhanian}},\ }\href {https://doi.org/10.1088/1361-6382/ac1618} {\bibfield
  {journal} {\bibinfo  {journal} {Class. Quant. Grav.}\ }\textbf {\bibinfo
  {volume} {38}},\ \bibinfo {pages} {175014} (\bibinfo {year} {2021})},\
  \Eprint {https://arxiv.org/abs/2010.15202} {arXiv:2010.15202 [gr-qc]}
  \BibitemShut {NoStop}%
\bibitem [{\citenamefont {Cutler}(1998)}]{Cutler:1997ta}%
  \BibitemOpen
  \bibfield  {author} {\bibinfo {author} {\bibfnamefont {C.}~\bibnamefont
  {Cutler}},\ }\href {https://doi.org/10.1103/PhysRevD.57.7089} {\bibfield
  {journal} {\bibinfo  {journal} {Phys. Rev. D}\ }\textbf {\bibinfo {volume}
  {57}},\ \bibinfo {pages} {7089} (\bibinfo {year} {1998})},\ \Eprint
  {https://arxiv.org/abs/gr-qc/9703068} {arXiv:gr-qc/9703068} \BibitemShut
  {NoStop}%
\bibitem [{\citenamefont {Harms}\ \emph {et~al.}(2021)\citenamefont {Harms}
  \emph {et~al.}}]{LGWA:2020mma}%
  \BibitemOpen
  \bibfield  {author} {\bibinfo {author} {\bibfnamefont {J.}~\bibnamefont
  {Harms}} \emph {et~al.} (\bibinfo {collaboration} {LGWA}),\ }\href
  {https://doi.org/10.3847/1538-4357/abe5a7} {\bibfield  {journal} {\bibinfo
  {journal} {Astrophys. J.}\ }\textbf {\bibinfo {volume} {910}},\ \bibinfo
  {pages} {1} (\bibinfo {year} {2021})},\ \Eprint
  {https://arxiv.org/abs/2010.13726} {arXiv:2010.13726 [gr-qc]} \BibitemShut
  {NoStop}%
\bibitem [{\citenamefont {Iacovelli}\ \emph {et~al.}(2025)\citenamefont
  {Iacovelli}, \citenamefont {Tissino}, \citenamefont {Harms},\ and\
  \citenamefont {Berti}}]{Iacovelli:2025kwn}%
  \BibitemOpen
  \bibfield  {author} {\bibinfo {author} {\bibfnamefont {F.}~\bibnamefont
  {Iacovelli}}, \bibinfo {author} {\bibfnamefont {J.}~\bibnamefont {Tissino}},
  \bibinfo {author} {\bibfnamefont {J.}~\bibnamefont {Harms}},\ and\ \bibinfo
  {author} {\bibfnamefont {E.}~\bibnamefont {Berti}},\ }\href
  {https://doi.org/10.1103/d2rh-btf9} {\bibinfo {title} {{Gravitational-wave
  parameter estimation to the Moon and back: massive binaries and the case of
  GW231123}}} (\bibinfo {year} {2025}),\ \Eprint
  {https://arxiv.org/abs/2512.09978} {arXiv:2512.09978 [gr-qc]} \BibitemShut
  {NoStop}%
\bibitem [{\citenamefont {Barausse}\ \emph {et~al.}(2016)\citenamefont
  {Barausse}, \citenamefont {Yunes},\ and\ \citenamefont
  {Chamberlain}}]{Barausse:2016eii}%
  \BibitemOpen
  \bibfield  {author} {\bibinfo {author} {\bibfnamefont {E.}~\bibnamefont
  {Barausse}}, \bibinfo {author} {\bibfnamefont {N.}~\bibnamefont {Yunes}},\
  and\ \bibinfo {author} {\bibfnamefont {K.}~\bibnamefont {Chamberlain}},\
  }\href {https://doi.org/10.1103/PhysRevLett.116.241104} {\bibfield  {journal}
  {\bibinfo  {journal} {Phys. Rev. Lett.}\ }\textbf {\bibinfo {volume} {116}},\
  \bibinfo {pages} {241104} (\bibinfo {year} {2016})},\ \Eprint
  {https://arxiv.org/abs/1603.04075} {arXiv:1603.04075 [gr-qc]} \BibitemShut
  {NoStop}%
\bibitem [{\citenamefont {Carson}\ and\ \citenamefont
  {Yagi}(2020)}]{Carson:2019rda}%
  \BibitemOpen
  \bibfield  {author} {\bibinfo {author} {\bibfnamefont {Z.}~\bibnamefont
  {Carson}}\ and\ \bibinfo {author} {\bibfnamefont {K.}~\bibnamefont {Yagi}},\
  }\href {https://doi.org/10.1088/1361-6382/ab5c9a} {\bibfield  {journal}
  {\bibinfo  {journal} {Class. Quant. Grav.}\ }\textbf {\bibinfo {volume}
  {37}},\ \bibinfo {pages} {02LT01} (\bibinfo {year} {2020})},\ \Eprint
  {https://arxiv.org/abs/1905.13155} {arXiv:1905.13155 [gr-qc]} \BibitemShut
  {NoStop}%
\bibitem [{\citenamefont {Gnocchi}\ \emph {et~al.}(2019)\citenamefont
  {Gnocchi}, \citenamefont {Maselli}, \citenamefont {Abdelsalhin},
  \citenamefont {Giacobbo},\ and\ \citenamefont {Mapelli}}]{Gnocchi:2019jzp}%
  \BibitemOpen
  \bibfield  {author} {\bibinfo {author} {\bibfnamefont {G.}~\bibnamefont
  {Gnocchi}}, \bibinfo {author} {\bibfnamefont {A.}~\bibnamefont {Maselli}},
  \bibinfo {author} {\bibfnamefont {T.}~\bibnamefont {Abdelsalhin}}, \bibinfo
  {author} {\bibfnamefont {N.}~\bibnamefont {Giacobbo}},\ and\ \bibinfo
  {author} {\bibfnamefont {M.}~\bibnamefont {Mapelli}},\ }\href
  {https://doi.org/10.1103/PhysRevD.100.064024} {\bibfield  {journal} {\bibinfo
   {journal} {Phys. Rev. D}\ }\textbf {\bibinfo {volume} {100}},\ \bibinfo
  {pages} {064024} (\bibinfo {year} {2019})},\ \Eprint
  {https://arxiv.org/abs/1905.13460} {arXiv:1905.13460 [gr-qc]} \BibitemShut
  {NoStop}%
\bibitem [{\citenamefont {Toubiana}\ \emph {et~al.}(2020)\citenamefont
  {Toubiana}, \citenamefont {Marsat}, \citenamefont {Barausse}, \citenamefont
  {Babak},\ and\ \citenamefont {Baker}}]{Toubiana:2020vtf}%
  \BibitemOpen
  \bibfield  {author} {\bibinfo {author} {\bibfnamefont {A.}~\bibnamefont
  {Toubiana}}, \bibinfo {author} {\bibfnamefont {S.}~\bibnamefont {Marsat}},
  \bibinfo {author} {\bibfnamefont {E.}~\bibnamefont {Barausse}}, \bibinfo
  {author} {\bibfnamefont {S.}~\bibnamefont {Babak}},\ and\ \bibinfo {author}
  {\bibfnamefont {J.}~\bibnamefont {Baker}},\ }\href
  {https://doi.org/10.1103/PhysRevD.101.104038} {\bibfield  {journal} {\bibinfo
   {journal} {Phys. Rev. D}\ }\textbf {\bibinfo {volume} {101}},\ \bibinfo
  {pages} {104038} (\bibinfo {year} {2020})},\ \Eprint
  {https://arxiv.org/abs/2004.03626} {arXiv:2004.03626 [gr-qc]} \BibitemShut
  {NoStop}%
\bibitem [{\citenamefont {Gupta}\ \emph {et~al.}(2020)\citenamefont {Gupta},
  \citenamefont {Datta}, \citenamefont {Kastha}, \citenamefont {Borhanian},
  \citenamefont {Arun},\ and\ \citenamefont {Sathyaprakash}}]{Gupta:2020lxa}%
  \BibitemOpen
  \bibfield  {author} {\bibinfo {author} {\bibfnamefont {A.}~\bibnamefont
  {Gupta}}, \bibinfo {author} {\bibfnamefont {S.}~\bibnamefont {Datta}},
  \bibinfo {author} {\bibfnamefont {S.}~\bibnamefont {Kastha}}, \bibinfo
  {author} {\bibfnamefont {S.}~\bibnamefont {Borhanian}}, \bibinfo {author}
  {\bibfnamefont {K.~G.}\ \bibnamefont {Arun}},\ and\ \bibinfo {author}
  {\bibfnamefont {B.~S.}\ \bibnamefont {Sathyaprakash}},\ }\href
  {https://doi.org/10.1103/PhysRevLett.125.201101} {\bibfield  {journal}
  {\bibinfo  {journal} {Phys. Rev. Lett.}\ }\textbf {\bibinfo {volume} {125}},\
  \bibinfo {pages} {201101} (\bibinfo {year} {2020})},\ \Eprint
  {https://arxiv.org/abs/2005.09607} {arXiv:2005.09607 [gr-qc]} \BibitemShut
  {NoStop}%
\bibitem [{\citenamefont {Datta}\ \emph {et~al.}(2021)\citenamefont {Datta},
  \citenamefont {Gupta}, \citenamefont {Kastha}, \citenamefont {Arun},\ and\
  \citenamefont {Sathyaprakash}}]{Datta:2020vcj}%
  \BibitemOpen
  \bibfield  {author} {\bibinfo {author} {\bibfnamefont {S.}~\bibnamefont
  {Datta}}, \bibinfo {author} {\bibfnamefont {A.}~\bibnamefont {Gupta}},
  \bibinfo {author} {\bibfnamefont {S.}~\bibnamefont {Kastha}}, \bibinfo
  {author} {\bibfnamefont {K.~G.}\ \bibnamefont {Arun}},\ and\ \bibinfo
  {author} {\bibfnamefont {B.~S.}\ \bibnamefont {Sathyaprakash}},\ }\href
  {https://doi.org/10.1103/PhysRevD.103.024036} {\bibfield  {journal} {\bibinfo
   {journal} {Phys. Rev. D}\ }\textbf {\bibinfo {volume} {103}},\ \bibinfo
  {pages} {024036} (\bibinfo {year} {2021})},\ \Eprint
  {https://arxiv.org/abs/2006.12137} {arXiv:2006.12137 [gr-qc]} \BibitemShut
  {NoStop}%
\bibitem [{\citenamefont {Miller}\ and\ \citenamefont
  {Colbert}(2004)}]{Miller:2003sc}%
  \BibitemOpen
  \bibfield  {author} {\bibinfo {author} {\bibfnamefont {M.~C.}\ \bibnamefont
  {Miller}}\ and\ \bibinfo {author} {\bibfnamefont {E.~J.~M.}\ \bibnamefont
  {Colbert}},\ }\href {https://doi.org/10.1142/S0218271804004426} {\bibfield
  {journal} {\bibinfo  {journal} {Int. J. Mod. Phys. D}\ }\textbf {\bibinfo
  {volume} {13}},\ \bibinfo {pages} {1} (\bibinfo {year} {2004})},\ \Eprint
  {https://arxiv.org/abs/astro-ph/0308402} {arXiv:astro-ph/0308402}
  \BibitemShut {NoStop}%
\bibitem [{\citenamefont {{McKernan}}\ \emph {et~al.}(2012)\citenamefont
  {{McKernan}}, \citenamefont {{Ford}}, \citenamefont {{Lyra}},\ and\
  \citenamefont {{Perets}}}]{2012MNRAS.425..460M}%
  \BibitemOpen
  \bibfield  {author} {\bibinfo {author} {\bibfnamefont {B.}~\bibnamefont
  {{McKernan}}}, \bibinfo {author} {\bibfnamefont {K.~E.~S.}\ \bibnamefont
  {{Ford}}}, \bibinfo {author} {\bibfnamefont {W.}~\bibnamefont {{Lyra}}},\
  and\ \bibinfo {author} {\bibfnamefont {H.~B.}\ \bibnamefont {{Perets}}},\
  }\href {https://doi.org/10.1111/j.1365-2966.2012.21486.x} {\bibfield
  {journal} {\bibinfo  {journal} {Mon. Not. R. Astron. Soc.}\ }\textbf
  {\bibinfo {volume} {425}},\ \bibinfo {pages} {460} (\bibinfo {year}
  {2012})},\ \Eprint {https://arxiv.org/abs/1206.2309} {arXiv:1206.2309
  [astro-ph.GA]} \BibitemShut {NoStop}%
\bibitem [{\citenamefont {McKernan}\ \emph {et~al.}(2014)\citenamefont
  {McKernan}, \citenamefont {Ford}, \citenamefont {Kocsis}, \citenamefont
  {Lyra},\ and\ \citenamefont {Winter}}]{McKernan:2014oxa}%
  \BibitemOpen
  \bibfield  {author} {\bibinfo {author} {\bibfnamefont {B.}~\bibnamefont
  {McKernan}}, \bibinfo {author} {\bibfnamefont {K.~E.~S.}\ \bibnamefont
  {Ford}}, \bibinfo {author} {\bibfnamefont {B.}~\bibnamefont {Kocsis}},
  \bibinfo {author} {\bibfnamefont {W.}~\bibnamefont {Lyra}},\ and\ \bibinfo
  {author} {\bibfnamefont {L.~M.}\ \bibnamefont {Winter}},\ }\href
  {https://doi.org/10.1093/mnras/stu553} {\bibfield  {journal} {\bibinfo
  {journal} {Mon. Not. Roy. Astron. Soc.}\ }\textbf {\bibinfo {volume} {441}},\
  \bibinfo {pages} {900} (\bibinfo {year} {2014})},\ \Eprint
  {https://arxiv.org/abs/1403.6433} {arXiv:1403.6433 [astro-ph.GA]}
  \BibitemShut {NoStop}%
\bibitem [{\citenamefont {Fragione}\ \emph {et~al.}(2022)\citenamefont
  {Fragione}, \citenamefont {Loeb}, \citenamefont {Kocsis},\ and\ \citenamefont
  {Rasio}}]{Fragione:2022avp}%
  \BibitemOpen
  \bibfield  {author} {\bibinfo {author} {\bibfnamefont {G.}~\bibnamefont
  {Fragione}}, \bibinfo {author} {\bibfnamefont {A.}~\bibnamefont {Loeb}},
  \bibinfo {author} {\bibfnamefont {B.}~\bibnamefont {Kocsis}},\ and\ \bibinfo
  {author} {\bibfnamefont {F.~A.}\ \bibnamefont {Rasio}},\ }\href
  {https://doi.org/10.3847/1538-4357/ac75d0} {\bibfield  {journal} {\bibinfo
  {journal} {Astrophys. J.}\ }\textbf {\bibinfo {volume} {933}},\ \bibinfo
  {pages} {170} (\bibinfo {year} {2022})},\ \Eprint
  {https://arxiv.org/abs/2204.03745} {arXiv:2204.03745 [astro-ph.HE]}
  \BibitemShut {NoStop}%
\bibitem [{\citenamefont {Syer}\ \emph {et~al.}(1991)\citenamefont {Syer},
  \citenamefont {Clarke},\ and\ \citenamefont {Rees}}]{syer_1991}%
  \BibitemOpen
  \bibfield  {author} {\bibinfo {author} {\bibfnamefont {D.}~\bibnamefont
  {Syer}}, \bibinfo {author} {\bibfnamefont {C.~J.}\ \bibnamefont {Clarke}},\
  and\ \bibinfo {author} {\bibfnamefont {M.~J.}\ \bibnamefont {Rees}},\ }\href
  {https://doi.org/10.1093/mnras/250.3.505} {\bibfield  {journal} {\bibinfo
  {journal} {Monthly Notices of the Royal Astronomical Society}\ }\textbf
  {\bibinfo {volume} {250}},\ \bibinfo {pages} {505} (\bibinfo {year}
  {1991})},\ \Eprint
  {https://arxiv.org/abs/https://academic.oup.com/mnras/article-pdf/250/3/505/3189276/mnras250-0505.pdf}
  {https://academic.oup.com/mnras/article-pdf/250/3/505/3189276/mnras250-0505.pdf}
  \BibitemShut {NoStop}%
\bibitem [{\citenamefont {Artymowicz}\ \emph {et~al.}(1993)\citenamefont
  {Artymowicz}, \citenamefont {Lin},\ and\ \citenamefont
  {Wampler}}]{Artymowicz:1993xz}%
  \BibitemOpen
  \bibfield  {author} {\bibinfo {author} {\bibfnamefont {P.}~\bibnamefont
  {Artymowicz}}, \bibinfo {author} {\bibfnamefont {D.~N.~C.}\ \bibnamefont
  {Lin}},\ and\ \bibinfo {author} {\bibfnamefont {E.~J.}\ \bibnamefont
  {Wampler}},\ }\href {https://doi.org/10.1086/172690} {\bibfield  {journal}
  {\bibinfo  {journal} {Astrophys. J.}\ }\textbf {\bibinfo {volume} {409}},\
  \bibinfo {pages} {592} (\bibinfo {year} {1993})}\BibitemShut {NoStop}%
\bibitem [{\citenamefont {Goodman}\ and\ \citenamefont
  {Tan}(2004)}]{Goodman:2003sf}%
  \BibitemOpen
  \bibfield  {author} {\bibinfo {author} {\bibfnamefont {J.}~\bibnamefont
  {Goodman}}\ and\ \bibinfo {author} {\bibfnamefont {J.~C.}\ \bibnamefont
  {Tan}},\ }\href {https://doi.org/10.1086/386360} {\bibfield  {journal}
  {\bibinfo  {journal} {Astrophys. J.}\ }\textbf {\bibinfo {volume} {608}},\
  \bibinfo {pages} {108} (\bibinfo {year} {2004})},\ \Eprint
  {https://arxiv.org/abs/astro-ph/0307361} {arXiv:astro-ph/0307361}
  \BibitemShut {NoStop}%
\bibitem [{\citenamefont {Yang}\ \emph {et~al.}(2019)\citenamefont {Yang} \emph
  {et~al.}}]{Yang:2019cbr}%
  \BibitemOpen
  \bibfield  {author} {\bibinfo {author} {\bibfnamefont {Y.}~\bibnamefont
  {Yang}} \emph {et~al.},\ }\href
  {https://doi.org/10.1103/PhysRevLett.123.181101} {\bibfield  {journal}
  {\bibinfo  {journal} {Phys. Rev. Lett.}\ }\textbf {\bibinfo {volume} {123}},\
  \bibinfo {pages} {181101} (\bibinfo {year} {2019})},\ \Eprint
  {https://arxiv.org/abs/1906.09281} {arXiv:1906.09281 [astro-ph.HE]}
  \BibitemShut {NoStop}%
\bibitem [{\citenamefont {Levin}(2007)}]{Levin:2006uc}%
  \BibitemOpen
  \bibfield  {author} {\bibinfo {author} {\bibfnamefont {Y.}~\bibnamefont
  {Levin}},\ }\href {https://doi.org/10.1111/j.1365-2966.2006.11155.x}
  {\bibfield  {journal} {\bibinfo  {journal} {Mon. Not. Roy. Astron. Soc.}\
  }\textbf {\bibinfo {volume} {374}},\ \bibinfo {pages} {515} (\bibinfo {year}
  {2007})},\ \Eprint {https://arxiv.org/abs/astro-ph/0603583}
  {arXiv:astro-ph/0603583} \BibitemShut {NoStop}%
\bibitem [{\citenamefont {Stone}\ \emph {et~al.}(2017)\citenamefont {Stone},
  \citenamefont {Metzger},\ and\ \citenamefont {Haiman}}]{Stone:2016wzz}%
  \BibitemOpen
  \bibfield  {author} {\bibinfo {author} {\bibfnamefont {N.~C.}\ \bibnamefont
  {Stone}}, \bibinfo {author} {\bibfnamefont {B.~D.}\ \bibnamefont {Metzger}},\
  and\ \bibinfo {author} {\bibfnamefont {Z.}~\bibnamefont {Haiman}},\ }\href
  {https://doi.org/10.1093/mnras/stw2260} {\bibfield  {journal} {\bibinfo
  {journal} {Mon. Not. Roy. Astron. Soc.}\ }\textbf {\bibinfo {volume} {464}},\
  \bibinfo {pages} {946} (\bibinfo {year} {2017})},\ \Eprint
  {https://arxiv.org/abs/1602.04226} {arXiv:1602.04226 [astro-ph.GA]}
  \BibitemShut {NoStop}%
\bibitem [{\citenamefont {Bellovary}\ \emph {et~al.}(2016)\citenamefont
  {Bellovary}, \citenamefont {Mac~Low}, \citenamefont {McKernan},\ and\
  \citenamefont {Ford}}]{Bellovary:2015ifg}%
  \BibitemOpen
  \bibfield  {author} {\bibinfo {author} {\bibfnamefont {J.~M.}\ \bibnamefont
  {Bellovary}}, \bibinfo {author} {\bibfnamefont {M.-M.}\ \bibnamefont
  {Mac~Low}}, \bibinfo {author} {\bibfnamefont {B.}~\bibnamefont {McKernan}},\
  and\ \bibinfo {author} {\bibfnamefont {K.~E.~S.}\ \bibnamefont {Ford}},\
  }\href {https://doi.org/10.3847/2041-8205/819/2/L17} {\bibfield  {journal}
  {\bibinfo  {journal} {Astrophys. J. Lett.}\ }\textbf {\bibinfo {volume}
  {819}},\ \bibinfo {pages} {L17} (\bibinfo {year} {2016})},\ \Eprint
  {https://arxiv.org/abs/1511.00005} {arXiv:1511.00005 [astro-ph.GA]}
  \BibitemShut {NoStop}%
\bibitem [{\citenamefont {Bartos}\ \emph {et~al.}(2017)\citenamefont {Bartos},
  \citenamefont {Kocsis}, \citenamefont {Haiman},\ and\ \citenamefont
  {M{\'a}rka}}]{Bartos:2016dgn}%
  \BibitemOpen
  \bibfield  {author} {\bibinfo {author} {\bibfnamefont {I.}~\bibnamefont
  {Bartos}}, \bibinfo {author} {\bibfnamefont {B.}~\bibnamefont {Kocsis}},
  \bibinfo {author} {\bibfnamefont {Z.}~\bibnamefont {Haiman}},\ and\ \bibinfo
  {author} {\bibfnamefont {S.}~\bibnamefont {M{\'a}rka}},\ }\href
  {https://doi.org/10.3847/1538-4357/835/2/165} {\bibfield  {journal} {\bibinfo
   {journal} {Astrophys. J.}\ }\textbf {\bibinfo {volume} {835}},\ \bibinfo
  {pages} {165} (\bibinfo {year} {2017})},\ \Eprint
  {https://arxiv.org/abs/1602.03831} {arXiv:1602.03831 [astro-ph.HE]}
  \BibitemShut {NoStop}%
\bibitem [{\citenamefont {Ma}\ \emph {et~al.}(2025)\citenamefont {Ma},
  \citenamefont {Wang}, \citenamefont {Wu},\ and\ \citenamefont
  {Wang}}]{Ma:2024bfo}%
  \BibitemOpen
  \bibfield  {author} {\bibinfo {author} {\bibfnamefont {Z.-P.}\ \bibnamefont
  {Ma}}, \bibinfo {author} {\bibfnamefont {K.}~\bibnamefont {Wang}}, \bibinfo
  {author} {\bibfnamefont {Q.}~\bibnamefont {Wu}},\ and\ \bibinfo {author}
  {\bibfnamefont {J.-M.}\ \bibnamefont {Wang}},\ }\href
  {https://doi.org/10.1103/PhysRevD.111.083033} {\bibfield  {journal} {\bibinfo
   {journal} {Phys. Rev. D}\ }\textbf {\bibinfo {volume} {111}},\ \bibinfo
  {pages} {083033} (\bibinfo {year} {2025})},\ \Eprint
  {https://arxiv.org/abs/2409.18567} {arXiv:2409.18567 [astro-ph.HE]}
  \BibitemShut {NoStop}%
\bibitem [{\citenamefont {Wang}\ \emph {et~al.}(2021)\citenamefont {Wang},
  \citenamefont {Liu}, \citenamefont {Ho},\ and\ \citenamefont
  {Du}}]{Wang:2021tfd}%
  \BibitemOpen
  \bibfield  {author} {\bibinfo {author} {\bibfnamefont {J.-M.}\ \bibnamefont
  {Wang}}, \bibinfo {author} {\bibfnamefont {J.-R.}\ \bibnamefont {Liu}},
  \bibinfo {author} {\bibfnamefont {L.~C.}\ \bibnamefont {Ho}},\ and\ \bibinfo
  {author} {\bibfnamefont {P.~U.}\ \bibnamefont {Du}},\ }\href
  {https://doi.org/10.3847/2041-8213/abee81} {\bibfield  {journal} {\bibinfo
  {journal} {Astrophys. J. Lett.}\ }\textbf {\bibinfo {volume} {911}},\
  \bibinfo {pages} {L14} (\bibinfo {year} {2021})},\ \Eprint
  {https://arxiv.org/abs/2103.07708} {arXiv:2103.07708 [astro-ph.HE]}
  \BibitemShut {NoStop}%
\bibitem [{\citenamefont {Kimura}\ \emph {et~al.}(2021)\citenamefont {Kimura},
  \citenamefont {Murase},\ and\ \citenamefont {Bartos}}]{Kimura:2021xxu}%
  \BibitemOpen
  \bibfield  {author} {\bibinfo {author} {\bibfnamefont {S.~S.}\ \bibnamefont
  {Kimura}}, \bibinfo {author} {\bibfnamefont {K.}~\bibnamefont {Murase}},\
  and\ \bibinfo {author} {\bibfnamefont {I.}~\bibnamefont {Bartos}},\ }\href
  {https://doi.org/10.3847/1538-4357/ac0535} {\bibfield  {journal} {\bibinfo
  {journal} {Astrophys. J.}\ }\textbf {\bibinfo {volume} {916}},\ \bibinfo
  {pages} {111} (\bibinfo {year} {2021})},\ \Eprint
  {https://arxiv.org/abs/2103.02461} {arXiv:2103.02461 [astro-ph.HE]}
  \BibitemShut {NoStop}%
\bibitem [{\citenamefont {Rodr{\'\i}guez-Ram{\'\i}rez}\ \emph
  {et~al.}(2025)\citenamefont {Rodr{\'\i}guez-Ram{\'\i}rez}, \citenamefont
  {Nemmen},\ and\ \citenamefont {Bom}}]{Rodriguez-Ramirez:2024ikd}%
  \BibitemOpen
  \bibfield  {author} {\bibinfo {author} {\bibfnamefont {J.~C.}\ \bibnamefont
  {Rodr{\'\i}guez-Ram{\'\i}rez}}, \bibinfo {author} {\bibfnamefont
  {R.}~\bibnamefont {Nemmen}},\ and\ \bibinfo {author} {\bibfnamefont {C.~R.}\
  \bibnamefont {Bom}},\ }\href {https://doi.org/10.1103/PhysRevD.111.083020}
  {\bibfield  {journal} {\bibinfo  {journal} {Phys. Rev. D}\ }\textbf {\bibinfo
  {volume} {111}},\ \bibinfo {pages} {083020} (\bibinfo {year} {2025})},\
  \Eprint {https://arxiv.org/abs/2407.09945} {arXiv:2407.09945 [astro-ph.HE]}
  \BibitemShut {NoStop}%
\bibitem [{\citenamefont {Tagawa}\ \emph {et~al.}(2023)\citenamefont {Tagawa},
  \citenamefont {Kimura}, \citenamefont {Haiman}, \citenamefont {Perna},\ and\
  \citenamefont {Bartos}}]{Tagawa:2023uqa}%
  \BibitemOpen
  \bibfield  {author} {\bibinfo {author} {\bibfnamefont {H.}~\bibnamefont
  {Tagawa}}, \bibinfo {author} {\bibfnamefont {S.~S.}\ \bibnamefont {Kimura}},
  \bibinfo {author} {\bibfnamefont {Z.}~\bibnamefont {Haiman}}, \bibinfo
  {author} {\bibfnamefont {R.}~\bibnamefont {Perna}},\ and\ \bibinfo {author}
  {\bibfnamefont {I.}~\bibnamefont {Bartos}},\ }\href
  {https://doi.org/10.3847/1538-4357/acc4bb} {\bibfield  {journal} {\bibinfo
  {journal} {Astrophys. J.}\ }\textbf {\bibinfo {volume} {950}},\ \bibinfo
  {pages} {13} (\bibinfo {year} {2023})},\ \Eprint
  {https://arxiv.org/abs/2301.07111} {arXiv:2301.07111 [astro-ph.HE]}
  \BibitemShut {NoStop}%
\bibitem [{\citenamefont {Rodr{\'\i}guez-Ram{\'\i}rez}\ \emph
  {et~al.}(2023)\citenamefont {Rodr{\'\i}guez-Ram{\'\i}rez}, \citenamefont
  {Bom}, \citenamefont {Fraga},\ and\ \citenamefont
  {Nemmen}}]{Rodriguez-Ramirez:2023ejf}%
  \BibitemOpen
  \bibfield  {author} {\bibinfo {author} {\bibfnamefont {J.~C.}\ \bibnamefont
  {Rodr{\'\i}guez-Ram{\'\i}rez}}, \bibinfo {author} {\bibfnamefont {C.~R.}\
  \bibnamefont {Bom}}, \bibinfo {author} {\bibfnamefont {B.}~\bibnamefont
  {Fraga}},\ and\ \bibinfo {author} {\bibfnamefont {R.}~\bibnamefont
  {Nemmen}},\ }\href {https://doi.org/10.1093/mnras/stad3575} {\bibfield
  {journal} {\bibinfo  {journal} {Mon. Not. Roy. Astron. Soc.}\ }\textbf
  {\bibinfo {volume} {527}},\ \bibinfo {pages} {6076} (\bibinfo {year}
  {2023})},\ \Eprint {https://arxiv.org/abs/2304.10567} {arXiv:2304.10567
  [astro-ph.HE]} \BibitemShut {NoStop}%
\bibitem [{\citenamefont {Graham}\ \emph {et~al.}(2020)\citenamefont {Graham}
  \emph {et~al.}}]{Graham:2020gwr}%
  \BibitemOpen
  \bibfield  {author} {\bibinfo {author} {\bibfnamefont {M.~J.}\ \bibnamefont
  {Graham}} \emph {et~al.},\ }\href
  {https://doi.org/10.1103/PhysRevLett.124.251102} {\bibfield  {journal}
  {\bibinfo  {journal} {Phys. Rev. Lett.}\ }\textbf {\bibinfo {volume} {124}},\
  \bibinfo {pages} {251102} (\bibinfo {year} {2020})},\ \Eprint
  {https://arxiv.org/abs/2006.14122} {arXiv:2006.14122 [astro-ph.HE]}
  \BibitemShut {NoStop}%
\bibitem [{\citenamefont {Morton}\ \emph {et~al.}(2023)\citenamefont {Morton},
  \citenamefont {Rinaldi}, \citenamefont {Torres-Orjuela}, \citenamefont
  {Derdzinski}, \citenamefont {Vaccaro},\ and\ \citenamefont
  {Del~Pozzo}}]{Morton:2023wxg}%
  \BibitemOpen
  \bibfield  {author} {\bibinfo {author} {\bibfnamefont {S.~L.}\ \bibnamefont
  {Morton}}, \bibinfo {author} {\bibfnamefont {S.}~\bibnamefont {Rinaldi}},
  \bibinfo {author} {\bibfnamefont {A.}~\bibnamefont {Torres-Orjuela}},
  \bibinfo {author} {\bibfnamefont {A.}~\bibnamefont {Derdzinski}}, \bibinfo
  {author} {\bibfnamefont {M.~P.}\ \bibnamefont {Vaccaro}},\ and\ \bibinfo
  {author} {\bibfnamefont {W.}~\bibnamefont {Del~Pozzo}},\ }\href
  {https://doi.org/10.1103/PhysRevD.108.123039} {\bibfield  {journal} {\bibinfo
   {journal} {Phys. Rev. D}\ }\textbf {\bibinfo {volume} {108}},\ \bibinfo
  {pages} {123039} (\bibinfo {year} {2023})},\ \Eprint
  {https://arxiv.org/abs/2310.16025} {arXiv:2310.16025 [gr-qc]} \BibitemShut
  {NoStop}%
\bibitem [{\citenamefont {Ashton}\ \emph {et~al.}(2018)\citenamefont {Ashton},
  \citenamefont {Burns}, \citenamefont {Canton}, \citenamefont {Dent},
  \citenamefont {Eggenstein}, \citenamefont {Nielsen}, \citenamefont {Prix},
  \citenamefont {Was},\ and\ \citenamefont {Zhu}}]{Ashton:2017ykh}%
  \BibitemOpen
  \bibfield  {author} {\bibinfo {author} {\bibfnamefont {G.}~\bibnamefont
  {Ashton}}, \bibinfo {author} {\bibfnamefont {E.}~\bibnamefont {Burns}},
  \bibinfo {author} {\bibfnamefont {T.~D.}\ \bibnamefont {Canton}}, \bibinfo
  {author} {\bibfnamefont {T.}~\bibnamefont {Dent}}, \bibinfo {author}
  {\bibfnamefont {H.~B.}\ \bibnamefont {Eggenstein}}, \bibinfo {author}
  {\bibfnamefont {A.~B.}\ \bibnamefont {Nielsen}}, \bibinfo {author}
  {\bibfnamefont {R.}~\bibnamefont {Prix}}, \bibinfo {author} {\bibfnamefont
  {M.}~\bibnamefont {Was}},\ and\ \bibinfo {author} {\bibfnamefont {S.~J.}\
  \bibnamefont {Zhu}},\ }\href {https://doi.org/10.3847/1538-4357/aabfd2}
  {\bibfield  {journal} {\bibinfo  {journal} {Astrophys. J.}\ }\textbf
  {\bibinfo {volume} {860}},\ \bibinfo {pages} {6} (\bibinfo {year} {2018})},\
  \Eprint {https://arxiv.org/abs/1712.05392} {arXiv:1712.05392 [astro-ph.HE]}
  \BibitemShut {NoStop}%
\bibitem [{\citenamefont {Abbott}\ \emph
  {et~al.}(2017{\natexlab{e}})\citenamefont {Abbott} \emph
  {et~al.}}]{LIGOScientific:2017zic}%
  \BibitemOpen
  \bibfield  {author} {\bibinfo {author} {\bibfnamefont {B.~P.}\ \bibnamefont
  {Abbott}} \emph {et~al.} (\bibinfo {collaboration} {LIGO Scientific, Virgo,
  Fermi-GBM, INTEGRAL}),\ }\href {https://doi.org/10.3847/2041-8213/aa920c}
  {\bibfield  {journal} {\bibinfo  {journal} {Astrophys. J. Lett.}\ }\textbf
  {\bibinfo {volume} {848}},\ \bibinfo {pages} {L13} (\bibinfo {year}
  {2017}{\natexlab{e}})},\ \Eprint {https://arxiv.org/abs/1710.05834}
  {arXiv:1710.05834 [astro-ph.HE]} \BibitemShut {NoStop}%
\bibitem [{\citenamefont {Coulter}\ \emph {et~al.}(2017)\citenamefont {Coulter}
  \emph {et~al.}}]{Coulter:2017wya}%
  \BibitemOpen
  \bibfield  {author} {\bibinfo {author} {\bibfnamefont {D.~A.}\ \bibnamefont
  {Coulter}} \emph {et~al.},\ }\href {https://doi.org/10.1126/science.aap9811}
  {\bibfield  {journal} {\bibinfo  {journal} {Science}\ }\textbf {\bibinfo
  {volume} {358}},\ \bibinfo {pages} {1556} (\bibinfo {year} {2017})},\ \Eprint
  {https://arxiv.org/abs/1710.05452} {arXiv:1710.05452 [astro-ph.HE]}
  \BibitemShut {NoStop}%
\bibitem [{\citenamefont {Margutti}\ \emph {et~al.}(2017)\citenamefont
  {Margutti} \emph {et~al.}}]{Margutti:2017cjl}%
  \BibitemOpen
  \bibfield  {author} {\bibinfo {author} {\bibfnamefont {R.}~\bibnamefont
  {Margutti}} \emph {et~al.},\ }\href
  {https://doi.org/10.3847/2041-8213/aa9057} {\bibfield  {journal} {\bibinfo
  {journal} {Astrophys. J. Lett.}\ }\textbf {\bibinfo {volume} {848}},\
  \bibinfo {pages} {L20} (\bibinfo {year} {2017})},\ \Eprint
  {https://arxiv.org/abs/1710.05431} {arXiv:1710.05431 [astro-ph.HE]}
  \BibitemShut {NoStop}%
\bibitem [{\citenamefont {Hajela}\ \emph {et~al.}(2022)\citenamefont {Hajela}
  \emph {et~al.}}]{Hajela:2021faz}%
  \BibitemOpen
  \bibfield  {author} {\bibinfo {author} {\bibfnamefont {A.}~\bibnamefont
  {Hajela}} \emph {et~al.},\ }\href {https://doi.org/10.3847/2041-8213/ac504a}
  {\bibfield  {journal} {\bibinfo  {journal} {Astrophys. J. Lett.}\ }\textbf
  {\bibinfo {volume} {927}},\ \bibinfo {pages} {L17} (\bibinfo {year}
  {2022})},\ \Eprint {https://arxiv.org/abs/2104.02070} {arXiv:2104.02070
  [astro-ph.HE]} \BibitemShut {NoStop}%
\bibitem [{\citenamefont {Hallinan}\ \emph {et~al.}(2017)\citenamefont
  {Hallinan} \emph {et~al.}}]{Hallinan:2017woc}%
  \BibitemOpen
  \bibfield  {author} {\bibinfo {author} {\bibfnamefont {G.}~\bibnamefont
  {Hallinan}} \emph {et~al.},\ }\href {https://doi.org/10.1126/science.aap9855}
  {\bibfield  {journal} {\bibinfo  {journal} {Science}\ }\textbf {\bibinfo
  {volume} {358}},\ \bibinfo {pages} {1579} (\bibinfo {year} {2017})},\ \Eprint
  {https://arxiv.org/abs/1710.05435} {arXiv:1710.05435 [astro-ph.HE]}
  \BibitemShut {NoStop}%
\bibitem [{\citenamefont {Balasubramanian}\ \emph {et~al.}(2022)\citenamefont
  {Balasubramanian}, \citenamefont {Corsi}, \citenamefont {Mooley},
  \citenamefont {Hotokezaka}, \citenamefont {Kaplan}, \citenamefont {Frail},
  \citenamefont {Hallinan}, \citenamefont {Lazzati},\ and\ \citenamefont
  {Murphy}}]{Balasubramanian:2022sie}%
  \BibitemOpen
  \bibfield  {author} {\bibinfo {author} {\bibfnamefont {A.}~\bibnamefont
  {Balasubramanian}}, \bibinfo {author} {\bibfnamefont {A.}~\bibnamefont
  {Corsi}}, \bibinfo {author} {\bibfnamefont {K.~P.}\ \bibnamefont {Mooley}},
  \bibinfo {author} {\bibfnamefont {K.}~\bibnamefont {Hotokezaka}}, \bibinfo
  {author} {\bibfnamefont {D.~L.}\ \bibnamefont {Kaplan}}, \bibinfo {author}
  {\bibfnamefont {D.~A.}\ \bibnamefont {Frail}}, \bibinfo {author}
  {\bibfnamefont {G.}~\bibnamefont {Hallinan}}, \bibinfo {author}
  {\bibfnamefont {D.}~\bibnamefont {Lazzati}},\ and\ \bibinfo {author}
  {\bibfnamefont {E.~J.}\ \bibnamefont {Murphy}},\ }\href
  {https://doi.org/10.3847/1538-4357/ac9133} {\bibfield  {journal} {\bibinfo
  {journal} {Astrophys. J.}\ }\textbf {\bibinfo {volume} {938}},\ \bibinfo
  {pages} {12} (\bibinfo {year} {2022})},\ \Eprint
  {https://arxiv.org/abs/2205.14788} {arXiv:2205.14788 [astro-ph.HE]}
  \BibitemShut {NoStop}%
\bibitem [{\citenamefont {Gupta}(2023{\natexlab{b}})}]{gupta_2023_8087733}%
  \BibitemOpen
  \bibfield  {author} {\bibinfo {author} {\bibfnamefont {I.}~\bibnamefont
  {Gupta}},\ }\href {https://doi.org/10.5281/zenodo.8087733}
  {10.5281/zenodo.8087733} (\bibinfo {year} {2023}{\natexlab{b}})\BibitemShut
  {NoStop}%
\bibitem [{\citenamefont {{Farr}}\ and\ \citenamefont
  {{Chatziioannou}}(2020)}]{2020RNAAS...4...65F}%
  \BibitemOpen
  \bibfield  {author} {\bibinfo {author} {\bibfnamefont {W.~M.}\ \bibnamefont
  {{Farr}}}\ and\ \bibinfo {author} {\bibfnamefont {K.}~\bibnamefont
  {{Chatziioannou}}},\ }\href {https://doi.org/10.3847/2515-5172/ab9088}
  {\bibfield  {journal} {\bibinfo  {journal} {Research Notes of the American
  Astronomical Society}\ }\textbf {\bibinfo {volume} {4}},\ \bibinfo {eid} {65}
  (\bibinfo {year} {2020})},\ \Eprint {https://arxiv.org/abs/2005.00032}
  {arXiv:2005.00032 [astro-ph.GA]} \BibitemShut {NoStop}%
\bibitem [{\citenamefont {Akmal}\ \emph {et~al.}(1998)\citenamefont {Akmal},
  \citenamefont {Pandharipande},\ and\ \citenamefont
  {Ravenhall}}]{Akmal:1998cf}%
  \BibitemOpen
  \bibfield  {author} {\bibinfo {author} {\bibfnamefont {A.}~\bibnamefont
  {Akmal}}, \bibinfo {author} {\bibfnamefont {V.~R.}\ \bibnamefont
  {Pandharipande}},\ and\ \bibinfo {author} {\bibfnamefont {D.~G.}\
  \bibnamefont {Ravenhall}},\ }\href {https://doi.org/10.1103/PhysRevC.58.1804}
  {\bibfield  {journal} {\bibinfo  {journal} {Phys. Rev. C}\ }\textbf {\bibinfo
  {volume} {58}},\ \bibinfo {pages} {1804} (\bibinfo {year} {1998})},\ \Eprint
  {https://arxiv.org/abs/nucl-th/9804027} {arXiv:nucl-th/9804027} \BibitemShut
  {NoStop}%
\bibitem [{\citenamefont {Abbott}\ \emph {et~al.}(2021)\citenamefont {Abbott}
  \emph {et~al.}}]{LIGOScientific:2020kqk}%
  \BibitemOpen
  \bibfield  {author} {\bibinfo {author} {\bibfnamefont {R.}~\bibnamefont
  {Abbott}} \emph {et~al.} (\bibinfo {collaboration} {LIGO Scientific,
  Virgo}),\ }\href {https://doi.org/10.3847/2041-8213/abe949} {\bibfield
  {journal} {\bibinfo  {journal} {Astrophys. J. Lett.}\ }\textbf {\bibinfo
  {volume} {913}},\ \bibinfo {pages} {L7} (\bibinfo {year} {2021})},\ \Eprint
  {https://arxiv.org/abs/2010.14533} {arXiv:2010.14533 [astro-ph.HE]}
  \BibitemShut {NoStop}%
\bibitem [{\citenamefont {Cutler}\ and\ \citenamefont
  {Flanagan}(1994)}]{Cutler:1994ys}%
  \BibitemOpen
  \bibfield  {author} {\bibinfo {author} {\bibfnamefont {C.}~\bibnamefont
  {Cutler}}\ and\ \bibinfo {author} {\bibfnamefont {E.~E.}\ \bibnamefont
  {Flanagan}},\ }\href {https://doi.org/10.1103/PhysRevD.49.2658} {\bibfield
  {journal} {\bibinfo  {journal} {Phys. Rev. D}\ }\textbf {\bibinfo {volume}
  {49}},\ \bibinfo {pages} {2658} (\bibinfo {year} {1994})},\ \Eprint
  {https://arxiv.org/abs/gr-qc/9402014} {arXiv:gr-qc/9402014} \BibitemShut
  {NoStop}%
\bibitem [{\citenamefont {Poisson}\ and\ \citenamefont
  {Will}(1995)}]{Poisson:1995}%
  \BibitemOpen
  \bibfield  {author} {\bibinfo {author} {\bibfnamefont {E.}~\bibnamefont
  {Poisson}}\ and\ \bibinfo {author} {\bibfnamefont {C.~M.}\ \bibnamefont
  {Will}},\ }\href {https://doi.org/10.1103/PhysRevD.52.848} {\bibfield
  {journal} {\bibinfo  {journal} {Phys. Rev. D}\ }\textbf {\bibinfo {volume}
  {52}},\ \bibinfo {pages} {848} (\bibinfo {year} {1995})},\ \Eprint
  {https://arxiv.org/abs/gr-qc/9502040} {arXiv:gr-qc/9502040} \BibitemShut
  {NoStop}%
\bibitem [{\citenamefont {Pratten}\ \emph {et~al.}(2021)\citenamefont {Pratten}
  \emph {et~al.}}]{Pratten:2020ceb}%
  \BibitemOpen
  \bibfield  {author} {\bibinfo {author} {\bibfnamefont {G.}~\bibnamefont
  {Pratten}} \emph {et~al.},\ }\href
  {https://doi.org/10.1103/PhysRevD.103.104056} {\bibfield  {journal} {\bibinfo
   {journal} {Phys. Rev. D}\ }\textbf {\bibinfo {volume} {103}},\ \bibinfo
  {pages} {104056} (\bibinfo {year} {2021})},\ \Eprint
  {https://arxiv.org/abs/2004.06503} {arXiv:2004.06503 [gr-qc]} \BibitemShut
  {NoStop}%
\bibitem [{\citenamefont {Dietrich}\ \emph {et~al.}(2019)\citenamefont
  {Dietrich}, \citenamefont {Samajdar}, \citenamefont {Khan}, \citenamefont
  {Johnson-McDaniel}, \citenamefont {Dudi},\ and\ \citenamefont
  {Tichy}}]{Dietrich:2019kaq}%
  \BibitemOpen
  \bibfield  {author} {\bibinfo {author} {\bibfnamefont {T.}~\bibnamefont
  {Dietrich}}, \bibinfo {author} {\bibfnamefont {A.}~\bibnamefont {Samajdar}},
  \bibinfo {author} {\bibfnamefont {S.}~\bibnamefont {Khan}}, \bibinfo {author}
  {\bibfnamefont {N.~K.}\ \bibnamefont {Johnson-McDaniel}}, \bibinfo {author}
  {\bibfnamefont {R.}~\bibnamefont {Dudi}},\ and\ \bibinfo {author}
  {\bibfnamefont {W.}~\bibnamefont {Tichy}},\ }\href
  {https://doi.org/10.1103/PhysRevD.100.044003} {\bibfield  {journal} {\bibinfo
   {journal} {Phys. Rev. D}\ }\textbf {\bibinfo {volume} {100}},\ \bibinfo
  {pages} {044003} (\bibinfo {year} {2019})},\ \Eprint
  {https://arxiv.org/abs/1905.06011} {arXiv:1905.06011 [gr-qc]} \BibitemShut
  {NoStop}%
\bibitem [{\citenamefont {Ferri}\ \emph {et~al.}(2025)\citenamefont {Ferri},
  \citenamefont {Tashiro}, \citenamefont {Abramo}, \citenamefont {Matos},
  \citenamefont {Quartin},\ and\ \citenamefont {Sturani}}]{Ferri:2024amc}%
  \BibitemOpen
  \bibfield  {author} {\bibinfo {author} {\bibfnamefont {J.}~\bibnamefont
  {Ferri}}, \bibinfo {author} {\bibfnamefont {I.~L.}\ \bibnamefont {Tashiro}},
  \bibinfo {author} {\bibfnamefont {L.~R.}\ \bibnamefont {Abramo}}, \bibinfo
  {author} {\bibfnamefont {I.}~\bibnamefont {Matos}}, \bibinfo {author}
  {\bibfnamefont {M.}~\bibnamefont {Quartin}},\ and\ \bibinfo {author}
  {\bibfnamefont {R.}~\bibnamefont {Sturani}},\ }\href
  {https://doi.org/10.1088/1475-7516/2025/04/008} {\bibfield  {journal}
  {\bibinfo  {journal} {JCAP}\ }\textbf {\bibinfo {volume} {04}},\ \bibinfo
  {pages} {008}},\ \Eprint {https://arxiv.org/abs/2412.00202} {arXiv:2412.00202
  [astro-ph.CO]} \BibitemShut {NoStop}%
\bibitem [{\citenamefont {Afroz}\ and\ \citenamefont
  {Mukherjee}(2024)}]{Afroz:2024joi}%
  \BibitemOpen
  \bibfield  {author} {\bibinfo {author} {\bibfnamefont {S.}~\bibnamefont
  {Afroz}}\ and\ \bibinfo {author} {\bibfnamefont {S.}~\bibnamefont
  {Mukherjee}},\ }\href {https://doi.org/10.1093/mnras/stae2139} {\bibfield
  {journal} {\bibinfo  {journal} {Mon. Not. Roy. Astron. Soc.}\ }\textbf
  {\bibinfo {volume} {534}},\ \bibinfo {pages} {1283} (\bibinfo {year}
  {2024})},\ \Eprint {https://arxiv.org/abs/2407.09262} {arXiv:2407.09262
  [astro-ph.CO]} \BibitemShut {NoStop}%
\bibitem [{\citenamefont {Ezquiaga}\ and\ \citenamefont
  {Holz}(2022)}]{Ezquiaga:2022zkx}%
  \BibitemOpen
  \bibfield  {author} {\bibinfo {author} {\bibfnamefont {J.~M.}\ \bibnamefont
  {Ezquiaga}}\ and\ \bibinfo {author} {\bibfnamefont {D.~E.}\ \bibnamefont
  {Holz}},\ }\href {https://doi.org/10.1103/PhysRevLett.129.061102} {\bibfield
  {journal} {\bibinfo  {journal} {Phys. Rev. Lett.}\ }\textbf {\bibinfo
  {volume} {129}},\ \bibinfo {pages} {061102} (\bibinfo {year} {2022})},\
  \Eprint {https://arxiv.org/abs/2202.08240} {arXiv:2202.08240 [astro-ph.CO]}
  \BibitemShut {NoStop}%
\bibitem [{\citenamefont {Zhang}\ \emph {et~al.}(2025)\citenamefont {Zhang},
  \citenamefont {Yu}, \citenamefont {Li}, \citenamefont {Kazempour},
  \citenamefont {Li},\ and\ \citenamefont {Sun}}]{Zhang:2025eeh}%
  \BibitemOpen
  \bibfield  {author} {\bibinfo {author} {\bibfnamefont {X.}~\bibnamefont
  {Zhang}}, \bibinfo {author} {\bibfnamefont {C.}~\bibnamefont {Yu}}, \bibinfo
  {author} {\bibfnamefont {H.}~\bibnamefont {Li}}, \bibinfo {author}
  {\bibfnamefont {S.}~\bibnamefont {Kazempour}}, \bibinfo {author}
  {\bibfnamefont {M.}~\bibnamefont {Li}},\ and\ \bibinfo {author}
  {\bibfnamefont {S.}~\bibnamefont {Sun}},\ }\href@noop {} {\bibinfo {title}
  {{The new generation lunar gravitational wave detectors: sky map resolution
  and joint analysis}}} (\bibinfo {year} {2025}),\ \Eprint
  {https://arxiv.org/abs/2512.23556} {arXiv:2512.23556 [gr-qc]} \BibitemShut
  {NoStop}%
\bibitem [{\citenamefont {Song}\ \emph {et~al.}(2026)\citenamefont {Song},
  \citenamefont {Dong}, \citenamefont {Jin}, \citenamefont {Xiao},
  \citenamefont {Zhang},\ and\ \citenamefont {Zhang}}]{Song:2026kii}%
  \BibitemOpen
  \bibfield  {author} {\bibinfo {author} {\bibfnamefont {J.-Y.}\ \bibnamefont
  {Song}}, \bibinfo {author} {\bibfnamefont {Y.-Y.}\ \bibnamefont {Dong}},
  \bibinfo {author} {\bibfnamefont {S.-J.}\ \bibnamefont {Jin}}, \bibinfo
  {author} {\bibfnamefont {S.-R.}\ \bibnamefont {Xiao}}, \bibinfo {author}
  {\bibfnamefont {J.-F.}\ \bibnamefont {Zhang}},\ and\ \bibinfo {author}
  {\bibfnamefont {X.}~\bibnamefont {Zhang}},\ }\href@noop {} {\bibinfo {title}
  {{Three-band dark-siren cosmology with intermediate-mass black hole binaries:
  synergy of Taiji, LGWA, and Einstein Telescope}}} (\bibinfo {year} {2026}),\
  \Eprint {https://arxiv.org/abs/2603.13080} {arXiv:2603.13080 [astro-ph.CO]}
  \BibitemShut {NoStop}%
\bibitem [{\citenamefont {de~Mattia}\ \emph {et~al.}(2021)\citenamefont
  {de~Mattia} \emph {et~al.}}]{eBOSS:2020fvk}%
  \BibitemOpen
  \bibfield  {author} {\bibinfo {author} {\bibfnamefont {A.}~\bibnamefont
  {de~Mattia}} \emph {et~al.} (\bibinfo {collaboration} {eBOSS}),\ }\href
  {https://doi.org/10.1093/mnras/staa3891} {\bibfield  {journal} {\bibinfo
  {journal} {Mon. Not. Roy. Astron. Soc.}\ }\textbf {\bibinfo {volume} {501}},\
  \bibinfo {pages} {5616} (\bibinfo {year} {2021})},\ \Eprint
  {https://arxiv.org/abs/2007.09008} {arXiv:2007.09008 [astro-ph.CO]}
  \BibitemShut {NoStop}%
\bibitem [{\citenamefont {Tamanini}\ \emph {et~al.}(2016)\citenamefont
  {Tamanini}, \citenamefont {Caprini}, \citenamefont {Barausse}, \citenamefont
  {Sesana}, \citenamefont {Klein},\ and\ \citenamefont
  {Petiteau}}]{Tamanini:2016zlh}%
  \BibitemOpen
  \bibfield  {author} {\bibinfo {author} {\bibfnamefont {N.}~\bibnamefont
  {Tamanini}}, \bibinfo {author} {\bibfnamefont {C.}~\bibnamefont {Caprini}},
  \bibinfo {author} {\bibfnamefont {E.}~\bibnamefont {Barausse}}, \bibinfo
  {author} {\bibfnamefont {A.}~\bibnamefont {Sesana}}, \bibinfo {author}
  {\bibfnamefont {A.}~\bibnamefont {Klein}},\ and\ \bibinfo {author}
  {\bibfnamefont {A.}~\bibnamefont {Petiteau}},\ }\href
  {https://doi.org/10.1088/1475-7516/2016/04/002} {\bibfield  {journal}
  {\bibinfo  {journal} {JCAP}\ }\textbf {\bibinfo {volume} {04}},\ \bibinfo
  {pages} {002}},\ \Eprint {https://arxiv.org/abs/1601.07112} {arXiv:1601.07112
  [astro-ph.CO]} \BibitemShut {NoStop}%
\bibitem [{\citenamefont {Sun}\ \emph {et~al.}(2020)\citenamefont {Sun} \emph
  {et~al.}}]{Sun:2020wke}%
  \BibitemOpen
  \bibfield  {author} {\bibinfo {author} {\bibfnamefont {L.}~\bibnamefont
  {Sun}} \emph {et~al.},\ }\href {https://doi.org/10.1088/1361-6382/abb14e}
  {\bibfield  {journal} {\bibinfo  {journal} {Class. Quant. Grav.}\ }\textbf
  {\bibinfo {volume} {37}},\ \bibinfo {pages} {225008} (\bibinfo {year}
  {2020})},\ \Eprint {https://arxiv.org/abs/2005.02531} {arXiv:2005.02531
  [astro-ph.IM]} \BibitemShut {NoStop}%
\bibitem [{\citenamefont {Dhani}\ \emph
  {et~al.}(2025{\natexlab{a}})\citenamefont {Dhani}, \citenamefont
  {V{\"o}lkel}, \citenamefont {Buonanno}, \citenamefont {Estelles},
  \citenamefont {Gair}, \citenamefont {Pfeiffer}, \citenamefont {Pompili},\
  and\ \citenamefont {Toubiana}}]{Dhani:2024jja}%
  \BibitemOpen
  \bibfield  {author} {\bibinfo {author} {\bibfnamefont {A.}~\bibnamefont
  {Dhani}}, \bibinfo {author} {\bibfnamefont {S.~H.}\ \bibnamefont
  {V{\"o}lkel}}, \bibinfo {author} {\bibfnamefont {A.}~\bibnamefont
  {Buonanno}}, \bibinfo {author} {\bibfnamefont {H.}~\bibnamefont {Estelles}},
  \bibinfo {author} {\bibfnamefont {J.}~\bibnamefont {Gair}}, \bibinfo {author}
  {\bibfnamefont {H.~P.}\ \bibnamefont {Pfeiffer}}, \bibinfo {author}
  {\bibfnamefont {L.}~\bibnamefont {Pompili}},\ and\ \bibinfo {author}
  {\bibfnamefont {A.}~\bibnamefont {Toubiana}},\ }\href
  {https://doi.org/10.1103/5pks-qz6b} {\bibfield  {journal} {\bibinfo
  {journal} {Phys. Rev. X}\ }\textbf {\bibinfo {volume} {15}},\ \bibinfo
  {pages} {031036} (\bibinfo {year} {2025}{\natexlab{a}})},\ \Eprint
  {https://arxiv.org/abs/2404.05811} {arXiv:2404.05811 [gr-qc]} \BibitemShut
  {NoStop}%
\bibitem [{\citenamefont {Dhani}\ \emph
  {et~al.}(2025{\natexlab{b}})\citenamefont {Dhani}, \citenamefont {Gair},\
  and\ \citenamefont {Buonanno}}]{Dhani:2025xgt}%
  \BibitemOpen
  \bibfield  {author} {\bibinfo {author} {\bibfnamefont {A.}~\bibnamefont
  {Dhani}}, \bibinfo {author} {\bibfnamefont {J.}~\bibnamefont {Gair}},\ and\
  \bibinfo {author} {\bibfnamefont {A.}~\bibnamefont {Buonanno}},\ }\href@noop
  {} {\bibinfo {title} {{The fault in our sirens: Hierarchical diagnosis of
  waveform systematics in Hubble-Lema{\^\i}tre constant measurements}}}
  (\bibinfo {year} {2025}{\natexlab{b}}),\ \Eprint
  {https://arxiv.org/abs/2507.11278} {arXiv:2507.11278 [gr-qc]} \BibitemShut
  {NoStop}%
\bibitem [{\citenamefont {Saleem}\ \emph {et~al.}(2022)\citenamefont {Saleem}
  \emph {et~al.}}]{Saleem:2021iwi}%
  \BibitemOpen
  \bibfield  {author} {\bibinfo {author} {\bibfnamefont {M.}~\bibnamefont
  {Saleem}} \emph {et~al.},\ }\href {https://doi.org/10.1088/1361-6382/ac3b99}
  {\bibfield  {journal} {\bibinfo  {journal} {Class. Quant. Grav.}\ }\textbf
  {\bibinfo {volume} {39}},\ \bibinfo {pages} {025004} (\bibinfo {year}
  {2022})},\ \Eprint {https://arxiv.org/abs/2105.01716} {arXiv:2105.01716
  [gr-qc]} \BibitemShut {NoStop}%
\bibitem [{\citenamefont {Luo}\ \emph {et~al.}(2016)\citenamefont {Luo} \emph
  {et~al.}}]{TianQin:2015yph}%
  \BibitemOpen
  \bibfield  {author} {\bibinfo {author} {\bibfnamefont {J.}~\bibnamefont
  {Luo}} \emph {et~al.} (\bibinfo {collaboration} {TianQin}),\ }\href
  {https://doi.org/10.1088/0264-9381/33/3/035010} {\bibfield  {journal}
  {\bibinfo  {journal} {Class. Quant. Grav.}\ }\textbf {\bibinfo {volume}
  {33}},\ \bibinfo {pages} {035010} (\bibinfo {year} {2016})},\ \Eprint
  {https://arxiv.org/abs/1512.02076} {arXiv:1512.02076 [astro-ph.IM]}
  \BibitemShut {NoStop}%
\bibitem [{\citenamefont {Hu}\ and\ \citenamefont {Wu}(2017)}]{Hu:2017mde}%
  \BibitemOpen
  \bibfield  {author} {\bibinfo {author} {\bibfnamefont {W.-R.}\ \bibnamefont
  {Hu}}\ and\ \bibinfo {author} {\bibfnamefont {Y.-L.}\ \bibnamefont {Wu}},\
  }\href {https://doi.org/10.1093/nsr/nwx116} {\bibfield  {journal} {\bibinfo
  {journal} {Natl. Sci. Rev.}\ }\textbf {\bibinfo {volume} {4}},\ \bibinfo
  {pages} {685} (\bibinfo {year} {2017})}\BibitemShut {NoStop}%
\bibitem [{\citenamefont {Kawamura}\ \emph {et~al.}(2021)\citenamefont
  {Kawamura} \emph {et~al.}}]{Kawamura:2020pcg}%
  \BibitemOpen
  \bibfield  {author} {\bibinfo {author} {\bibfnamefont {S.}~\bibnamefont
  {Kawamura}} \emph {et~al.},\ }\href {https://doi.org/10.1093/ptep/ptab019}
  {\bibfield  {journal} {\bibinfo  {journal} {PTEP}\ }\textbf {\bibinfo
  {volume} {2021}},\ \bibinfo {pages} {05A105} (\bibinfo {year} {2021})},\
  \Eprint {https://arxiv.org/abs/2006.13545} {arXiv:2006.13545 [gr-qc]}
  \BibitemShut {NoStop}%
\bibitem [{\citenamefont {Hunter}(2007)}]{Hunter:2007}%
  \BibitemOpen
  \bibfield  {author} {\bibinfo {author} {\bibfnamefont {J.~D.}\ \bibnamefont
  {Hunter}},\ }\href {https://doi.org/10.1109/MCSE.2007.55} {\bibfield
  {journal} {\bibinfo  {journal} {Computing in Science \& Engineering}\
  }\textbf {\bibinfo {volume} {9}},\ \bibinfo {pages} {90} (\bibinfo {year}
  {2007})}\BibitemShut {NoStop}%
\bibitem [{\citenamefont {Harris}\ \emph {et~al.}(2020)\citenamefont {Harris}
  \emph {et~al.}}]{Harris:2020xlr}%
  \BibitemOpen
  \bibfield  {author} {\bibinfo {author} {\bibfnamefont {C.~R.}\ \bibnamefont
  {Harris}} \emph {et~al.},\ }\href {https://doi.org/10.1038/s41586-020-2649-2}
  {\bibfield  {journal} {\bibinfo  {journal} {Nature}\ }\textbf {\bibinfo
  {volume} {585}},\ \bibinfo {pages} {357} (\bibinfo {year} {2020})},\ \Eprint
  {https://arxiv.org/abs/2006.10256} {arXiv:2006.10256 [cs.MS]} \BibitemShut
  {NoStop}%
\bibitem [{\citenamefont {pandas~development team}(2020)}]{reback2020pandas}%
  \BibitemOpen
  \bibfield  {author} {\bibinfo {author} {\bibfnamefont {T.}~\bibnamefont
  {pandas~development team}},\ }\href {https://doi.org/10.5281/zenodo.3509134}
  {\bibinfo {title} {pandas-dev/pandas: Pandas}} (\bibinfo {year}
  {2020})\BibitemShut {NoStop}%
\bibitem [{\citenamefont {{W}es
  {M}c{K}inney}(2010)}]{mckinney-proc-scipy-2010}%
  \BibitemOpen
  \bibfield  {author} {\bibinfo {author} {\bibnamefont {{W}es {M}c{K}inney}},\
  }in\ \href {https://doi.org/10.25080/Majora-92bf1922-00a} {\emph {\bibinfo
  {booktitle} {{P}roceedings of the 9th {P}ython in {S}cience {C}onference}}},\
  \bibinfo {editor} {edited by\ \bibinfo {editor} {\bibnamefont {{S}t\'efan
  van~der {W}alt}}\ and\ \bibinfo {editor} {\bibnamefont {{J}arrod
  {M}illman}}}\ (\bibinfo {year} {2010})\ pp.\ \bibinfo {pages} {56 --
  61}\BibitemShut {NoStop}%
\bibitem [{\citenamefont {{Astropy Collaboration}}\ \emph
  {et~al.}(2022)\citenamefont {{Astropy Collaboration}}, \citenamefont
  {{Price-Whelan}} \emph {et~al.}}]{astropy:2022}%
  \BibitemOpen
  \bibfield  {author} {\bibinfo {author} {\bibnamefont {{Astropy
  Collaboration}}}, \bibinfo {author} {\bibfnamefont {A.~M.}\ \bibnamefont
  {{Price-Whelan}}}, \emph {et~al.},\ }\href
  {https://doi.org/10.3847/1538-4357/ac7c74} {\bibfield  {journal} {\bibinfo
  {journal} {\apj}\ }\textbf {\bibinfo {volume} {935}},\ \bibinfo {eid} {167}
  (\bibinfo {year} {2022})},\ \Eprint {https://arxiv.org/abs/2206.14220}
  {arXiv:2206.14220 [astro-ph.IM]} \BibitemShut {NoStop}%
\bibitem [{\citenamefont {{Astropy Collaboration}}\ \emph
  {et~al.}(2018)\citenamefont {{Astropy Collaboration}}, \citenamefont
  {{Price-Whelan}} \emph {et~al.}}]{astropy:2018}%
  \BibitemOpen
  \bibfield  {author} {\bibinfo {author} {\bibnamefont {{Astropy
  Collaboration}}}, \bibinfo {author} {\bibfnamefont {A.~M.}\ \bibnamefont
  {{Price-Whelan}}}, \emph {et~al.},\ }\href
  {https://doi.org/10.3847/1538-3881/aabc4f} {\bibfield  {journal} {\bibinfo
  {journal} {\aj}\ }\textbf {\bibinfo {volume} {156}},\ \bibinfo {eid} {123}
  (\bibinfo {year} {2018})},\ \Eprint {https://arxiv.org/abs/1801.02634}
  {arXiv:1801.02634 [astro-ph.IM]} \BibitemShut {NoStop}%
\bibitem [{\citenamefont {{Astropy Collaboration}}\ \emph
  {et~al.}(2013)\citenamefont {{Astropy Collaboration}}, \citenamefont
  {{Robitaille}} \emph {et~al.}}]{astropy:2013}%
  \BibitemOpen
  \bibfield  {author} {\bibinfo {author} {\bibnamefont {{Astropy
  Collaboration}}}, \bibinfo {author} {\bibfnamefont {T.~P.}\ \bibnamefont
  {{Robitaille}}}, \emph {et~al.},\ }\href
  {https://doi.org/10.1051/0004-6361/201322068} {\bibfield  {journal} {\bibinfo
   {journal} {\aap}\ }\textbf {\bibinfo {volume} {558}},\ \bibinfo {eid} {A33}
  (\bibinfo {year} {2013})},\ \Eprint {https://arxiv.org/abs/1307.6212}
  {arXiv:1307.6212 [astro-ph.IM]} \BibitemShut {NoStop}%
\bibitem [{\citenamefont {Virtanen}\ \emph {et~al.}(2020)\citenamefont
  {Virtanen}, \citenamefont {Gommers}, \citenamefont {Oliphant}, \citenamefont
  {Haberland}, \citenamefont {Reddy}, \citenamefont {Cournapeau}, \citenamefont
  {Burovski}, \citenamefont {Peterson}, \citenamefont {Weckesser},
  \citenamefont {Bright}, \citenamefont {{van der Walt}}, \citenamefont
  {Brett}, \citenamefont {Wilson}, \citenamefont {Millman}, \citenamefont
  {Mayorov}, \citenamefont {Nelson}, \citenamefont {Jones}, \citenamefont
  {Kern}, \citenamefont {Larson}, \citenamefont {Carey}, \citenamefont {Polat},
  \citenamefont {Feng}, \citenamefont {Moore}, \citenamefont {{VanderPlas}},
  \citenamefont {Laxalde}, \citenamefont {Perktold}, \citenamefont {Cimrman},
  \citenamefont {Henriksen}, \citenamefont {Quintero}, \citenamefont {Harris},
  \citenamefont {Archibald}, \citenamefont {Ribeiro}, \citenamefont
  {Pedregosa}, \citenamefont {{van Mulbregt}},\ and\ \citenamefont {{SciPy 1.0
  Contributors}}}]{2020SciPy-NMeth}%
  \BibitemOpen
  \bibfield  {author} {\bibinfo {author} {\bibfnamefont {P.}~\bibnamefont
  {Virtanen}}, \bibinfo {author} {\bibfnamefont {R.}~\bibnamefont {Gommers}},
  \bibinfo {author} {\bibfnamefont {T.~E.}\ \bibnamefont {Oliphant}}, \bibinfo
  {author} {\bibfnamefont {M.}~\bibnamefont {Haberland}}, \bibinfo {author}
  {\bibfnamefont {T.}~\bibnamefont {Reddy}}, \bibinfo {author} {\bibfnamefont
  {D.}~\bibnamefont {Cournapeau}}, \bibinfo {author} {\bibfnamefont
  {E.}~\bibnamefont {Burovski}}, \bibinfo {author} {\bibfnamefont
  {P.}~\bibnamefont {Peterson}}, \bibinfo {author} {\bibfnamefont
  {W.}~\bibnamefont {Weckesser}}, \bibinfo {author} {\bibfnamefont
  {J.}~\bibnamefont {Bright}}, \bibinfo {author} {\bibfnamefont {S.~J.}\
  \bibnamefont {{van der Walt}}}, \bibinfo {author} {\bibfnamefont
  {M.}~\bibnamefont {Brett}}, \bibinfo {author} {\bibfnamefont
  {J.}~\bibnamefont {Wilson}}, \bibinfo {author} {\bibfnamefont {K.~J.}\
  \bibnamefont {Millman}}, \bibinfo {author} {\bibfnamefont {N.}~\bibnamefont
  {Mayorov}}, \bibinfo {author} {\bibfnamefont {A.~R.~J.}\ \bibnamefont
  {Nelson}}, \bibinfo {author} {\bibfnamefont {E.}~\bibnamefont {Jones}},
  \bibinfo {author} {\bibfnamefont {R.}~\bibnamefont {Kern}}, \bibinfo {author}
  {\bibfnamefont {E.}~\bibnamefont {Larson}}, \bibinfo {author} {\bibfnamefont
  {C.~J.}\ \bibnamefont {Carey}}, \bibinfo {author} {\bibfnamefont
  {{\.I}.}~\bibnamefont {Polat}}, \bibinfo {author} {\bibfnamefont
  {Y.}~\bibnamefont {Feng}}, \bibinfo {author} {\bibfnamefont {E.~W.}\
  \bibnamefont {Moore}}, \bibinfo {author} {\bibfnamefont {J.}~\bibnamefont
  {{VanderPlas}}}, \bibinfo {author} {\bibfnamefont {D.}~\bibnamefont
  {Laxalde}}, \bibinfo {author} {\bibfnamefont {J.}~\bibnamefont {Perktold}},
  \bibinfo {author} {\bibfnamefont {R.}~\bibnamefont {Cimrman}}, \bibinfo
  {author} {\bibfnamefont {I.}~\bibnamefont {Henriksen}}, \bibinfo {author}
  {\bibfnamefont {E.~A.}\ \bibnamefont {Quintero}}, \bibinfo {author}
  {\bibfnamefont {C.~R.}\ \bibnamefont {Harris}}, \bibinfo {author}
  {\bibfnamefont {A.~M.}\ \bibnamefont {Archibald}}, \bibinfo {author}
  {\bibfnamefont {A.~H.}\ \bibnamefont {Ribeiro}}, \bibinfo {author}
  {\bibfnamefont {F.}~\bibnamefont {Pedregosa}}, \bibinfo {author}
  {\bibfnamefont {P.}~\bibnamefont {{van Mulbregt}}},\ and\ \bibinfo {author}
  {\bibnamefont {{SciPy 1.0 Contributors}}},\ }\href
  {https://doi.org/10.1038/s41592-019-0686-2} {\bibfield  {journal} {\bibinfo
  {journal} {Nature Methods}\ }\textbf {\bibinfo {volume} {17}},\ \bibinfo
  {pages} {261} (\bibinfo {year} {2020})}\BibitemShut {NoStop}%
\bibitem [{\citenamefont {Collette}(2013)}]{collette_python_hdf5_2014}%
  \BibitemOpen
  \bibfield  {author} {\bibinfo {author} {\bibfnamefont {A.}~\bibnamefont
  {Collette}},\ }\href@noop {} {\emph {\bibinfo {title} {Python and HDF5}}}\
  (\bibinfo  {publisher} {O'Reilly},\ \bibinfo {year} {2013})\BibitemShut
  {NoStop}%
\end{thebibliography}%

\end{document}